\documentclass[iop]{emulateapj} 
\usepackage{apjfonts}
\usepackage{psfig}
\usepackage{mathrsfs}
\usepackage{rotating}
\usepackage{longtable}
\newcommand{\ibvs}{IBVS}
\newcommand{\rmxaa}{RevMexAA}

\shorttitle{Variable Stars in M62}
\shortauthors{Contreras et al.}
\slugcomment{AJ, in press}

\begin{document}

 \title{Time-Series Photometry of Globular Clusters: M62 (NGC~6266), 
 \\the Most RR Lyrae-Rich Globular Cluster in the Galaxy?}
 
\author{R. Contreras,\altaffilmark{1,2} M. Catelan,\altaffilmark{2} H. A. Smith,\altaffilmark{3}
B. J. Pritzl,\altaffilmark{4} J. Borissova,\altaffilmark{5} C. A. Kuehn\altaffilmark{3}}

\altaffiltext{1}{INAF--Osservatorio Astronomico di Bologna, via Ranzani 1,
40127, Bologna, Italy} %\email{rodrigo.contreras@oabo.inaf.it}
\altaffiltext{2}{Pontificia Universidad Cat\'olica de Chile, Departamento de 
       Astronom\'\i a y Astrof\'\i sica, Av. Vicu\~{n}a Mackenna 4860, 
      782-0436 Macul, Santiago, Chile}
\altaffiltext{3}{Department of Physics and Astronomy, Michigan State University, East Lansing, MI 48824}
\altaffiltext{4}{Department of Physics and Astronomy, University of Wisconsin, Oshkosh, WI 54901}
\altaffiltext{5}{Departamento de F\'{i}sica y Astronom\'{i}a, Facultad de Ciencias, Universidad de Valpara\'{i}so, Ave. Gran Breta\~na 1111, Playa Ancha, Casilla 5030, Valpara\'{i}so, Chile}


\begin{abstract}
We present new time-series CCD photometry, in the $B$ and $V$ bands,
for the moderately metal-rich 
(${\rm [Fe/H]} \simeq -1.3$)
Galactic globular cluster M62 (NGC~6266).
The present dataset is the largest obtained so far for this cluster,
and consists of 168 
images per filter, obtained with the Warsaw 1.3m telescope at the Las
Campanas Observatory (LCO) and the 1.3m telescope of the Cerro Tololo
Inter-American Observatory (CTIO), in two separate runs over the time span of
three months. The procedure adopted to detect the variable stars was
the optimal image subtraction method (ISIS v2.2), as implemented by
Alard. The photometry was performed using both ISIS and Stetson's 
DAOPHOT/ALLFRAME package. We have identified 245
variable stars in the cluster fields that have been analyzed 
so far, of which 179 are new discoveries. 
Of these variables, 133 are fundamental mode RR Lyrae stars (RRab),  
76 are first overtone (RRc) pulsators, 4 are type II 
Cepheids, 25 are long-period variables (LPV), 1 is an eclipsing
binary, and 6 are not yet well classified. Such a
large number of RR
Lyrae stars places M62 among the top two most RR Lyrae-rich 
(in the sense of total number of RR Lyrae stars present) 
globular clusters known in the Galaxy, second only to M3 (NGC~5272) 
with a total of 230 known RR Lyrae stars. 
Since this study covers most but not all of the cluster area, 
it is not unlikely that M62 is in fact the most RR Lyrae-rich
globular cluster in the Galaxy. In like vein, thanks to the time coverage of 
our datasets, we were also able to detect the largest sample of LPV's 
known so far in a Galactic globular cluster. 

We analyze a variety of Oosterhoff type indicators for the cluster, including 
mean periods, period distribution, Bailey diagrams, and Fourier decomposition 
parameters (as well as the physical parameters derived therefrom). All of these 
indicators clearly show that M62 is an Oosterhoff type I system. This is in good 
agreement with the moderately high metallicity of the cluster, in spite of its 
predominantly blue horizontal branch morphology~-- which is more typical of 
Oosterhoff type II systems. We thus conclude that metallicity plays a key role 
in defining Oosterhoff type. Finally, based on an application of the ``$A$-method,'' 
we conclude that the cluster RR Lyrae stars have a similar He abundance as M3, 
although more work on the temperatures of the M62 RR Lyrae is needed before
this result can be conclusively established.  

\end{abstract}



\section{Introduction}
By the early 1990's, it was widely perceived that ``most variable stars
that belong to Galactic globular clusters have by now been
discovered'' \citep*{nsea91}. Indeed, \citeauthor{nsea91} 
estimated that only a few percent of the total population of RR Lyrae 
variable stars remained to be discovered in globular clusters. 
However, most of the pre-1990 studies based their results on 
photographic photometry, which in many cases appears not to have been precise
enough to detect small-amplitude variables. On the other hand, new techniques, 
based on image subtraction algorithms, have been developed in the 
last years, which are capable of quickly, efficiently and automatically 
detecting star variations even in the most crowded fields 
\citep[e.g.,][]{ca00,db08}. In fact, several studies have reported substantial 
increases in the reported globular cluster RR Lyrae populations using these 
techniques \citep*[e.g.,][]{jkea01,mcea04,mzea09}. It seems that, contrary 
to what was previously thought, the sample of RR Lyrae variables identified 
so far in Galactic globular clusters is significantly incomplete, thus 
rendering further analyses, based on high-quality CCD observations and 
image subtraction techniques, well worth the while.

NGC~6266 (M62) is a high-density ($\log\rho_{c} = 5.34 \, L_{\odot} {\rm pc}^3$), 
highly reddened [$E(\bv) = 0.47$] cluster, and is also one of the most massive in 
our galaxy, with $M_{V} = -9.19$ \citep{wh96}. Located at just 1.7~kpc from the 
Galactic center, it has also been classified as a possible post-core collapse
globular cluster by \citet*{stea93,stea95}~-- a possibility which however was not 
confirmed by \citet{gbea06} in their study of the radial density profile of 
the cluster. Also worth mentioning is the fact that the cluster currently ranks
fifth in the number of millisecond pulsars \citep[][ and references therein]{gcea08}.

The morphology of the cluster's horizontal branch (HB) shows a prominent blue
component, in addition to a very extended blue tail, reaching down to at least 
the main sequence turnoff level \citep[e.g.,][]{vcea87,gpea02}. 
The cluster is also known to be rich in RR Lyrae variables 
\citep[][ and references therein]{ccea01}. 
These features are strikingly similar to those of M15 (NGC~7078), perhaps the  
best known Oosterhoff type II globular cluster. Yet, M62 is more  
metal-rich by about 1~dex, with an ${\rm [Fe/H]} = -1.29$, compared with 
${\rm [Fe/H]} = -2.26$ for M15 \citep{wh96}. Since there is significant 
debate as to whether metallicity or HB morphology play the dominant role 
in defining the Oosterhoff types of globular clusters (e.g., Clement 2000;
Pritzl et al. 2002), M62 can provide a very important constraint on whether
metallicity differences, at a fixed HB morphology, can by themselves change 
the classification of an object from Oosterhoff type II (as in the case 
of M15) to Oosterhoff type I (as is typical of globular clusters with 
metallicity similar to M62's, but with significantly redder HB types). 

The main time-series study of M62 available in the literature so far was 
carried out by \citet{vao59}, where extensive photographic observations 
were presented, and periods derived for a total of 74 (out of 83) stars. More recently, 
\citet{yumea97} provided a list of 43 additional RR Lyrae star candidates 
in the cluster, but without determining their periods nor their detailed 
variability status. As a consequence, 126 variable star candidates have 
been listed for the cluster, 74 of which have known periods 
of variability \citep[][ and references therein]{ccea01}. 

Given the availability of high-quality CCD observations and state-of-the-art 
image subtraction techniques, we expected to find many new variable stars 
in the course of our new time-series study of M62. Indeed, we were able to 
find more than 200 RR Lyrae stars in M62, in addition to a large number of 
long-period variables (LPV's) and 
type II Cepheids (CpII). 

As reported in Contreras et al. (2005), the newly detected 
RR Lyrae stars in M62 offer us important insight into the role played by 
metallicity in defining Oosterhoff type, suggesting that M62 is indeed an 
Oosterhoff type I (OoI) object, in spite of its predominantly blue HB 
morphology, but in accord with its fairly high metallicity. 

The main purpose of the present paper is to provide the new, extensive 
variability data for M62, upon which the preliminary results by Contreras
et al. (2005) were based. We begin in \S2 by describing our data and reduction
procedures. In \S3 we discuss how the variable stars were detected in our data. 
In \S4 we describe the results of a Fourier decomposition analysis of the 
measured light curves. A CMD is produced in \S5, where our approach to account
for the effects of differential reddening is also described. In \S6 we revisit
the Oosterhoff type determination for the cluster. In \S7 we apply the 
``$A$-method'' to study the He abundance in the cluster, and in \S8 we 
provide a summary of our results.

%\begin{deluxetable}{lcccccccccl}
\begin{longtable*}{lcccccccccl}[t]
\tablewidth{0pc} 
\tabletypesize{\scriptsize}
\tablecaption{Photometric Parameters for M62 Variables}
\tablehead{\colhead{ID} & \colhead{RA (J2000)} & \colhead{DEC (J2000)} & \colhead{$P$ (d)} & \colhead{$A_B$} & \colhead{$A_V$} &
\colhead{$\langle B \rangle$} & \colhead{$\langle V \rangle$}&
\colhead{$(\bv)_{\rm mag}$} & \colhead{$E(\bv)$} & \colhead{Type}}
\startdata
V1  & 255.317419  & $-$30.113051  & 0.5047  & ...    & ...	& ...     & ...     & ...    & ...     & RRab\\
V2  & 255.295900  & $-$30.134135  & 10.59   & 1.164  & 1.054	& 14.408  & 13.418  & 1.036  & ...     & CpII\\
V3  & 255.275491  & $-$30.116931  & 0.4913  & ...    & ...	& ...     & ...     & ...    & ...     & RRab\\
V4  & 255.273855  & $-$30.126022  & 0.54113 & 1.382  & 1.068	& 16.889  & 16.109  & 0.820  & 0.511  & RRab\\
V6  & 255.277571  & $-$30.105578  & 0.4920  & 1.363  & 1.082	& 16.861  & 16.121  & 0.772  & 0.442  & RRab\\
V7  & 255.310374  & $-$30.068129  & 0.5640  & 1.235  & 0.946	& 16.799  & 16.044  & 0.785  & 0.462  & RRab\\
V8  & 255.273122  & $-$30.070173  & 0.5327  & 1.315  & 1.009	& 16.779  & 16.034  & 0.780  & 0.462  & RRab\\
V10 & 255.157448  & $-$30.071484  & 0.53259 & 1.531  & 1.208	& 16.616  & 15.948  & 0.710  & 0.387  & RRab\\
V11 & 255.156348  & $-$30.080112  & 0.59823 & 0.994  & 0.749	& 16.651  & 15.921  & 0.746  & 0.369  & RRab\\
V13 & 255.302296  & $-$30.105842  & 0.3033  & ...    & ...	& ...     & ...     & ...    & ...    & RRc\\
V16 & 255.279503  & $-$30.088975  & 0.594   & 1.441  & 1.105	& 16.620  & 15.883  & 0.774  & 0.434  & RRab\\
V17 & 255.296272  & $-$30.086534  & 0.529   & 1.475  & 1.135	& 16.873  & 16.109  & 0.803  & 0.480  & RRab\\
V18 & 255.292820  & $-$30.089474  & 0.5241  & 1.285  & 1.027	& 16.937  & 16.156  & 0.806  & 0.461  & RRab\\
V19 & 255.298855  & $-$30.096809  & 0.5227  & ...    & ...	& ...     & ...     & ...    & ...    & RRab\\
V20 & 255.345754  & $-$30.070601  & 0.47201 & 1.590  & 1.252	& 16.842  & 16.099  & 0.792  & 0.499  & RRab\\
V21 & 255.337973  & $-$30.092659  & 0.4502  & 1.651  & 1.301	& 16.990  & 16.267  & 0.781  & 0.497  & RRab\\
V22 & 255.324073  & $-$30.111547  & 0.5013  & ...    & ...	& ...     & ...     & ...    & ...    & RRab\\
V23 & 255.280529  & $-$30.125258  & 0.44821 & 0.693  & 0.306	& 16.011  & 14.699  & 1.329  & ...      & RRab\\
V24 & 255.323323  & $-$30.125505  & 0.5223  & 1.480  & 1.121	& 17.309  & 16.422  & 0.932  & 0.620  & RRab\\
V25 & 255.353541  & $-$30.134815  & 0.4459  & 1.691  & 1.292	& 17.463  & 16.626  & 0.890  & 0.596  & RRab\\
V26 & 255.245931  & $-$30.198522  & 0.3717  & 1.680  & 1.210	& 14.343  & 13.547  & 0.862  & 0.607  & RRab\\
V27 & 255.302024  & $-$30.131513  & 0.44916 & 1.703  & 1.382	& 17.183  & 16.361  & 0.869  & 0.562  & RRab\\
V28 & 255.353523  & $-$30.109435  & 0.4978  & 1.424  & 1.123	& 17.348  & 16.610  & 0.774  & ...    & RRab\\
V29 & 255.353458  & $-$30.110630  & 0.5653  & 1.551  & 1.184	& 17.238  & 16.396  & 0.888  & 0.574  & RRab\\
V30 & 255.285077  & $-$30.165287  & 0.3041  & 0.565  & 0.422	& 17.308  & 16.502  & 0.813  & ...    & RRc	\\
V31 & 255.289800  & $-$30.154549  & 0.4855  & 1.577  & 1.271	& 17.496  & 16.590  & 0.954  & 0.655  & RRab\\
V32 & 255.304371  & $-$30.152469  & 0.5479  & 1.315  & 0.953	& 17.499  & 16.571  & 0.966  & 0.688  & RRab\\
V33 & 255.300167  & $-$30.147727  & 0.5736  & 1.273  & 0.973	& 17.600  & 16.620  & 1.009  & 0.688  & RRab\\
V34 & 255.284474  & $-$30.116560  & 0.5834  & 1.402  & 1.053	& 16.789  & 15.955  & 0.876  & 0.559  & RRab\\
V35 & 255.267377  & $-$30.108630  & 0.5292  & 1.271  & 0.975	& 16.942  & 16.153  & 0.822  & 0.499  & RRab\\
V36 & 255.290154  & $-$30.080264  & 0.6527  & 0.819  & 0.659	& 16.730  & 15.933  & 0.810  & 0.442  & RRab\\
V37 & 255.286882  & $-$30.113188  & 0.5844  & ...    & ...	& ...     & ...     & ...    & ...    & RRab\\
V38 & 255.297099  & $-$30.127381  & 0.77083 & ...    & ...	& ...     & ...     & ...    & ...    & RRab\\
V39 & 255.264623  & $-$30.098901  & 0.6401  & 0.607  & 0.479	& 16.897  & 16.065  & 0.840  & 0.470  & RRab\\
V40 & 255.264299  & $-$30.102617  & 0.3012  & 0.622  & 0.531	& 16.790  & 16.124  & 0.673  & ...    & RRc\\	
V41 & 255.265477  & $-$30.103925  & 0.5590  & 1.078  & 0.801	& 16.909  & 16.112  & 0.821  & 0.482  & RRab\\
V42 & 255.261784  & $-$30.101339  & 0.2469  & 0.434  & 0.351	& 16.689  & 16.085  & 0.608  & ...    & RRc	\\
V43 & 255.285001  & $-$30.171389  & 0.56356 & 1.159  & 0.859	& 17.377  & 16.481  & 0.925  & 0.832  & RRab\\
V44 & 255.289436  & $-$30.148943  & 0.4456  & 1.471  & 1.122	& 17.631  & 16.724  & 0.954  & 0.681  & RRab\\
V45 & 255.324166  & $-$30.166686  & 0.51688 & ...    & ...	& ...     & ...     & ...    & ...    & RRab\\
V48 & 255.276867  & $-$30.151301  & 0.7432  & 0.934  & 0.725	& 17.108  & 16.185  & 0.940  & 0.571  & RRab\\
V49 & 255.349296  & $-$30.143648  & 0.5434  & 1.239  & 0.962	& 17.414  & 16.524  & 0.918  & 0.599  & RRab\\
V50 & 255.394967  & $-$30.123945  & 0.50264 & ...    & ...	& ...     & ...     & ...    & ...    & RRab\\
V51 & 255.398036  & $-$30.060953  & 0.2618  & 0.617  & ...	& 16.883  & ...     & ...    & ...    & RRc	\\
V52 & 255.328949  & $-$30.165131  & 0.50538 & ...    & ...	& ...     & ...     & ...    & ...    & RRab\\
V53 & 255.268551  & $-$30.143154  & 0.2731  & 0.654  & 0.502	& 17.296  & 16.512  & 0.794  & ...    & RRc	\\
V56 & 255.315460  & $-$30.081934  & 0.5616  & 1.164  & 0.857	& 17.089  & 16.249  & 0.864  & 0.520  & RRab\\
V57 & 255.319997  & $-$30.081330  & 0.5564  & 1.096  & 0.854	& 17.074  & 16.235  & 0.864  & 0.535  & RRab\\
V58 & 255.272245  & $-$30.106241  & 0.481   & 1.159  & 0.996	& 16.832  & ...     & ...    & ...    & RRab\\
V59 & 255.343055  & $-$30.088719  & 0.5791  & 1.148  & 0.871	& 17.093  & 16.250  & 0.870  & 0.550  & RRab\\
V61 & 255.372598  & $-$30.061961  & 0.2660  & 0.655  & ...	& 17.095  & ...     & ...    & ...    & RRc	\\
V62 & 255.380443  & $-$30.085530  & 0.54807 & 1.235  & 0.958	& 17.048  & 16.230  & 0.845  & 0.519  & RRab\\
V63 & 255.338601  & $-$30.143041  & 0.6421  & 0.831  & 0.612	& 17.533  & 16.547  & 1.001  & 0.642  & RRab\\
V64 & 255.361610  & $-$30.195524  & 0.47299 & 0.785  & 0.592	& 17.077  & 16.157  & 0.937  & 0.614  & RRab\\
V65 & 255.275462  & $-$30.077039  & 0.2523  & 0.485  & 0.397	& 16.705  & 16.086  & 0.623  & ...    & RRc \\
V66 & 255.201993  & $-$30.110297  & 0.33383 & 0.570  & 0.420	& 16.676  & 16.011  & 0.674  & ...    & RRc	\\
V69 & 255.343090  & $-$30.084393  & 0.3136  & 0.556  & 0.417	& 16.996  & 16.249  & 0.754  & ...    & RRc	\\
V72 & 255.245694  & $-$30.144019  & 0.468   & 1.422  & 1.053	& 17.034  & 16.286  & 0.795  & 0.529  & RRab\\
V73 & 255.238968  & $-$30.144283  & 1.70    & 1.036  & 0.788	& 16.147  & 15.243  & 0.923  & ...    & CpII\\
V74 & 255.297471  & $-$30.129784  & 0.4667  & ...    & ...	& ...     & ...     & ...    & ...    & RRab\\
V77 & 255.392629  & $-$30.105263  & 0.319   & 0.531  & 0.405	& 17.240  & 16.427  & 0.820  & ...    & RRc	\\
V78 & 255.412564  & $-$30.066713  & 0.62170 & 0.880  & 0.613	& 17.326  & 16.362  & 0.979  & 0.619  & RRab\\
V80 & 255.276084  & $-$30.089919  & 0.5962  & 0.914  & 0.645	& 16.985  & 16.090  & 0.911  & 0.739  & RRab\\
V81 & 255.267912  & $-$30.088047  & 0.5309  & 1.325  & 1.042	& 16.821  & 16.078  & 0.774  & 0.449  & RRab\\
V82 & 255.291579  & $-$30.133857  & 0.5648  & 0.835  & 0.642	& 17.291  & 16.479  & 0.826  & 0.483  & RRab\\
V83 & 255.309339  & $-$30.120442  & 0.4676  & ...    & ...	& ...     & ...     & ...    & ...    & RRab\\
NV84 & 255.271467  & $-$30.134929  & 0.7312  & 0.636  & 0.461	& 17.003  & 16.111  & 0.900  & 0.523  & RRab\\
NV85 & 255.276401  & $-$30.139297  & 0.3196  & 0.484  & 0.432	& 17.189  & 16.365  & 0.826  & ...    & RRc	\\
NV86 & 255.282427  & $-$30.095331  & 0.2913  & 0.565  & 0.433	& 16.795  & 16.112  & 0.690  & ...    & RRc	\\
NV87 & 255.284966  & $-$30.088009  & 0.6424  & 0.401  & 0.296	& 16.876  & 16.049  & 0.830  & 0.449  & RRab\\
NV88 & 255.285771  & $-$30.091009  & 0.5807  & 0.922  & 0.737	& 16.924  & 16.129  & 0.810  & 0.437  & RRab\\
NV89 &  255.288146 & $-$30.130509  & 0.5581  & ...	 & ...	& ...     & ...     & ...    & ...    & RRab\\
NV90 &  255.290712 & $-$30.138332  & 0.3273  & 0.602	 & 0.462 & 17.391  & 16.534  & 0.865  &  ...    & RRc\\
NV91 &  255.293574 & $-$30.132221  & 0.3167  & 0.521	 & 0.421 & 17.090  & 16.337  & 0.758  &  ...    & RRc\\
NV92 &  255.294328 & $-$30.135833  & 0.5256  & ...	 & ...	 & ...     & ...     & ...    &  ...    & RRab\\
NV93 &  255.298447 & $-$30.097374  & 0.552   & 1.205	 & 0.966 & 17.249  & 16.340  & 0.935  &  0.600  & RRab\\
NV94 &  255.306828 & $-$30.090652  & 0.3181  & ...	 & ...	 & ...     & ...     & ...    &  ...    & RRc \\
NV95 &  255.311633 & $-$30.142402  & 0.4941  & 1.618	 & 1.226 & 17.533  & 16.609  & 0.970  &  0.668  & RRab\\
NV96 &  255.315438 & $-$30.140551  & 0.4663  & 1.582	 & 1.257 & 17.452  & 16.615  & 0.880  &  0.572  & RRab\\
NV97 &  255.323047 & $-$30.096211  & 0.5510  & 1.166	 & 0.883 & 17.096  & 16.262  & 0.861  &  0.534  & RRab\\
NV98 &  255.321234 & $-$30.147140  & 0.5620  & 1.113	 & 0.878 & 17.611  & 16.670  & 0.969  &  0.645  & RRab\\
NV99 &  255.323772 & $-$30.144410  & 0.6028  & 0.565	 & 0.410 & 17.661  & 16.644  & 1.023  &  0.653  & RRab\\
NV100 & 255.332677 & $-$30.100980  & 0.2665  & 0.612	 & 0.490 & 17.097  & 16.370  & 0.735  &  ...    & RRc\\	 
NV101 & 255.331020 & $-$30.146758  & 0.3055  & 0.493	 & 0.397 & 17.333  & 16.516  & 0.821  &  ...    & RRc\\	 
NV102 & 255.334042 & $-$30.094895  & 0.6307  & 0.737	 & 0.541 & 17.048  & 16.174  & 0.885  &  0.523  & RRab\\ 
NV103 & 255.331861 & $-$30.150089  & 0.4836  & 0.825	 & 0.611 & 17.588  & 16.655  & 0.949  &  0.602  & RRab\\
NV104 & 255.335589 & $-$30.111854  & 0.6307  & 0.306	 & 0.221 & 17.290  & 16.380  & 0.913  &  ...    & RRc?\\
NV105 & 255.337117 & $-$30.101376  & 0.5205  & 1.517	 & 1.162 & 17.181  & 16.351  & 0.871  &  0.565  & RRab\\
NV106 & 255.342344 & $-$30.102395  & 0.5037  & 1.504	 & 1.153 & 17.175  & 16.359  & 0.857  &  0.549  & RRab\\
NV107 & 255.342654 & $-$30.121002  & 0.5728  & 0.898	 & 0.680 & 17.472  & 16.541  & 0.947  &  0.593  & RRab\\
NV108 & 255.343116 & $-$30.112112  & 0.2988  & 0.539	 & 0.442 & 17.238  & 16.468  & 0.774  &  ...    & RRc\\
NV109 & 255.348015 & $-$30.124260  & 0.6078  & 1.015	 & 0.700 & 17.464  & 16.525  & 0.959  &  0.602  & RRab\\
NV110 & 255.353263 & $-$30.116493  & 0.3354  & 0.584	 & 0.417 & 17.320  & 16.483  & 0.844  &  ...    & RRc\\
NV111 & 255.264451 & $-$30.188755  & 0.2494  & 0.462	 & 0.311 & 17.012  & 16.317  & 0.701  &  ...    & RRc\\
NV112 & 255.191976 & $-$30.186995  & 0.503   & ...	 & ...	 & ...     & ...     & ...    &  ...    & RRab\\
NV113 & 255.323088 & $-$30.186512  & 0.478   & 1.632	 & 1.221 & 17.226  & 16.404  & 0.873  &  0.578  & RRab\\
NV114 & 255.427287 & $-$30.178179  & 0.398   & 1.063	 & 1.010 & 18.423  & 17.135  & 1.294  &  ...    & EB	\\
NV115 & 255.349885 & $-$30.176159  & 0.2690  & 0.671	 & 0.519 & 17.276  & 16.529  & 0.757  &  ...    & RRc\\
NV116 & 255.356841 & $-$30.175140  & 0.615   & 0.558	 & 0.390 & 17.290  & 16.384  & 0.913  &  0.551  & RRab\\
NV117 & 255.247513 & $-$30.170192  & 0.321   & 0.236	 & ...	 & 17.246  & ...     & ...    &  ...    & RRc\\
NV118 & 255.353800 & $-$30.166045  & 0.2987  & 0.558	 & 0.455 & 17.221  & 16.468  & 0.760  &  ...    & RRc\\
NV119 & 255.312109 & $-$30.163121  & 0.3191  & 0.525	 & 0.397 & 17.552  & 16.678  & 0.880  &  ...    & RRc\\
NV120 & 255.413338 & $-$30.160585  & 0.489   & 1.658	 & 1.271 & ...     & ...     & ...    &  ...    & RRab\\
NV121 & 255.299162 & $-$30.159912  & 0.2805  & 0.498	 & 0.378 & 18.060  & 17.251  & 0.814  &  ...    & RRc\\
NV122 & 255.222519 & $-$30.139789  & 0.373   & 0.116	 & ...	 & 16.845  & ...     & ...    &  ...    & RRc\\
NV123 & 255.222694 & $-$30.138895  & 0.3043  & 0.535	 & 0.439 & 16.990  & 16.262  & 0.733  &  ...    & RRc\\
NV124 & 255.287656 & $-$30.127940  & 0.4855  & 1.539	 & 1.115 & 16.817  & ...     & ...    &  ...    & RRab\\
NV125 & 255.292043 & $-$30.127814  & 0.2737  & ...	 & ...	 & ...     & ...     & ...    &  ...    & RRc\\
NV126 & 255.310969 & $-$30.127268  & 0.5133  & ...	 & ...	 & ...     & ...     & ...    &  ...    & RRab\\
NV127 & 255.288948 & $-$30.127077  & 0.5328  & 1.677	 & 1.307 & 16.676  & 15.942  & 0.792  &  0.502  & RRab\\
NV128 & 255.328694 & $-$30.127022  & 0.2504  & 0.538	 & 0.359 & 17.519  & 16.692  & 0.836  &  ...    & RRc\\
NV129 & 255.311556 & $-$30.126846  & 0.3292  & ...	 & ...	 & ...     & ...     & ...    &  ...    & RRc\\
NV130 & 255.299567 & $-$30.125709  & 0.2613  & ...	 & ...	 & ...     & ...     & ...    &  ...    & RRc\\
NV131 & 255.327951 & $-$30.125428  & 0.3132  & 0.538	 & 0.431 & 17.371  & 16.530  & 0.848  &  ...    & RRc\\
NV132 & 255.300206 & $-$30.124985  & 0.2842  & 0.764	 & 0.628 & 16.919  & 16.189  & 0.741  &  ...    & RRc\\
NV133 & 255.303920 & $-$30.124868  & 0.3156  & 0.565	 & 0.479 & 16.988  & 16.190  & 0.802  &  ...    & RRc\\
NV134 & 255.304938 & $-$30.124915  & 0.336   & ...	 & ...	 & ...     & ...     & ...    &  ...    & RRc\\       
NV135 & 255.212612 & $-$30.124641  & 0.593   & 0.924	 & 0.698 & 17.099  & 16.263  & 0.853  &  0.504  & RRab\\
NV136 & 255.313600 & $-$30.124235  & 0.6258  & 1.160	 & 0.878 & 17.105  & 16.227  & 0.904  &  0.553  & RRab\\
NV137 & 255.300694 & $-$30.123833  & 0.515   & ...	 & ...	 & ...     & ...     & ...    &  ...    & RRab\\
NV138 & 255.303221 & $-$30.123809  & 0.6051  & ...	 & ...	 & ...     & ...     & ...    &  ...    & RRab\\
NV139 & 255.304235 & $-$30.123830  & 0.5405  & ...	 & ...	 & ...     & ...     & ...    &  ...    & RRab\\
NV140 & 255.313693 & $-$30.123622  & 0.3830  & 0.541	 & 0.427 & 16.905  & 16.097  & 0.813  &  ...    & RRc\\
NV141 & 255.302897 & $-$30.123377  & 0.2994  & ...	 & ...	 & ...     & ...     & ...    &  ...    & RRc\\
NV142 & 255.324870 & $-$30.122729  & 0.2758  & ...	 & ...	 & ...     & ...     & ...    &  ...    & RRc\\
NV143 & 255.299682 & $-$30.121951  & 0.2959  & ...	 & ...	 & ...     & ...     & ...    &  ...    & RRc\\
NV144 & 255.297520 & $-$30.121835  & 0.6105  & ...	 & ...	 & ...     & ...     & ...    &  ...    & RRab\\
NV145 & 255.310817 & $-$30.121214  & 0.5665  & ...	 & ...	 & ...     & ...     & ...    &  ...    & RRab\\
NV146 & 255.296802 & $-$30.121044  & 0.4660  & ...	 & ...	 & ...     & ...     & ...    &  ...    & RRab\\
NV147 & 255.301092 & $-$30.120878  & 0.3259  & ...	 & ...	 & ...     & ...     & ...    &  ...    & RRc\\
NV148 & 255.307127 & $-$30.120699  & 0.5606  & ...	 & ...	 & ...     & ...     & ...    &  ...    & RRab\\
NV149 & 255.300972 & $-$30.120147  & 0.319   & ...	 & ...	 & ...     & ...     & ...    &  ...    & RRc\\
NV150 & 255.290041 & $-$30.119827  & 0.5480  & 0.910	 & 0.686 & 17.001  & 16.159  & 0.862  &  0.538  & RRab\\
NV151 & 255.294859 & $-$30.119358  & 0.3153  & ...	 & ...	 & ...     & ...     & ...    &  ...    & RRc\\
NV152 & 255.321269 & $-$30.119151  & 0.3032  & ...	 & ...	 & ...     & ...     & ...    &  ...    & RRc\\
NV153 & 255.266009 & $-$30.118983  & 0.3129  & 0.596	 & 0.462 & 16.807  & 16.109  & 0.706  &  ...    & RRc\\
NV154 & 255.290525 & $-$30.118882  & 0.3151  & ...	 & ...	 & ...     & ...     & ...    &  ...    & RRc\\
NV155 & 255.304738 & $-$30.118780  & 0.271   & ...	 & ...	 & ...     & ...     & ...    &  ...    & RRc\\
NV156 & 255.301102 & $-$30.118553  & 0.577   & ...	 & ...	 & ...     & ...     & ...    &  ...    & RRab\\
NV157 & 255.307981 & $-$30.118448  & 0.7195  & ...	 & ...	 & ...     & ...     & ...    &  ...    & RRab\\
NV158 & 255.300638 & $-$30.118371  & 0.264   & ...	 & ...	 & ...     & ...     & ...    &  ...    & RRc\\
NV159 & 255.289372 & $-$30.118087  & 0.3763  & ...	 & ...	 & ...     & ...     & ...    &  ...    & RRc\\
NV160 & 255.292290 & $-$30.118020  & 0.5437  & 1.482	 & 1.218 & 16.933  & 16.147  & 0.819  &  0.487  & RRab\\
NV161 & 255.310399 & $-$30.117484  & 0.5568  & ...	 & ...	 & ...     & ...     & ...    &  ...    & RRab\\
NV162 & 255.304420 & $-$30.117429  & 0.6032  & ...	 & ...	 & ...     & ...     & ...    &  ...    & RRab\\
NV163 & 255.306762 & $-$30.117204  & 0.5943  & ...	 & ...	 & ...     & ...     & ...    &  ...    & RRab\\
NV164 & 255.303820 & $-$30.116505  & 7.6     & ...	 & ...	 & ...     & ...     & ...    &  ...    & CpII\\
NV165 & 255.302432 & $-$30.116402  & 0.4520  & ...	 & ...	 & ...     & ...     & ...    &  ...    & RRab\\
NV166 & 255.304920 & $-$30.116240  & 0.290   & ...	 & ...	 & ...     & ...     & ...    &  ...    & RRc\\
NV167 & 255.314500 & $-$30.116126  & 0.630   & ...	 & ...	 & ...     & ...     & ...    &  ...    & RRab\\
NV168 & 255.270280 & $-$30.115961  & 0.5754  & 0.538	 & 0.417 & 16.965  & 16.143  & 0.827  &  0.489  & RRab\\
NV169 & 255.291652 & $-$30.115633  & 0.5144  & ...	 & ...	 & ...     & ...     & ...    &  ...    & RRab\\
NV170 & 255.306553 & $-$30.115631  & 0.635   & ...	 & ...	 & ...     & ...     & ...    &  ...    & ?	\\
NV171 & 255.318789 & $-$30.115528  & 0.6482  & 0.322	 & 0.254 & 17.159  & 16.246  & 0.914  &  ...    & RRc?	\\
NV172 & 255.308393 & $-$30.115490  & 0.3132  & ...	 & ...	 & ...     & ...     & ...    &  ...    & RRc \\
NV173 & 255.305549 & $-$30.115363  & 0.347   & ...	 & ...	 & ...     & ...     & ...    &  ...    & RRc\\
NV174 & 255.301761 & $-$30.115313  & 0.3175  & ...	 & ...	 & ...     & ...     & ...    &  ...    & RRc\\
NV175 & 255.321256 & $-$30.115305  & 0.2816  & 0.643	 & 0.516 &  17.08  & 16.372  & 0.725  &  ...    & RRc\\
NV176 & 255.305474 & $-$30.114681  & 0.2628  & ...	 & ...	 & ...     & ...     & ...    &  ...    & RRc\\
NV177 & 255.291754 & $-$30.114502  & 0.6851  & ...	 & ...	 & ...     & ...     & ...    &  ...    & RRab\\
NV178 & 255.304743 & $-$30.114507  & 0.407   & ...	 & ...	 & ...     & ...     & ...    &  ...    & RRc\\
NV179 & 255.301488 & $-$30.114427  & 0.5483  & ...	 & ...	 & ...     & ...     & ...    &  ...    & RRab\\
NV180 & 255.296731 & $-$30.114401  & 1.376   & ...	 & ...	 & ...     & ...     & ...    &  ...    & CpII\\
NV181 & 255.295084 & $-$30.113733  & 0.5895  & ...	 & ...	 & ...     & ...     & ...    &  ...    & RRab\\
NV182 & 255.300463 & $-$30.113602  & 0.505   & ...	 & ...	 & ...     & ...     & ...    &  ...    & RRab\\
NV183 & 255.305893 & $-$30.113635  & 0.5686  & ...	 & ...	 & ...     & ...     & ...    &  ...    & RRab\\
NV184 & 255.302188 & $-$30.113549  & 0.5722  & ...	 & ...	 & ...     & ...     & ...    &  ...    & RRab\\
NV185 & 255.313926 & $-$30.113517  & 0.591   & ...	 & ...	 & ...     & ...     & ...    &  ...    & RRab\\
NV186 & 255.307716 & $-$30.113428  & 0.3025  & ...	 & ...	 & ...     & ...     & ...    &  ...    & RRc\\
NV187 & 255.298718 & $-$30.113290  & 0.491   & ...	 & ...	 & ...     & ...     & ...    &  ...    & RRab\\
NV188 & 255.303150 & $-$30.113177  & 0.436   & ...	 & ...	 & ...     & ...     & ...    &  ...    & RRab\\
NV189 & 255.315829 & $-$30.113097  & 0.5617  & ...	 & ...	 & ...     & ...     & ...    &  ...    & RRab\\
NV190 & 255.302756 & $-$30.112618  & 0.554   & ...	 & ...	 & ...     & ...     & ...    &  ...    & RRab\\
NV191 & 255.305548 & $-$30.112558  & 0.5857  & ...	 & ...	 & ...     & ...     & ...    &  ...    & RRab\\
NV192 & 255.353057 & $-$30.112569  & 0.2652  & 0.406	 & 0.314 & 17.250  & 16.510  & 0.743  &  ...    & RRc\\
NV193 & 255.314500 & $-$30.112496  & 0.3198  & ...	 & ...	 & ...     & ...     & ...    &  ...    & RRc\\
NV194 & 255.317154 & $-$30.112403  & 0.505   & ...	 & ...	 & ...     & ...     & ...    &  ...    & RRab\\
NV195 & 255.298155 & $-$30.112409  & 0.47    & ...	 & ...	 & ...     & ...     & ...    &  ...    & RRc\\
NV196 & 255.288121 & $-$30.112265  & 0.3109  & ...	 & ...	 & ...     & ...     & ...    &  ...    & RRc\\
NV197 & 255.304747 & $-$30.111837  & 0.3183  & ...	 & ...	 & ...     & ...     & ...    &  ...    & RRc\\
NV198 & 255.289817 & $-$30.111245  & 0.5763  & 1.102	 & 0.782 & 16.922  & 16.059  & 0.889  &  0.559  & RRab\\
NV199 & 255.315134 & $-$30.110762  & 0.5568  & ...	 & ...	 & ...     & ...     & ...    &  ...    & RRab\\
NV200 & 255.293965 & $-$30.110272  & 0.487   & ...	 & ...	 & ...     & ...     & ...    &  ...    & RRab\\
NV201 & 255.314913 & $-$30.108397  & 0.2507  & ...	 & ...	 & ...     & ...     & ...    &  ...    & RRc\\
NV202 & 255.307417 & $-$30.108318  & 0.2709  & 0.517	 & 0.439 & 16.974  & 16.258  & 0.720  &  ...    & RRc\\
NV203 & 255.301002 & $-$30.108253  & 0.491   & ...	 & ...	 & ...     & ...     & ...    &  ...    & RRab\\
NV204 & 255.284814 & $-$30.108090  & 0.263   & 0.415	 & 0.329 & 16.922  & 16.224  & 0.701  &  ...    & RRc\\
NV205 & 255.295998 & $-$30.107995  & 0.552   & ...	 & ...	 & ...     & ...     & ...    &  ...    & RRab\\
NV206 & 255.300336 & $-$30.107351  & 0.640   & ...	 & ...	 & ...     & ...     & ...    &  ...    & RRab\\       
NV207 & 255.299857 & $-$30.106901  & 0.265   & ...	 & ...	 & ...     & ...     & ...    &  ...    & RRc\\
NV208 & 255.298778 & $-$30.106808  & 0.5410  & ...	 & ...	 & ...     & ...     & ...    &  ...    & RRab\\
NV209 & 255.307591 & $-$30.106368  & 0.2883  & ...	 & ...	 & ...     & ...     & ...    &  ...    & RRc\\
NV210 & 255.309869 & $-$30.106245  & 0.4370  & 1.902	 & 1.647 & 17.015  & 16.207  & 0.861  &  0.553  & RRab\\
NV211 & 255.319518 & $-$30.106219  & 0.3327  & ...	 & ...	 & ...     & ...     & ...    &  ...    & RRc\\
NV212 & 255.314221 & $-$30.105641  & 0.6072  & ...	 & ...	 & ...     & ...     & ...    &  ...    & RRab\\
NV213 & 255.303032 & $-$30.104438  & 0.5853  & ...	 & ...	 & ...     & ...     & ...    &  ...    & RRab\\
NV214 & 255.318374 & $-$30.103102  & 0.587   & 1.004	 & 0.776 & 17.154  & 16.299  & 0.875  &  0.538  & RRab\\
NV215 & 255.305126 & $-$30.102857  & 0.4616  & ...	 & ...	 & ...     & ...     & ...    &  ...    & RRab?\\
NV216 & 255.298332 & $-$30.102701  & 0.2666  & ...	 & ...	 & ...     & ...     & ...    &  ...    & RRc\\
NV217 & 255.283447 & $-$30.102377  & 0.3203  & ...	 & ...	 & ...     & ...     & ...    &  ...    & RRc\\
NV218 & 255.307103 & $-$30.101639  & 0.4447  & 1.485	 & 1.249 & 16.867  & 16.095  & 0.804  &  0.492  & RRab\\
NV219 & 255.290591 & $-$30.101456  & 0.7177  & 0.862	 & 0.744 & 16.940  & 16.032  & 0.916  &  0.509  & RRab\\
NV220 & 255.303126 & $-$30.101010  & 0.496   & ...	 & ...	 & ...     & ...     & ...    &  ...    & RRab\\
NV221 & 255.265266 & $-$30.091025  & 0.3300  & 0.555	 & 0.425 & 16.802  & 16.085  & 0.725  &  ...    & RRc\\
NV222 & 255.250721 & $-$30.080969  & 0.460   & ...	 & ...	 & ...     & ...     & ...    &  ...    & RRab\\
NV223 & 255.323255 & $-$30.080828  & 0.5325  & 1.346	 & 1.026 & 17.123  & 16.305  & 0.852  &  0.529  & RRab\\
NV224 & 255.395528 & $-$30.079961  & 0.3196  & 0.546	 & 0.417 & 17.725  & 16.952  & 0.779  &  ...    & RRc\\
NV225 & 255.322496 & $-$30.079209  & 0.2884  & 0.396	 & 0.252 & 16.352  & 15.353  & 1.004  &  ...    & RRc\\
NV226 & 255.285658 & $-$30.071786  & 0.6301  & 1.073	 & 0.796 & 16.736  & 15.984  & 0.776  &  0.450  & RRab\\
NV227 & 255.450562 & $-$30.068877  & 0.456   & 1.725	 & 1.290 & 18.629  & 17.774  & 0.899  &  0.564  & RRab\\
NV228 & 255.334192 & $-$30.068270  & 0.6417  & 0.380	 & 0.322 & 17.132  & 16.253  & 0.881  &  0.479  & RRab\\
NV229 & 255.337618 & $-$30.066281  & 0.2773  & 0.295	 & 0.248 & 16.765  & 16.106  & 0.660  &  ...    & RRc\\
NV230 & 255.332977 & $-$30.096293  & ........  & ...	 & ...	 & ...     & ...     & ...    &  ...    & LP	\\
NV231 & 255.324541 & $-$30.097949  & $\sim$16  & ...	 & ...	 & ...     & ...     & ...    &  ...    & LP/CpII? \\
NV232 & 255.321150 & $-$30.136159  & ........  & ...	 & ...	 & ...     & ...     & ...    &  ...    & LP	\\
NV233 & 255.318726 & $-$30.108153  & $\sim$49  & ...	 & ...	 & ...     & ...     & ...    &  ...    & LP	\\
NV234 & 255.308946 & $-$30.113105  & ........  & ...	 & ...	 & ...     & ...     & ...    &  ...    & LP	\\
NV235 & 255.307303 & $-$30.111160  & ........  & ...	 & ...	 & ...     & ...     & ...    &  ...    & LP	\\
NV236 & 255.307158 & $-$30.110023  & $\sim$50  & ...	 & ...	 & ...     & ...     & ...    &  ...    & LP	\\
NV237 & 255.303834 & $-$30.130456  & ........  & ...	 & ...	 & ...     & ...     & ...    &  ...    & LP	\\
NV238 & 255.304566 & $-$30.106671  & ........  & ...	 & ...	 & ...     & ...     & ...    &  ...    & LP	\\
NV239 & 255.303782 & $-$30.113794  & $\sim$75  & ...	 & ...	 & ...     & ...     & ...    &  ...    & LP	\\
NV240 & 255.303710 & $-$30.118210  & ........  & ...	 & ...	 & ...     & ...     & ...    &  ...    & LP	\\
NV241 & 255.303521 & $-$30.116544  & 0.525     & ...	 & ...	 & ...     & ...     & ...    &  ...    & RRab?	\\
NV242 & 255.302059 & $-$30.108011  & $\sim$36  & ...	 & ...	 & ...     & ...     & ...    &  ...    & LP	\\
NV243 & 255.301054 & $-$30.113012  & 0.4911    & ...	 & ...	 & ...     & ...     & ...    &  ...    & RRab\\
NV244 & 255.298619 & $-$30.119052  & ........  & ...	 & ...	 & ...     & ...     & ...    &  ...    & LP  \\ 
NV245 & 255.297277 & $-$30.084141  & $\sim$88  & ...	 & ...	 & ...     & ...     & ...    &  ...    & LP	\\
NV246 & 255.295958 & $-$30.105957  & 0.5086    & ...	 & ...	 & ...     & ...     & ...    &  ...    & RRab\\
NV247 & 255.295155 & $-$30.120411  & 0.4928    & ...	 & ...	 & ...     & ...     & ...    &  ...    & RRab\\
NV248 & 255.294099 & $-$30.114300  & ........  & ...	 & ...	 & ...     & ...     & ...    &  ...    & LP	\\
NV249 & 255.292809 & $-$30.111699  & 0.2476    & ...	 & ...	 & ...     & ...     & ...    &  ...    & RRc\\
NV250 & 255.292034 & $-$30.119580  & $\sim$65  & ...	 & ...	 & ...     & ...     & ...    &  ...    & LP	\\
NV251 & 255.290391 & $-$30.115419  & ........  & ...	 & ...	 & ...     & ...     & ...    &  ...    & LP	\\
NV252 & 255.281460 & $-$30.119274  & ........  & ...	 & ...	 & ...     & ...     & ...    &  ...    & LP	\\
NV253 & 255.280065 & $-$30.102644  & ........  & ...	 & ...	 & ...     & ...     & ...    &  ...    & LP  \\
NV254 & 255.278510 & $-$30.126721  & $\sim$90  & ...	 & ...	 & ...     & ...     & ...    &  ...    & LP	\\
NV255 & 255.278235 & $-$30.114604  & ........  & ...	 & ...	 & ...     & ...     & ...    &  ...    & LP	\\
NV256 & 255.278240 & $-$30.102156  & ........  & ...	 & ...	 & ...     & ...     & ...    &  ...    & LP	\\
NV257 & 255.274444 & $-$30.135636  & $\sim$33  & ...	 & ...	 & ...     & ...     & ...    &  ...    & LP	\\
NV258 & 255.308126 & $-$30.120434  & ........  & ...	 & ...	 & ...     & ...     & ...    &  ...    & LP	\\
NV259 & 255.298877 & $-$30.063971  & 0.6704    & ...	 & ...	 & ...     & ...     & ...    &  ...    & RRab?	\\
NV260 & 255.305121 & $-$30.102731  & 0.2540    & ...	 & ...	 & ...     & ...     & ...    &  ...    & RRc \\
NV261 & 255.300681 & $-$30.123291  & 0.5041    & ...	 & ...	 & ...     & ...     & ...    &  ...    & RRab\\
NV262 & 255.261271 & $-$30.109493  & ........  & ...	 & ...	 & ...     & ...     & ...    &  ...    & LP 
%\enddata
\label{tab:efemer}
%\end{deluxetable}
\end{longtable*}
  
\section{Observations and data reductions}
M62 was observed in conjunction with M69 (NGC~6637; Escobar et
al. 2010, in preparation) and NGC~5286 \citep{mzea09,mzea10}
as part of a long-term project aimed at completing the census of (bright) variable stars in
Galactic globular clusters \citep{mcea06}. Time-series observations
in $B$ and $V$ were obtained with the Warsaw 1.3m telescope at the Las Campanas
Observatory (LCO), in the course of 7 consecutive nights over the period
April 6--13 2003. The camera used is the 8kMOSAIC camera, comprised
of eight $2040\times 4096$ chips, with a scale $0.26\arcsec/{\rm pixel}$ 
giving an observing area equal to $35\arcmin\times35\arcmin$. 
The cluster was roughly centered on chip 2, and so in this paper we focus our 
analysis on this chip (which covers a sky area of $8.8\arcmin\times17.8\arcmin$). 
The monitored field on chip 2 covers
most of the cluster area, as the tidal radius of M62 is estimated
at $r = 8.95$~arcmin \citep{stea95} or $r = 10.01$~arcmin \citep{gbea06}. 
The read out noise of the camera is 6 to 9~$e^{-}$ (depending on the chip) and the
gain is 6.3~$e^{-}$/ADU.
A total of 126 images in $B$ and 126 in $V$ were secured with this
setup. 
During the nights of the observations, the seeing was stable enough
with an average measured stellar point-spread funtion (PSF) on the frames
of about $0.98\arcsec$ FWHM.
Exposures times ranged from 100 s to 220 s for the $B$ frames and 30~s to 90~s 
for the $V$ frames. 

Observations of the standard fields PG+0918, PG+1323, PG+1525, 
PG+1528, PG+1633, PG+1657 and Ru 152 \citep{al92}
were obtained on the same nights, to calibrate the data to the
standard Johnson-Cousins photometric system. 
In order to provide better sampled light curves, the Warsaw data were
complemented by observations obtained with the 
Cerro Tololo Inter-American Observatory (CTIO) 1.3m telescope
in service mode, using the ANDICAM $1024\times 1024$ CCD, with a scale
$0.369\arcsec/{\rm pixel}$, over the timespan April 24 2003 to June 30 2003.
This additional dataset consists of 42 images in each of $B$ and 
$V$, and permitted us to extend the time interval spanned by our
observations up to about three months, thus resulting very useful to
pin down periods and to search for long-term variability.
The exposure times in this case were 145~s for the $B$ frames
and 40~s for the $V$ frames. 
The seeing during these observations was on average $\sim 1.3\arcsec$,
with stable and good photometric conditions. However, no standard fields 
were observed with the CTIO 1.3m telescope.
The LCO images were pre-processed with the Warsaw 1.3m pipeline, so
that no additional pre-reduction steps were necessary. The preliminary
reduction of the CTIO frames, including bias subtraction and flat
fielding, were carried out using the standard 
IRAF\footnote{IRAF is distributed by the National Optical Astronomy Observatories,
    which are operated by the Association of Universities for Research
    in Astronomy, Inc., under cooperative agreement with the National
    Science Foundation.} 
data reduction package.

Since no photometric calibration was obtained during the CTIO run, we
used the well calibrated LCO set to link the CTIO instrumental
magnitudes to the standard Johnson system. With this purpose in mind, 
we performed a cross correlation
between the LCO and the CTIO catalogues, and then selected the best 50
stars in common that covered a sufficiently wide range in color to
prevent any residual uncorrected color trend. These selected stars
were then used to calibrate the CTIO data by means of a least-squares fit.  

We will provide further detailed information regarding our calibration 
in Paper~II (Contreras et al. 2010, in preparation), when a detailed 
analysis of our derived cluster CMD will also be provided. In any case, 
we note that our derived calibration equations are well defined, and 
have zero point errors of only about 0.006~mag in $B$ and 0.011~mag in 
$V$, with similarly small errors in the derived color coefficients. 

%The equations relating the CTIO instrumental system to the LCO calibrated
%data set are as follows: 
%
%\begin{equation}
%B=b+0.3327295-2.736626\,(B-V)\,,
%\end{equation}
%\begin{equation}
%V=v+0.02472314-3.172442\,(B-V)\,.
%\end{equation}

\begin{deluxetable}{lccccc}
\tablewidth{0pc} 
\tabletypesize{\scriptsize}
\tablecaption{Photometry of the Variable Stars}
\tablehead{\colhead{Name} & \colhead{Filter} & \colhead{JD} & \colhead{Phase} & \colhead{Mag} & \colhead{e\_Mag}\\
\colhead{} & \colhead{} & \colhead{(d)} & \colhead{} & \colhead{(mag)} & \colhead{(mag)}}
\startdata
 V01 & $V$ & $2,452,736.54861$ & $0.0000$ & $16.2351$ & $0.0031$ \\ 
 V01 & $V$ & $2,452,736.55367$ & $0.0080$ & $16.2779$ & $0.0042$ \\ 
 V01 & $V$ & $2,452,736.55977$ & $0.0176$ & $16.2787$ & $0.0051$ \\ 
 V01 & $V$ & $2,452,736.57490$ & $0.0414$ & $16.3361$ & $0.0054$ \\ 
 V01 & $V$ & $2,452,736.58749$ & $0.0612$ & $16.3878$ & $0.0059$ \\
 V01 & $V$ & $2,452,736.59664$ & $0.0756$ & $16.3996$ & $0.0059$ 
\enddata
\tablecomments{This table is published in its entirety in the 
electronic edition of the {\it Astronomical Journal}.  A portion is 
shown here for guidance regarding its form and content.}
\label{tab:ascii}
\end{deluxetable}

\section{Variable Stars Identification and Period Determination}
Since the pioneering effort by \citet{tc96}, it became clear 
that the image subtraction technique is one of the best tools for identifying 
variable stars in crowded fields like globular clusters, due to its powerful capability 
of comparing images after all non-variable objects have been removed. We have 
selected the ISIS v2.2 package for this purposee \citep{ca00}.
The ISIS reduction procedure that we follow consists of several steps: 
(1)~We transform all the frames to a common coordinate grid, where the image 
taken with the best seeing was chosen as astrometric reference; 
(2)~We select 10\% of our frames with the best seeing conditions 
to construct a composite, reference photometric image;  
(3)~We then subtract each individual frame from the composite image, 
after convolving the latter so that both images end up having similar PSF's.
As the flux of non-variable stars on both 
images should be essentially identical, such objects will disappear when 
one image is subtracted from the other, and the remaining signal will (ideally) 
come exclusively from variable stars; (4)~We construct a median image
of all the subtracted images (known as ``var.fits'') in order to enhance these 
weak individual (residual) signals, and thus making variable stars 
candidates more easily identifiable as significant peaks in the median image; 
(5)~Finally, profile-fitting photometry was performed for each variable star 
candidate on the subtracted images.

\begin{figure}[t]
\includegraphics[width=8.5cm]{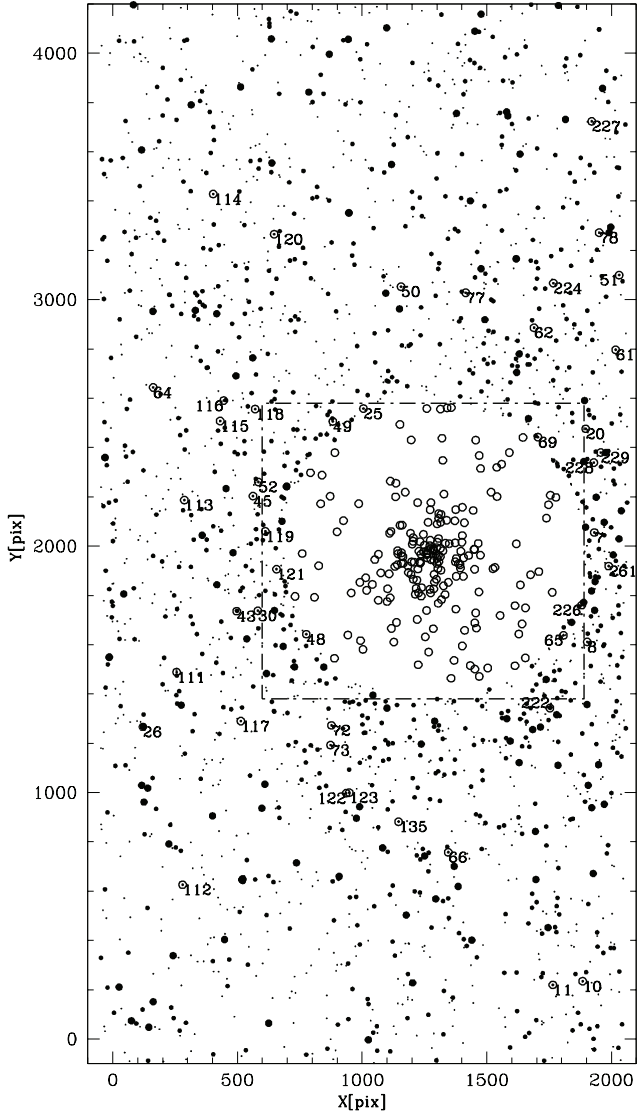}
\caption{Finding chart for the variable stars in M62, based on the Warsaw 1.3m chip containing
the cluster center. A zoom in around the dot-dashed region is shown in the next panel.}
\label{fig:fch1}
\end{figure}

Periods were determined using the phase dispertion minimization (PDM) method  
\citet{rs78}, as implemented in IRAF. PDM is a generalization of the \citet{lk65}, 
and essentially attempts to identify the phased light curve that produces the 
minimum scatter in phase. 
We were thus able to detect and confirm the existence
of at least 245 variable stars in the field of M62, including  209 RR Lyrae, 4 type II
Cepheids, 25 LPV's, 1 eclipsing binary, and 6 of uncertain nature. 
Among the detected RR Lyrae, 133 are fundamental-mode (RRab or RR0) pulsators, 
whereas 76 are first-overtone (RRc or RR1) stars. 
We must stress that just 1 out of the 8 Warsaw 1.3m chips were
analyzed, and so the total number of (undiscovered) variable stars in the cluster 
is almost certainly higher. 
Indeed, although the \citet{gbea06} profile suggests that we should find $\approx 98.6\%$
of all cluster stars within $185\arcsec$ of the cluster center, we find clear evidence for
an excess of variable stars whose properties are consistent with cluster membership further 
out. More specifically, in chip~2 we find about 4 times more variable stars outside 
$185\arcsec$ than predicted by the cluster's density profile. More details are provided 
in \S\ref{sec:cmd}. The remaining chips will be the subject of a future paper, where the 
possible existence of extra-tidal stars in this cluster will also be examined. 
Finding charts for the M62 variable stars are presented in Figure~\ref{fig:fch1}.

%Segun la tabla, dentro de 185'' esperariamos encontrar el 98.6% de las 
%estrellas variables. Considerando que dentro de este radio tenemos ~230, 
%(obviamente el calculo y el resultado es analogo si considero solo las RRL) 
%implicaria que el 100% de las detecciones corresponde a 233,6 estrellas 
%variables, y por lo tanto mas alla de 185" esperariamos encontrar como maximo  
%~4 variables mas y asi tener completa nuestra muestra, lo cual claramente no 
%se satisface pues solo en el chip 2 tenemos ~16 estrellas fuera de esta area, 
%y por lo que sabemos de Clement, en el chip 3 hay al menos 17 mas!!!

%Finding Chart 1

%Finding Chart 2
\begin{figure*}[t]
\figurenum{1, {\em cont}}
\includegraphics[width=17cm]{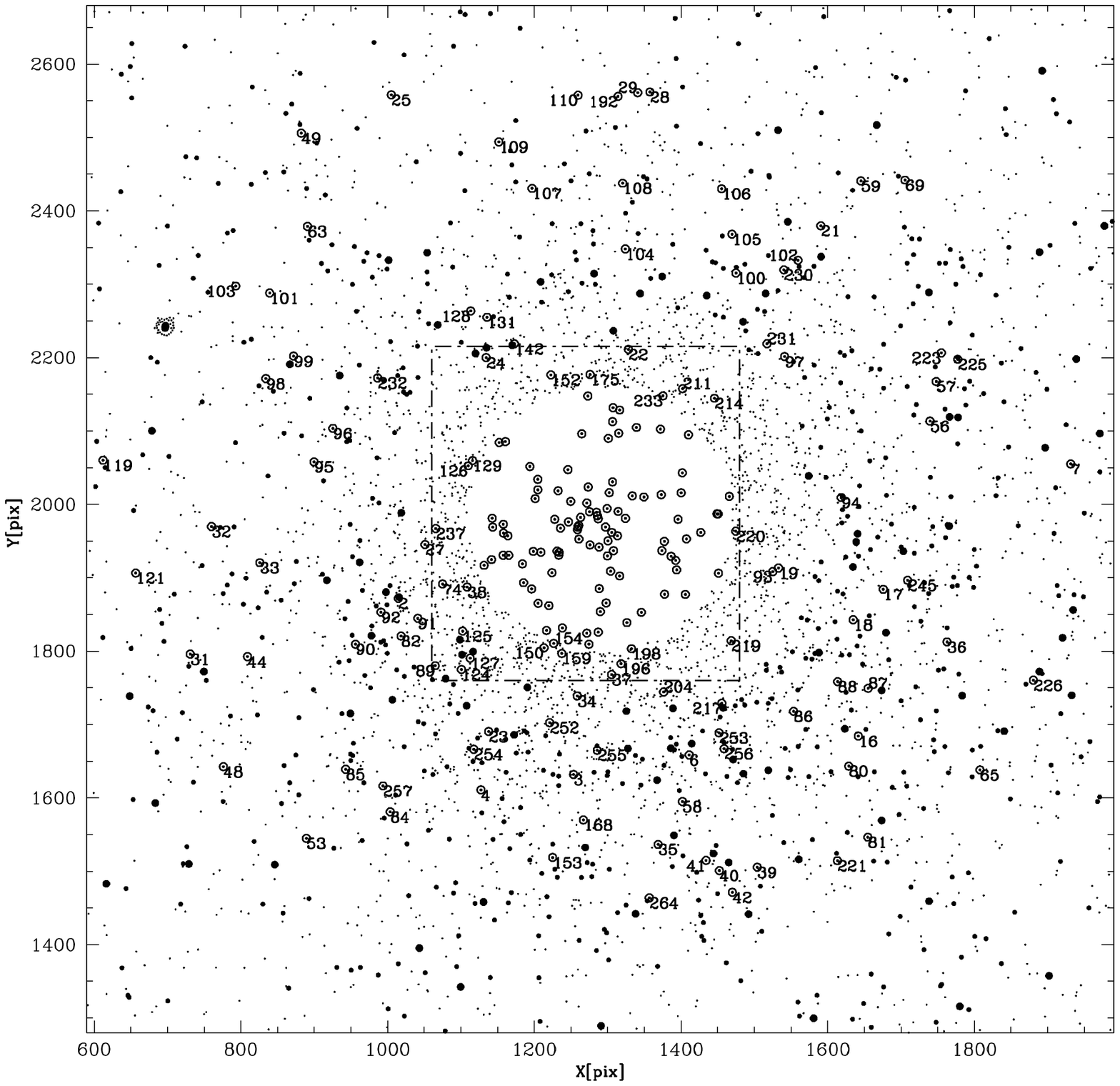}
\caption{Finding chart for the variable stars in M62. A zoom in around the dot-dashed region 
is shown in the next panel.} 
\label{fig:fch2}
\end{figure*}

%Finding Chart 3
\begin{figure*}[t]
\figurenum{1, {\em cont}}
\includegraphics[width=17cm]{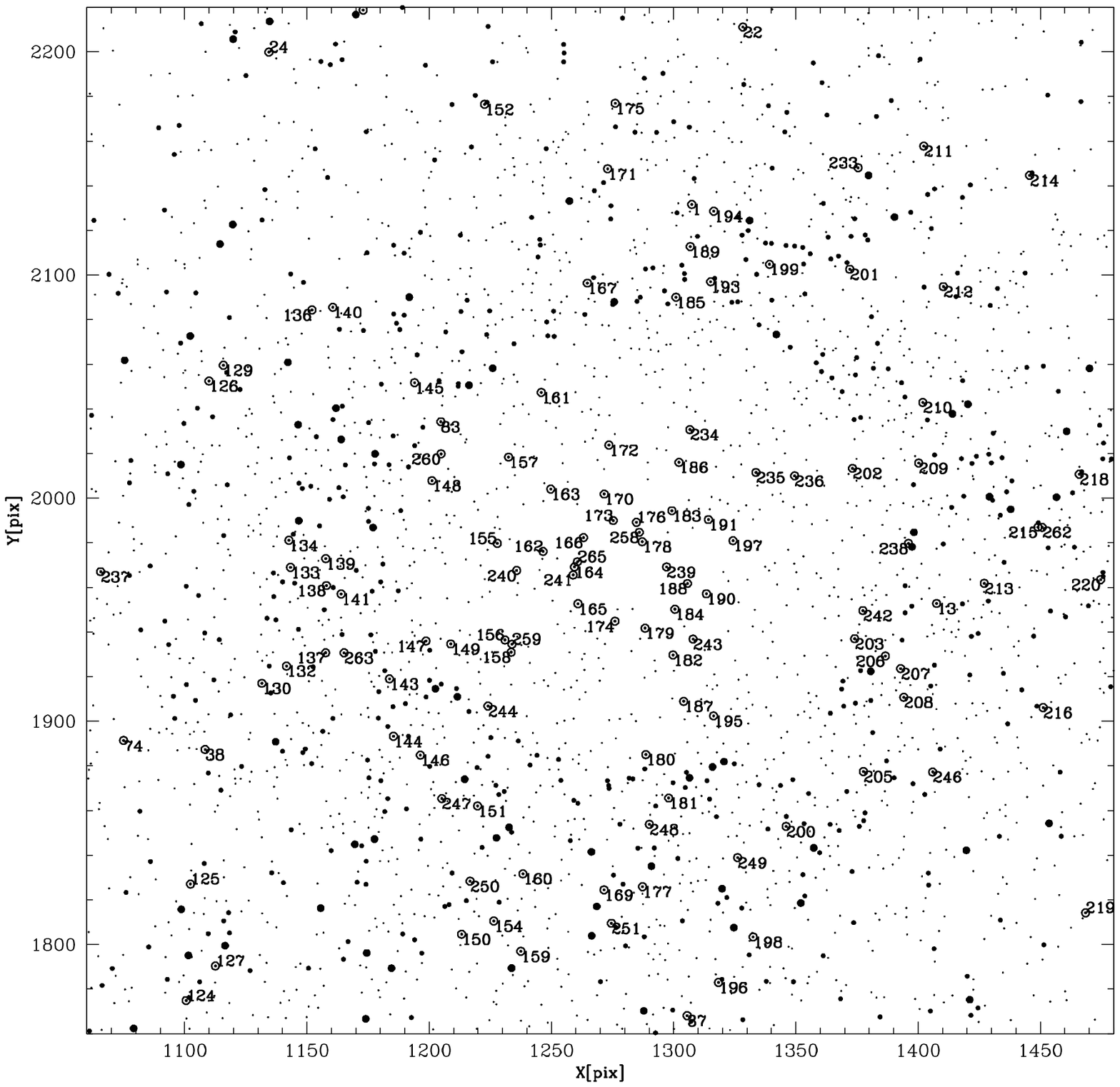}
\caption{Finding chart for the variable stars in M62 (innermost region).} 
\label{fig:fch3}
\end{figure*}

%Parametros de Fourier para 36 RRc
%\addtolength{\hoffset}{2cm}
\begin{deluxetable}{lcccccc}
\tablewidth{0pc} 
\tabletypesize{\scriptsize}
\tablecaption{Fourier Coefficients for RRc Variables in M62}
\tablehead{\colhead{ID  } & \colhead{$A_{21}$} & \colhead{$A_{31}$} & 
\colhead{$A_{41}$} & \colhead{$\phi_{21}$} &
\colhead{$\phi_{31}$} & \colhead{$\phi_{41}$} }
\startdata
V30        & 0.124  &0.068  &0.039  &4.514 & 3.298 $\pm$ 0.120 & 2.196\\
V40        & 0.144  &0.078  &0.057  &4.932 & 3.624 $\pm$ 0.083 & 2.188\\
V42        & 0.056  &0.023  &0.023  &4.159 & 0.825 $\pm$ 0.448 & 0.581\\
V53        & 0.163  &0.096  &0.080  &4.761 & 2.976 $\pm$ 0.080 & 1.709\\
V65        & 0.081  &0.038  &0.058  &4.954 & 4.192 $\pm$ 0.407 & 2.908\\
V66::      & 0.740  &0.647  &0.540  &6.262 & 6.243 $\pm$ 0.009 & 6.259\\
V69        & 0.067  &0.057  &0.032  &4.673 & 3.936 $\pm$ 0.154 & 2.960\\
V77:       & 0.133  &0.138  &0.073  &6.268 & 4.013 $\pm$ 0.166 & 2.574\\
NV85       & 0.074  &0.069  &0.020  &4.903 & 4.001 $\pm$ 0.120 & 1.797\\
NV86       & 0.133  &0.081  &0.037  &4.534 & 3.561 $\pm$ 0.106 & 2.417\\
NV90:      & 0.143  &0.028  &0.071  &5.017 & 3.616 $\pm$ 0.455 & 2.751\\
NV91       & 0.100  &0.074  &0.030  &4.875 & 3.972 $\pm$ 0.116 & 2.817\\
NV100      & 0.176  &0.077  &0.061  &4.641 & 2.794 $\pm$ 0.089 & 1.703\\
NV101      & 0.063  &0.054  &0.049  &5.314 & 3.592 $\pm$ 0.194 & 2.700\\
NV108      & 0.112  &0.089  &0.013  &4.542 & 3.602 $\pm$ 0.115 & 2.150\\
NV110      & 0.085  &0.087  &0.066  &5.124 & 4.118 $\pm$ 0.104 & 2.866\\
NV111::    & 0.046  &0.086  &0.093  &4.375 & 4.317 $\pm$ 0.106 & 0.925\\
NV115      & 0.184  &0.075  &0.052  &4.711 & 2.807 $\pm$ 0.079 & 1.773\\
NV118      & 0.115  &0.074  &0.046  &4.633 & 3.505 $\pm$ 0.106 & 2.446\\
NV119      & 0.090  &0.056  &0.043  &4.440 & 3.793 $\pm$ 0.149 & 2.296\\
NV121      & 0.064  &0.083  &0.036  &4.642 & 3.138 $\pm$ 0.133 & 2.480\\
NV123      & 0.101  &0.075  &0.041  &4.512 & 3.839 $\pm$ 0.130 & 2.705\\
NV128:     & 0.187  &0.025  &0.021  &3.889 & 1.897 $\pm$ 0.492 & 0.922\\
NV131      & 0.095  &0.079  &0.039  &5.002 & 3.764 $\pm$ 0.131 & 2.308\\
NV132      & 0.096  &0.042  &0.030  &4.058 & 4.687 $\pm$ 0.258 & 2.598\\
NV133:     & 0.115  &0.071  &0.050  &4.675 & 4.236 $\pm$ 0.162 & 2.321\\
NV140      & 0.035  &0.079  &0.028  &5.767 & 4.863 $\pm$ 0.134 & 3.072\\
NV153      & 0.113  &0.051  &0.037  &4.894 & 3.688 $\pm$ 0.150 & 2.227\\
NV175      & 0.136  &0.101  &0.049  &4.558 & 3.055 $\pm$ 0.083 & 1.833\\
NV192      & 0.090  &0.028  &0.026  &4.788 & 2.338 $\pm$ 0.398 & 1.210\\
NV202      & 0.131  &0.046  &0.037  &4.878 & 2.782 $\pm$ 0.413 & 0.751\\
NV204:     & 0.176  &0.084  &0.081  &4.028 & 2.769 $\pm$ 0.214 & 5.780\\
NV221:     & 0.120  &0.110  &0.071  &5.602 & 4.185 $\pm$ 0.121 & 3.345\\
NV224      & 0.109  &0.088  &0.036  &4.735 & 3.653 $\pm$ 0.131 & 1.639\\
NV225:     & 0.145  &0.091  &0.113  &4.341 & 4.403 $\pm$ 0.197 & 2.116\\
NV229:     & 0.060  &0.037  &0.005  &4.846 & 4.208 $\pm$ 0.450 & 0.562
\enddata
\label{tab:firrc}
\end{deluxetable}

We identified 66 of the 83 known variable stars discovered in previous 
studies \citep[see the online catalogue by][ for a listing]{ccea01}. Of the 
remaining 17 stars, we were unable to confirm variability for 2 of them,  
whereas the other 15 stars fall outside the fields that we have analyzed. 
On the other hand, \citet{yumea97} find an additional 43 stars that they 
claim to lie in the instability strip of the cluster, and which are 
accordingly RR Lyrae candidates. In our study we were able to confirm 
the variability of 27 among their 43 candidates, with the remaining 
16 stars being non-variable in our data. 

Taking into account the 209 RR Lyrae stars detected in our study and the 
15 additional RR Lyrae stars listed in \citet{ccea01} which fall outside 
our studied fields, this gives a total of 224 RR Lyrae stars that are known
so far in this cluster. For comparison, the most RR Lyrae-rich 
(in the sense of total number of RR Lyrae stars present) globular 
cluster known in our galaxy, M3 (NGC~5272), possesses a total of 230 
reported RR Lyrae stars \citep{gcea04}, being followed by $\omega$~Centauri
(NGC~5139), with a total of 178 RR Lyrae \citep{ccea01}. Clearly, our 
detections place M62 among the most RR Lyrae-rich globular clusters known, 
and further analysis of the outer fields not included in our study is not 
unlikely to give it the title of the most RR Lyrae-rich of all known 
globular star clusters \citep[see also][]{rcea05}. In terms of the 
specific frequency of RR Lyrae variables, given by 
$S_{\rm RR} = N_{\rm RR} \times 10^{0.4 (7.5 + M_V)}$, and using for the cluster a $M_V = -9.19$ 
(as given in the \citeauthor{wh96} \citeyear{wh96} catalog, Feb. 2003 update), 
one finds $S_{\rm RR} = 47.2$, which is very similar to the value 
$S_{\rm RR} \approx 46$ originally reported by \citet{rcea05}, and which 
again confirms the fact that M62 is an extremely RR Lyrae-rich object, 
since there are at present only 9 clusters with higher known $S_{\rm RR}$, 
again according to the \citeauthor{wh96} catalog.
In this sense, also noteworthy is the large number of LPV stars detected in 
the M62 field, with a total of 25 variables, 18 of which appear to 
be likely cluster members (see \S\ref{sec:cmd}). 
According to the \citeauthor{ccea01} catalogue, previously the most
LPV-rich of all globular clusters was $\omega$~Cen, with a total of 15 LPV
stars. This suggests that M62 may also be the most LPV-rich known of all 
globular clusters (again in the sense of total number of LPV stars 
present). 

While ISIS is very efficient in detecting variables stars in crowded fields, it
presents the drawback of providing light curves in flux values relative to 
the composite frame. For this reason, ISIS does not provide light curves in
standard magnitudes, and the composite image has to be processed independently
for this purpose. To put our light curves in standard magnitude units, we 
followed the procedure recommended by \citet{bmea01}, for those variable 
stars which could be reliably measured in the reference frame. More specifically, 
the variable stars detected by ISIS were counter-identified with the  $B$, $V$ 
master catalogue of the reference frame, as obtained with DAOPHOT/ALLFRAME 
\citep{pbs87,pbs94}. Then, following the same procedure as in \citeauthor{bmea01}, 
this allowed us to transform the light curves from differential fluxes into 
magnitude units. 

While DAOPHOT/ALLFRAME represents an excellent tool to perform absolute 
photometry in the crowded regions found in globular clusters, it is still often 
the case that the variable stars located in the very crowded cluster center, 
as well as those located near bright and/or saturated objects (or close to the 
edges of the frames) will lack reliable photometry, even in our best seeing 
(reference) images. Therefore, among our sample of variable stars, 110 objects 
have differential flux light curves only, either because we could not measure 
their magnitudes on the reference frames, or because we consider that they lack 
reliable DAOPHOT/ALLFRAME photometry due to one or more of the aforementioned 
reasons.

Photometric properties and basic elements for the 245 variable stars in our
study are presented in Table~\ref{tab:efemer}. 
Column 1 indicates the star's ID. 
Columns 2 and 3 provide the right ascension and declination (J2000 epoch), 
respectively, whereas column 4 gives our derived periods. 
Columns 5 and 6 list the derived 
amplitudes in the $B$ and $V$ bands, respectively. Columns 7 and 8,  
give the intensity-weighted mean $B$ and $V$ values, while
column 9 shows the magnitude-weighted mean color. 
In column 10 we provide our derived reddening values for individual 
RRab stars (see \S\ref{sec:cmd} for more details), and 
finally column 11 indicates the star's variability type. 
We assigned a prefix ``NV'' to the newly identified objects, including
the variable star candidates (NV84-NV110) from \citet{yumea97}. For the previously
known 66 confirmed variables in our field, we obtain revised periods based on our 
data alone, since these new periods produce less scattered light curves than
the old ones. The exception are variables V4, V10, V11, V20,
V23, V27, V38, V43, V45, V50, V52, V62, V62, V64, V66 and V78, where we
adopt periods from the \citet{ccea01} online catalog, since they provide good 
matches to our data. Sample light curves for the newly detected variables 
stars are shown in the Appendix, whereas the complete set of light curves can 
be found in the electronic version. The light 
curve data are provided, in machine-readable form, in Table~\ref{tab:ascii}. 

\subsection{Notes on Individual Variable Stars}\label{sec:notes}

\noindent {\bf V1, V3}: Periods for these stars are not provided in the online
\citet{ccea01} catalog. The derived periods are based on an analysis of the 
CTIO images, even though only the LCO photometry is shown in the electronic 
version of the Appendix.

\noindent {\bf V77, NV117, NV149, NV155, NV166, NV174, NV176, NV178, NV186, 
NV197, NV202, NV204, NV207, NV225, NV229}: These are short-period RRc stars 
with seemingly variable light curves. We have not been able to identify any 
source of spurious error in our photometry that could affect these stars in 
particular. 

\noindent {\bf NV92, NV103, NV124, NV126, NV129, NV167, NV182, NV200, NV203,
NV211, NV220}: The derived periods are based on an analysis of the CTIO images, 
even though only the LCO light curves, which present significantly less scatter 
but do not constrain these stars' periods as tightly, are shown in the 
Appendix.

\noindent {\bf NV112, NV120, NV137, NV187, NV194}: These stars present several 
aliases, and could not be detected in the CTIO data. The periods adopted are the 
ones that appear most consistent with an RRab type. 

\noindent {\bf NV134, NV149}: These stars present several aliases, and could not 
be detected in the CTIO data. The periods adopted are the ones that appear most 
consistent with an RRc type. 

\noindent {\bf NV159, NV169}: These stars show a curious mismatch in the light 
curves around phase 0.8 for the adopted periods. However, the latter are confirmed 
on the basis of the CTIO data.

\noindent {\bf NV170}: This star presents several aliases, and an uncertain 
classification. 

\noindent {\bf NV215}: This star's light curve presents an unusual behavior 
close to minimum light. A similar behavior is found in both the LCO and CTIO
datasets, though the latter is considerably more noisy and contains fewer 
datapoints.

\section{Fourier Decomposition}\label{sec:fourier}
In the Fourier decomposition method, the light curves of ab-type pulsating stars are
frequently fitted with a Fourier series of the form

\begin{equation}
mag=A_{0}+\sum_{j=1}^n A_{j}\,\sin(jwt+\phi_{j})\,,
\end{equation}

\noindent where $w \equiv 2\pi/P$. 
The light curve shape is then quantified 
in terms of the lower-order ($j = 2-4$) coefficients 
$A_{j1}=A_{j}/A_{1}$ and $\phi_{j1}=\phi_{j}-j\phi_{1}$. In the case of c-type RR Lyrae, 
a similar procedure is followed, but a cosine decomposition is frequently used instead. 
In our study, we performed such Fourier decomposition of the RR Lyrae light curves, 
using $n = 10$, and adopting a sine series for the RRab and a cosine series for the RRc 
stars. Amplitude ratios $A_{j1}$ and phase differences $\phi_{j1}$ for the lower-order
terms are provided in Tables~\ref{tab:firrc} and \ref{tab:firrab} for the RRc and 
RRab stars, respectively. For the RRab stars we also give the Jurcsik-Kov\'acs $D_m$ 
value \citep[][ computed on the basis of their eq.~6 and Table~6]{jk96}, which is 
intended to differentiate RRab stars 
with ``regular'' light curves from those with ``anomalous'' light curves, such as 
those presenting the Blazhko effect 
\citep*[but see][ for a critical discussion of $D_m$ as an indicator of the 
occurrence of the Blazhko phenomenon]{ccea05}. 
In these tables, a colon symbol (``:'') 
indicates an uncertain value, whereas a double 
colon (``::'') indicates a very uncertain value, the latter being provided for 
completeness only. 
The error in the $\phi_{31}$ coefficient was obtained from equation~(16d) of 
\citet{jp86}.

\subsection{RRc Variables}\label{sec:rrc}
\citet{sc93} demonstrated, based on hydrodynamical models, that Fourier 
decomposition of RRc light curves can potentially provide a very useful 
technique for determining physical parameters of these stars. As a matter
of fact, they have provided equations relating the masses, luminosities, 
temperatures, and even a ``helium abundance parameter'' of c-type RR Lyrae 
stars to their periods and $\phi_{31}$ values. Althought these equations 
have been widely use in literature they must be used with some caution, 
since a combination of their equations for the RR Lyrae masses and 
luminosities gives results that are inconsistent with the period-mean 
density equation of stellar pulsation theory \citep{mc04,ds09}.
Accordingly, while we still provide luminosities, masses and temperatures 
derived on the basis of the \citeauthor{sc93} relations, we warn the reader 
that these quantities cannot all be simultaneously valid, and should 
accordingly be used for comparison with similar work for other GCs only. 

Based on the \citet{sc93} relations, we find that an error of 0.2 in $\phi_{31}$ 
leads to an error of $\sim 0.03 \, M_{\odot}$ in mass and $\sim 0.03$~mag in 
(bolometric) magnitude, and so we apply this method only to RRc stars with 
errors in $\phi_{31}$ of 0.2 or less. We thus computed values of $M/M_{\odot}$, 
$\log  (L/L_{\odot})$, $\log  T_{\rm eff}$, and ``helium abundance parameter'' 
$y$ (which, as is well known, is not necessarily equal to the helium abundance $Y$; 
see, e.g., \citeauthor{mcea03} \citeyear{mcea03}); 
the resulting values are provided in Table~\ref{tab:pfrrc}, where we also 
provide ${\rm [Fe/H]}_{\rm ZW84}$ 
values \citep[in the][ scale]{zw84}, based on the calibration 
recently provided by \citet*{smea07}, and $M_V$ values, based 
on the calibration by \citet{gk98}.

%Parametros de Fourier para 80 RRab
%\begin{longtable*}{lccccccl}
\begin{deluxetable}{lccccccl}
\tablewidth{0pc} 
\tabletypesize{\scriptsize}
\tablecaption{Fourier Coefficients for RRab Variables in M62}
\tablehead{\colhead{ID  } & \colhead{$A_{21}$} & \colhead{$A_{31}$}
 & \colhead{$A_{41}$} & \colhead{$\phi_{21}$}&
\colhead{$\phi_{31}$} & \colhead{$\phi_{41}$} & \colhead{$D_{m}$} }
\startdata
V4       & 0.452 & 0.315 & 0.256 & 2.392 & 4.780 & 1.046  &   4.9   \\
V6:      & 0.455 & 0.301 & 0.220 & 2.351 & 4.929 & 1.237  &   2.7   \\
V7       & 0.523 & 0.341 & 0.200 & 2.638 & 5.468 & 2.141  &   2.2   \\
V8       & 0.520 & 0.356 & 0.218 & 2.354 & 5.130 & 1.563  &   3.5   \\
V10      & 0.511 & 0.351 & 0.234 & 2.363 & 5.101 & 1.526  &   1.2   \\
V11::    & 0.452 & 0.228 & 0.138 & 2.738 & 5.493 & 1.910  &   6.1   \\
V16      & 0.500 & 0.336 & 0.171 & 2.624 & 5.368 & 2.132  &   5.9   \\
V17      & 0.555 & 0.380 & 0.251 & 2.369 & 5.059 & 1.556  &    3.   \\
V18      & 0.492 & 0.340 & 0.221 & 2.378 & 5.153 & 1.645  &   3.1   \\
V20      & 0.476 & 0.394 & 0.257 & 2.281 & 4.878 & 1.087  &   3.4   \\
V21      & 0.476 & 0.364 & 0.221 & 2.218 & 4.772 & 1.032  &   2.4   \\
V23      & 0.547 & 0.392 & 0.238 & 2.084 & 4.474 & 0.628  &  42.9   \\
V24      & 0.508 & 0.344 & 0.264 & 2.377 & 5.204 & 1.608  &   6.2   \\
V25      & 0.473 & 0.343 & 0.216 & 2.225 & 4.796 & 1.142  &   1.7   \\
V26      & 0.569 & 0.288 & 0.186 & 2.519 & 5.227 & 1.701  &   2.9   \\
V27      & 0.478 & 0.353 & 0.224 & 2.288 & 4.891 & 1.261  &   2.5   \\
V28::    & 0.786 & 0.321 & 1.381 & 0.782 & 5.934 & 5.959  &1617.8   \\
V29      & 0.535 & 0.317 & 0.251 & 2.422 & 5.142 & 1.620  &   4.6   \\
V31:     & 0.523 & 0.381 & 0.236 & 2.268 & 4.947 & 1.152  &    4.   \\
V32      & 0.517 & 0.330 & 0.221 & 2.466 & 5.297 & 1.863  &   1.7   \\
V33      & 0.549 & 0.341 & 0.209 & 2.523 & 5.313 & 2.004  &   2.5   \\
V34      & 0.529 & 0.314 & 0.188 & 2.559 & 5.264 & 1.868  &   1.3   \\
V35      & 0.487 & 0.339 & 0.223 & 2.363 & 5.076 & 1.557  &   0.6   \\
V36      & 0.459 & 0.278 & 0.105 & 2.646 & 5.617 & 2.238  &   2.9   \\
V39      & 0.429 & 0.211 & 0.080 & 2.610 & 5.735 & 2.845  &   7.2   \\
V41      & 0.495 & 0.306 & 0.177 & 2.529 & 5.391 & 2.071  &   1.6   \\
V43      & 0.529 & 0.342 & 0.212 & 2.433 & 5.177 & 1.736  &   2.2   \\
V44      & 0.392 & 0.211 & 0.092 & 2.370 & 4.808 & 0.977  &  46.9   \\
V48      & 0.444 & 0.271 & 0.088 & 2.812 & 5.914 & 2.697  & 127.6   \\
V49      & 0.517 & 0.313 & 0.197 & 2.338 & 5.090 & 1.596  &    1.   \\
V50::    & 1.070 & 1.287 & 1.461 & 1.771 & 3.458 & 5.002  &  590.   \\
V52::    & 1.178 & 1.808 & 2.167 & 1.873 & 3.279 & 4.364  &3585.7   \\
V56      & 0.501 & 0.314 & 0.194 & 2.472 & 5.314 & 1.902  &   1.5   \\
V57      & 0.478 & 0.304 & 0.170 & 2.441 & 5.197 & 1.637  &   2.1   \\
V58::    & 1.058 & 0.888 & 0.674 & 4.873 & 3.491 & 2.047  & 204.9   \\
V59      & 0.531 & 0.334 & 0.191 & 2.567 & 5.404 & 2.086  &   2.8   \\
V62      & 0.516 & 0.330 & 0.212 & 2.423 & 5.184 & 1.693  &   1.5   \\
V63      & 0.461 & 0.264 & 0.102 & 2.741 & 5.692 & 2.740  &   5.6   \\
V64:     & 0.450 & 0.287 & 0.163 & 2.266 & 4.611 & 1.000  &   7.1   \\
V72      & 0.418 & 0.231 & 0.121 & 2.349 & 5.051 & 1.265  &  39.6   \\
V78:     & 0.491 & 0.318 & 0.168 & 2.682 & 5.546 & 2.297  &   3.9   \\
V80:     & 0.445 & 0.229 & 0.138 & 2.541 & 5.285 & 2.428  &   18.   \\
V81      & 0.528 & 0.360 & 0.216 & 2.411 & 5.153 & 1.634  &   1.7   \\
V82      & 0.489 & 0.292 & 0.145 & 2.536 & 5.400 & 2.278  &   2.7   \\
NV84     & 0.388 & 0.170 & 0.043 & 2.774 & 5.980 & 2.844  &  109.   \\
NV87     & 0.288 & 0.115 & 0.059 & 2.774 & 5.846 & 3.653  & 118.9   \\
NV88     & 0.496 & 0.301 & 0.137 & 2.608 & 5.508 & 2.222  &   1.7   \\
NV93     & 0.519 & 0.368 & 0.230 & 2.429 & 5.132 & 1.648  &    2.   \\
NV95:    & 0.466 & 0.342 & 0.203 & 2.138 & 4.648 & 0.846  &  45.6   \\
NV96     & 0.481 & 0.359 & 0.224 & 2.259 & 4.924 & 1.232  &   2.8   \\
NV97     & 0.509 & 0.313 & 0.203 & 2.445 & 5.274 & 1.820  &   2.1   \\
NV98     & 0.501 & 0.328 & 0.207 & 2.486 & 5.323 & 1.812  &   1.8   \\
NV99::   & 0.426 & 0.259 & 0.149 & 2.753 & 6.272 & 3.011  &  11.7   \\
NV102    & 0.421 & 0.229 & 0.103 & 2.646 & 5.465 & 2.292  &   1.3   \\
NV103::  & 0.872 & 0.726 & 0.556 & 4.808 & 3.251 & 1.729  & 298.9   \\
NV105    & 0.540 & 0.363 & 0.259 & 2.371 & 5.087 & 1.546  &   4.9   \\
NV106    & 0.424 & 0.302 & 0.220 & 2.336 & 4.881 & 1.328  &   4.2   \\
NV107    & 0.479 & 0.307 & 0.140 & 2.555 & 5.462 & 2.088  &   3.6   \\
NV109:   & 0.461 & 0.293 & 0.144 & 2.626 & 5.606 & 2.385  &   8.8   \\
NV112::  & 0.936 & 0.839 & 0.730 & 1.672 & 3.466 & 5.340  &2030.9   \\
NV113::  & 0.359 & 0.448 & 0.176 & 2.170 & 4.619 & 1.514  &  16.7   \\
NV116    & 0.375 & 0.165 & 0.066 & 2.643 & 5.776 & 2.954  &   7.8   \\
NV120::  & 0.872 & 0.741 & 0.649 & 1.613 & 3.284 & 4.894  &1632.9   \\
NV124::  & 0.815 & 0.698 & 0.548 & 4.734 & 3.247 & 1.697  & 276.1   \\
NV127    & 0.529 & 0.321 & 0.248 & 2.307 & 5.049 & 1.395  &   3.5   \\
NV135    & 0.460 & 0.300 & 0.107 & 2.636 & 5.420 & 2.623  & 108.4   \\
NV136    & 0.544 & 0.330 & 0.173 & 2.695 & 5.440 & 2.250  &   7.7   \\
NV150    & 0.343 & 0.146 & 0.019 & 2.535 & 5.053 & 1.546  &  43.9   \\
NV160    & 0.525 & 0.358 & 0.225 & 2.394 & 5.272 & 1.772  &   3.2   \\
NV168    & 0.261 & 0.088 & 0.023 & 2.864 & 5.627 & 5.322  & 115.6   \\
NV194::  & 0.924 & 0.775 & 0.623 & 1.644 & 3.280 & 4.853  & 762.4   \\
NV198:   & 0.561 & 0.319 & 0.168 & 2.317 & 5.101 & 1.824  &   8.9   \\
NV210    & 0.466 & 0.335 & 0.183 & 2.339 & 4.859 & 1.265  &   7.5   \\
NV214    & 0.502 & 0.328 & 0.167 & 2.489 & 5.325 & 1.987  &   1.8   \\
NV218:   & 0.443 & 0.271 & 0.144 & 2.253 & 4.759 & 0.771  &  41.9   \\
NV219    & 0.461 & 0.249 & 0.112 & 2.816 & 5.976 & 2.706  &   7.8   \\
NV223    & 0.519 & 0.337 & 0.208 & 2.386 & 5.101 & 1.584  &   1.1   \\
NV226    & 0.511 & 0.303 & 0.161 & 2.663 & 5.588 & 2.349  &   2.2   \\
NV227    & 0.486 & 0.326 & 0.195 & 2.304 & 4.796 & 1.174  &  44.8   \\
NV228:   & 0.329 & 0.113 & 0.031 & 2.680 & 5.805 & 3.165  & 132.3
\enddata
\label{tab:firrab}
\end{deluxetable}
%\end{longtable*}

%Para&metros Fisicos 21 RRc											
\begin{deluxetable*}{lcccccccc}
\tablewidth{0pc} 
\tabletypesize{\scriptsize}
\tablecaption{Physical Parameters derived for RRc Variables in M62}
\tablehead{\colhead{ID  } & \colhead{$M/M_{\odot}$} & \colhead{$\log(L/L_{\odot})$}
 & \colhead{$\log T_{\rm eff}$} & \colhead{$y$} & \colhead{${\rm [Fe/H]}$} & \colhead{$\langle M_{v} \rangle$}}
\startdata
 V30  & 0.573 &1.681 &3.868 &0.282 &-1.418 &0.733\\
 V40  & 0.525 &1.658 &3.871 &0.292 &-1.180 &0.686\\
 V53  & 0.588 &1.651 &3.873 &0.290 &-1.207 &0.699\\
 V69  & 0.496 &1.658 &3.870 &0.294 &-1.136 &0.725\\
NV85  & 0.492 &1.663 &3.869 &0.293 &-1.173 &0.719\\
NV86  & 0.524 &1.646 &3.872 &0.295 &-1.082 &0.746\\
NV91  & 0.494 &1.660 &3.869 &0.293 &-1.154 &0.714\\
NV100 & 0.608 &1.651 &3.874 &0.289 &-1.226 &0.736\\
NV101 & 0.534 &1.666 &3.870 &0.289 &-1.259 &0.689\\
NV108 & 0.526 &1.655 &3.871 &0.293 &-1.161 &0.760\\
NV110 & 0.490 &1.678 &3.867 &0.288 &-1.297 &0.652\\
NV115 & 0.609 &1.654 &3.873 &0.288 &-1.253 &0.736\\
NV118 & 0.539 &1.661 &3.870 &0.290 &-1.224 &0.723\\
NV119 & 0.519 &1.674 &3.868 &0.288 &-1.305 &0.719\\
NV121 & 0.573 &1.654 &3.872 &0.290 &-1.209 &0.756\\
NV123 & 0.500 &1.650 &3.871 &0.296 &-1.075 &0.730\\
NV131 & 0.517 &1.667 &3.869 &0.290 &-1.247 &0.703\\
NV140 & 0.435 &1.694 &3.862 &0.287 &-1.347 &0.613\\
NV153 & 0.527 &1.671 &3.869 &0.288 &-1.293 &0.708\\
NV175 & 0.586 &1.660 &3.872 &0.288 &-1.274 &0.734\\
NV224 & 0.538 &1.683 &3.867 &0.284 &-1.397 &0.713\\
mean  &0.533$\pm$0.043 &1.663$\pm$0.012 &3.870$\pm$0.002 &0.290$\pm$0.003 & 
-1.234$\pm$0.009 &0.714$\pm$0.033
\enddata
\label{tab:pfrrc}
\end{deluxetable*}

%Parametros Fisicos 40 RRab...la V26 no en los valores medios
\begin{deluxetable*}{lcccccccc}
\tablewidth{0pc} 
\tabletypesize{\scriptsize}
\tablecaption{Physical Parameters Derived for RRab Variables in M62}
\tablehead{\colhead{ID  } & 
\colhead{${\rm [Fe/H]_{J95}}$} & 
\colhead{$\langle M_{v} \rangle$}& \colhead{$\langle B-V \rangle$} & 
\colhead{$\log T_{\rm eff}^{\langle B-V \rangle}$}& \colhead{$\langle V-I \rangle$} &
\colhead{$\log T_{\rm eff}^{\langle V-I \rangle}$}& \colhead{$\langle V-K \rangle$} &
\colhead{$\log T_{\rm eff}^{\langle V-K \rangle}$} }
\startdata
  V4  &-1.527 & 0.782 & 0.328 & 3.814 & 0.479 & 3.814 & 1.110 & 3.810 \\
  V7  &-0.725 & 0.840 & 0.343 & 3.815 & 0.498 & 3.808 & 1.057 & 3.813 \\ 
  V8  &-1.011 & 0.843 & 0.337 & 3.815 & 0.491 & 3.810 & 1.058 & 3.814 \\
 V10  &-1.049 & 0.808 & 0.322 & 3.819 & 0.472 & 3.815 & 1.044 & 3.816 \\
 V17  &-1.089 & 0.830 & 0.334 & 3.815 & 0.487 & 3.811 & 1.064 & 3.814 \\
 V18  &-0.934 & 0.848 & 0.330 & 3.818 & 0.481 & 3.813 & 1.041 & 3.816 \\
 V20  &-1.023 & 0.867 & 0.316 & 3.822 & 0.464 & 3.817 & 0.991 & 3.822 \\
 V21  &-1.047 & 0.868 & 0.301 & 3.828 & 0.444 & 3.823 & 0.974 & 3.824 \\
 V25  &-0.993 & 0.875 & 0.296 & 3.830 & 0.439 & 3.824 & 0.969 & 3.824 \\
 V26  &-0.012 & 1.047 & 0.286 & 3.842 & 0.424 & 3.827 & 0.824 & 3.837 \\
 V27  &-0.882 & 0.868 & 0.293 & 3.831 & 0.435 & 3.825 & 0.952 & 3.826 \\
 V29  &-1.170 & 0.768 & 0.324 & 3.817 & 0.475 & 3.814 & 1.082 & 3.812 \\
 V32  &-0.869 & 0.843 & 0.338 & 3.815 & 0.493 & 3.809 & 1.059 & 3.814 \\
 V33  &-0.980 & 0.813 & 0.345 & 3.812 & 0.501 & 3.807 & 1.097 & 3.810 \\
 V34  &-1.104 & 0.772 & 0.334 & 3.814 & 0.487 & 3.811 & 1.101 & 3.810 \\
 V35  &-1.065 & 0.847 & 0.337 & 3.815 & 0.490 & 3.810 & 1.069 & 3.813 \\
 V36  &-1.004 & 0.774 & 0.369 & 3.803 & 0.532 & 3.799 & 1.164 & 3.802 \\
 V41  &-0.802 & 0.862 & 0.349 & 3.812 & 0.506 & 3.806 & 1.078 & 3.811 \\
 V43  &-1.114 & 0.828 & 0.350 & 3.809 & 0.508 & 3.806 & 1.110 & 3.809 \\
 V49  &-1.122 & 0.825 & 0.335 & 3.815 & 0.488 & 3.811 & 1.083 & 3.812 \\
 V56  &-0.919 & 0.837 & 0.344 & 3.813 & 0.500 & 3.808 & 1.081 & 3.811 \\
 V57  &-1.048 & 0.831 & 0.342 & 3.813 & 0.497 & 3.809 & 1.084 & 3.811 \\
 V59  &-0.892 & 0.828 & 0.351 & 3.810 & 0.509 & 3.805 & 1.096 & 3.809 \\
 V62  &-1.021 & 0.834 & 0.340 & 3.814 & 0.494 & 3.809 & 1.076 & 3.812 \\
 V81  &-0.970 & 0.843 & 0.335 & 3.816 & 0.488 & 3.811 & 1.052 & 3.815 \\
 V82  &-0.821 & 0.875 & 0.359 & 3.809 & 0.518 & 3.803 & 1.107 & 3.808 \\
NV88  &-0.762 & 0.854 & 0.357 & 3.809 & 0.516 & 3.803 & 1.093 & 3.809 \\
NV93  &-1.113 & 0.824 & 0.344 & 3.812 & 0.500 & 3.808 & 1.091 & 3.811 \\
NV96  &-0.930 & 0.874 & 0.309 & 3.826 & 0.454 & 3.820 & 0.979 & 3.823 \\
NV97  &-0.915 & 0.849 & 0.343 & 3.813 & 0.498 & 3.808 & 1.072 & 3.812 \\
NV98  &-0.909 & 0.836 & 0.345 & 3.813 & 0.501 & 3.807 & 1.073 & 3.812 \\
NV102 &-1.089 & 0.799 & 0.368 & 3.803 & 0.531 & 3.800 & 1.181 & 3.801 \\
NV105 &-1.002 & 0.837 & 0.327 & 3.818 & 0.478 & 3.813 & 1.040 & 3.816 \\
NV106 &-1.189 & 0.811 & 0.305 & 3.824 & 0.451 & 3.821 & 1.038 & 3.817 \\
NV107 &-0.781 & 0.868 & 0.360 & 3.809 & 0.519 & 3.803 & 1.090 & 3.810 \\
NV127 &-1.120 & 0.783 & 0.310 & 3.823 & 0.457 & 3.819 & 1.037 & 3.817 \\
NV160 &-0.879 & 0.810 & 0.325 & 3.820 & 0.476 & 3.814 & 1.033 & 3.816 \\
NV214 &-1.043 & 0.817 & 0.356 & 3.808 & 0.514 & 3.804 & 1.124 & 3.807 \\
NV223 &-1.048 & 0.836 & 0.333 & 3.816 & 0.485 & 3.812 & 1.063 & 3.814 \\
NV226 &-0.921 & 0.782 & 0.359 & 3.807 & 0.519 & 3.802 & 1.138 & 3.805 \\
mean\tablenotemark{a}  
      &-0.997$\pm$0.144 & 0.830$\pm$0.030 & 0.336$\pm$0.019 & 3.815$\pm$0.006 & 0.489$\pm$0.024 & 
3.811$\pm$0.006 & 1.068$\pm$0.049 & 3.813$\pm$0.005 
\enddata
\tablenotetext{a}{Excluding V26.}
\label{tab:pfrrab}
\end{deluxetable*}

As the reader will readily notice, many of the mass values given in 
Table~\ref{tab:pfrrc} are too low, approaching the mass of the degenerate 
helium core at the He flash 
\citep[$\simeq 0.5 \, M_{\odot}$; see][ for a recent review]{mc09}. Such low mass 
values, which are not uncommon in the literature 
\citep[e.g.,][ and reference therein]{mcea03} likely confirm the existence of a 
problem with the \citet{sc93} calibration equations. 

For the 21 retained RRc stars the unweighted mean values and standard deviations 
of the mass, log luminosity, effective temperature and helium parameter are 
($0.533 \pm 0.04)\, M/M_{\odot}$, $1.663 \pm 0.01$, ($7413 \pm 34$)~K, and 
$0.293 \pm 0.003$, respectively. The mean metallicity, in turn, is found to 
be ${\rm [Fe/H]}_{\rm ZW84} = -1.23 \pm 0.09$. 

According to the \citet{gk98} calibration, the mean absolute magnitude in $V$ of 
these RRc stars turns out to be $\langle M_V\rangle = 0.714 \pm 0.033$. Since 
for these stars we also have a $\langle V\rangle = 16.44 \pm 0.06$~mag 
(standard error of the mean), this gives for the 
cluster an apparent distance modulus of $(m-M)_V = 15.73 \pm 0.068$~mag.

\subsection{RRab Variables}\label{sec:rrab}
In a series of papers, the Hungarian team has provided a calibration of several 
physical parameters of ``well-behaved'' (as indicated by the aforementioned 
$D_m$ parameter) ab-type RR Lyrae stars as a function of their Fourier 
decomposition parameters \citep[e.g.,][]{jk96,jj98,kw99,kw01}. 
Unlike the approach adopted by \citet{sc93} for the RRc stars, 
their method does not rely on hydrodynamical models for the calibration.  
Following the same approach as described in detail in \citet{mcea03}, we obtain
the metallicities, mean colors, and associated temperatures that are listed 
in Table~\ref{tab:pfrrab}, for 40 RRab stars with $D_{m} \le 5$
\citep[see also][]{cs97}. Note that V26 is most likely a field star 
(see \S\ref{sec:cmd}), and therefore was not taken into account when 
computing the average values for the cluster, as indicated in this table. 

Note that the [Fe/H] values derived in this way are actually in the scale of 
\citet{jj95}. The latter is related to the more traditional \citet{zw84} scale 
by ${\rm [Fe/H]}_{\rm J95}=1.431 \, {\rm [Fe/H]}_{\rm ZW84} + 0.880$. Therefore, 
the mean metallicity ${\rm [Fe/H]}_{\rm J95} = -0.997$ that was derived for the 
cluster in Table~\ref{tab:pfrrab} translates into a metallicity value 
${\rm [Fe/H]}_{\rm ZW84} = -1.31$ in the \citeauthor{zw84} scale.
This agrees very well with the value adopted for the cluster by \citet{wh96}, 
namely ${\rm [Fe/H]}_{\rm ZW84} = -1.29$, in his catalog of globular cluster 
parameters (2003 update), as well as with the value derived from the RRc by 
using the \citet{smea07} calibration, namely ${\rm [Fe/H]}_{\rm ZW84} = -1.23$
(\S\ref{sec:rrc}). 

Likewise, we obtain a 
mean absolute magnitude of $\langle M_{V} \rangle = 0.83 \pm 0.03$~mag for 
the RRab stars in the cluster. The faint HB is a reflection of the adoption of 
the Baade-Wesselink luminosity zero point in the calibration of this method 
\citep[see][ for a discussion]{jk99}. 
For the same set of 39 RRab used to derive this 
value, we also find $\left\langle V_{RR} \right\rangle = 
16.260 \pm 0.03$~mag (standard error of the mean),  
which is also in very good agreement with the value of 16.25~mag adopted in 
the 2003 edition of the \citet{wh96} catalog. This implies 
an apparent distance modulus of $(m-M)_V = 15.43 \pm 0.04$~mag for M62,
which is significantly shorter (by 0.21~mag) than the value provided in the 
\citeauthor{wh96} catalog, and by an even wider margin (i.e., 0.3~mag) than 
the value obtained in \S\ref{sec:rrc} on the basis of the \citet{gk98} 
$M_V$ calibration for the c-type RR Lyrae. We ascribe these differences 
to the faint zero point adopted in the original $M_V$ calibrations. 
If we adopt instead the more recent calibration of the RR Lyrae absolute 
magnitude-metallicity relation provided by \citet{cc08}, and the metallicity 
value for M62 derived above (${\rm [Fe/H]}_{\rm ZW84}=-1.31$), we find 
$M_{V}(RR)=0.68 \pm 0.14$, and an apparent distance modulus of 
$(m-M)_V=15.58 \pm 0.14$, which is much more consistent with the value 
reported in the \citeauthor{wh96} catalog 
(being shorter by only 0.06~mag). Using a reddening value of $E(\bv) = 0.47$
\citep[from][]{wh96} and a standard extinction law with $A_V/E(\bv) = 3.1$, 
this implies a distance modulus $(m-M)_0=14.12\pm 0.14$, which corresponds to 
a distance of $6.7\pm 1$~kpc. 

Note that a distance modulus for the cluster may also be obtained on the basis 
of our detected type II Cepheids, namely V2 and V73, using equation~(3) in 
\citet{bpea03}. In this way, we obtain for distance moduli of $(m-M)_V=15.04$ 
and $(m-M)_V=15.57$~mag, respectively~-- giving an average distance modulus 
of $(m-M)_V \approx 15.31\pm 0.26$~mag. Given the large error bar, this value 
is not inconsistent with the one derived on the basis of the RR Lyrae stars. 

\begin{figure*}
\begin{center}
\includegraphics[height=16.0cm,width=16.0cm]{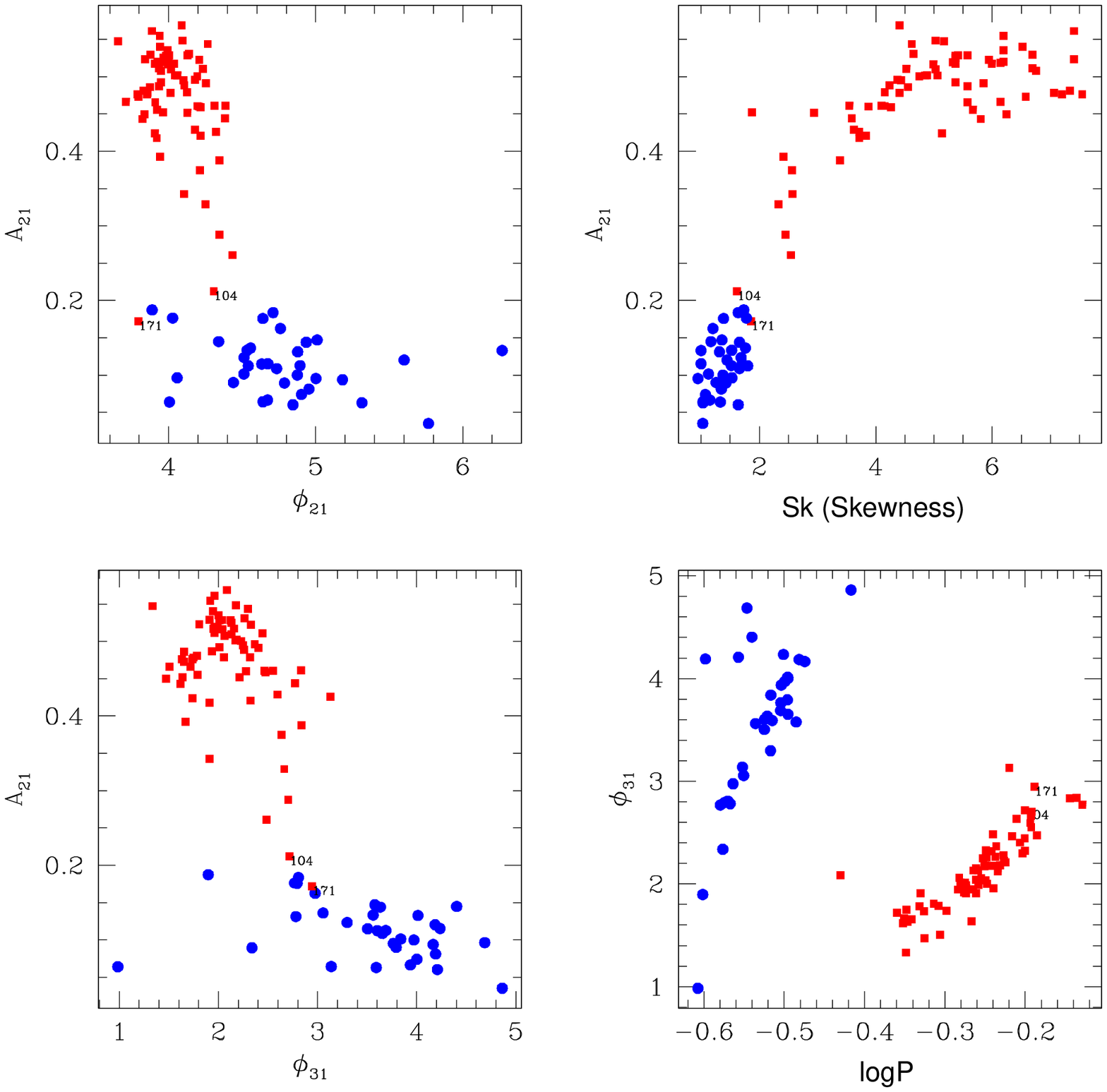}
\end{center}
\caption{Pulsation mode diagnostics for RR Lyrae stars. In all panels, circles 
 indicate c-type RR Lyrae stars, whereas squares indicate ab-type RR Lyrae. Some
 of the variables discussed in the text are indicated by their V (or NV) number.} 
\label{fig:fourier}
\end{figure*}

As noted by \citet{rcea05}, M62 may harbor long-period RRc's (see their Fig.~2), 
which are exceedingly rare among Galactic globular clusters \citep[see][ for a review]{mc04}.
In order to check the pulsation status of the two candidate long-period RRc stars that
we have found in the cluster, namely NV104 and NV171, we have used several diagnostics  
from \citet{st82} and \citet{cs97}, who have shown that the RRab and RRc stars 
occupy distinctly different positions in the $A_{21}$, $\phi_{21}$ plane in 
particular, as well as the Sk (skewness) parameter defined by \citet{sd87}. 
Figure~\ref{fig:fourier} shows that, for the RR Lyrae with clean 
light curves in our sample, most of the ab-type RR Lyrae do indeed have values 
of $A_{21} > 0.3$, and vice-versa for the RRc stars. Similarly, most of the RRc 
stars have ${\rm Sk} < 2$, whereas most of the ab-type RR Lyrae have ${\rm Sk} > 2$.
As can be seen, in all plots but the one showing $\phi_{31}$ as a function of 
$\log P$ one finds that the positions of these two stars are closer to the 
locus occupied by RRc than RRab stars. The atypical position of NV104 and NV171
in the $\phi_{31} - \log P$ plane is particularly intriguing, in view of the fact 
that, if these stars are indeed c-type RR Lyrae, their periods would clearly be 
longer than the vast majority of even the ab-type RR Lyrae in the cluster.

%Plots parametros Fourier

\section{CMD and Reddening}\label{sec:cmd}
\begin{figure*}[!t]
\includegraphics[width=17.0cm]{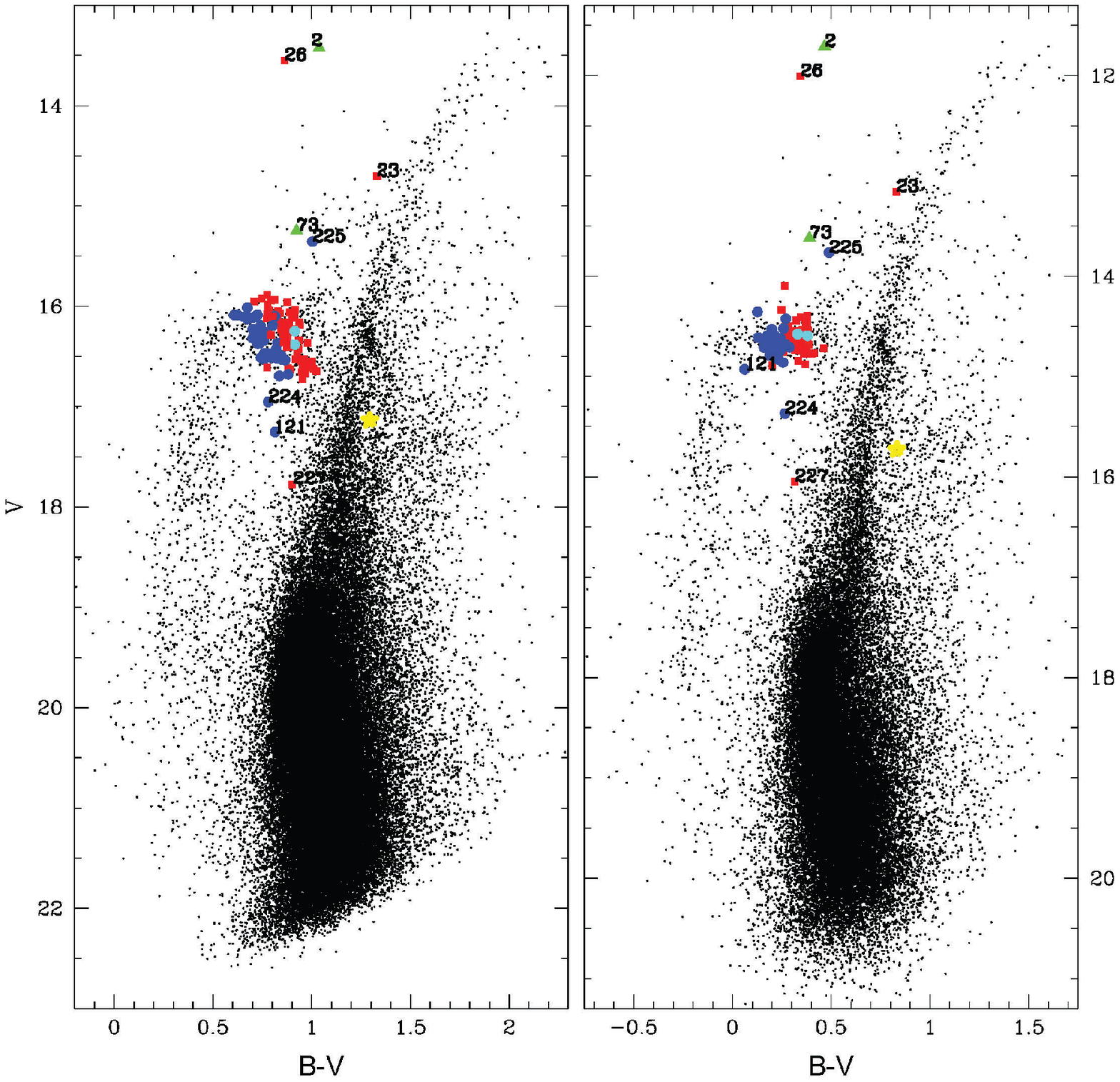}
\caption{Our derived CMD for the cluster, with the mean values for the detected variable
 stars overplotted. Circles indicate c-type RR Lyrae, squares ab-type RR Lyrae, 
 and triangles type II Cepheids. The diagram on the right is the same as the one 
 on the right, but with differential reddening accounted for as described in the text.} 
\label{fig:cmd}
\end{figure*}

On the basis of our ALLFRAME reductions, we were able to obtain a deep CMD for M62, 
which we show in Figure~\ref{fig:cmd}. The CMD properties will be discussed in detail 
in a forthcoming paper (Contreras et al. 2010, in preparation), and we show it here 
with the main purpose of verifying whether the positions of the variable stars that 
were detected in our field are consistent with cluster membership~-- which is 
clearly confirmed for the vast majority of the stars. One obvious exception is 
provided by the RRab star V26, which is clearly a foreground field RR Lyrae. That 
V26 is a field star is also suggested by the near-solar metallicity derived for it 
on the basis of its Fourier decomposition parameters (see Table~\ref{tab:pfrrab}). 
The membership status of NV224 and NV227, on the other hand, is less clear, for while
their CMD positions suggest that they may be RR Lyrae stars in the cluster background, 
their Fourier-based metallicities do not clearly point to them as being anomalous. 
In like vein, their metallicity values, as derived using the \citet{jj95} and 
\citet{smea07} techniques, suggest [Fe/H] values of $-1.40$ (for NV224) and 
$-1.35$ (for NV227), neither of which is clearly inconsistent with the cluster's 
metallicity. The derived [Fe/H] value for NV227 should be taken with due caution 
though, in view of the star's fairly large $D_m$ value (see Table~\ref{tab:firrab}).  

Unfortunately, as can be seen from Figure~\ref{fig:cmd} (left panel), the cluster CMD 
is severely affected by differential reddening, which is not unexpected in view of 
M62's large foreground reddening and low Galactic latitude. On the other hand, the 
presence of a large number of RR Lyrae variable stars across the face of the cluster 
can provide us with a handle of this problem, since RR Lyrae stars can themselves 
provide dependable reddening estimates, particularly on the basis of the colors of 
the ab-type RR Lyrae at minimum light \citep[e.g.,][]{bb92}.

%CMDcombine

\begin{figure*}[!t]
\includegraphics[width=16.75cm]{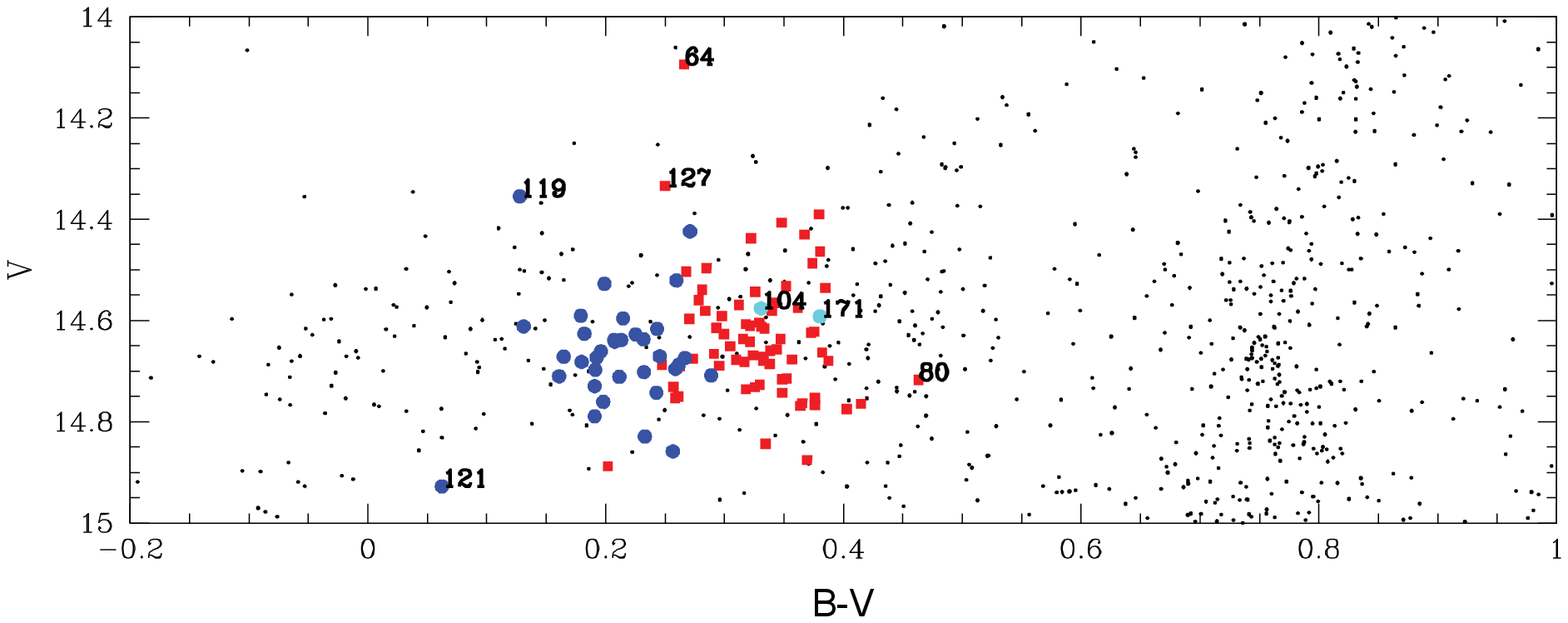}
\caption{As in Figure~\ref{fig:cmd} (right panel), but zooming in around the HB.} 
\label{fig:hbcmd}
\end{figure*}

We have applied the \citet{bb92} technique to 71 stars in our RRab sample, and thus 
obtained a two-dimensional reddening map across the face of the cluster. In this 
case, we adopted the same [Fe/H] value for all the RRab stars, namely 
${\rm [Fe/H]} = -1.31$, as derived from Fourier decomposition (\S\ref{sec:rrab}), 
and which is very similar to the value listed in the \citet{wh96} catalog, namely 
${\rm [Fe/H]} = -1.29$. We then 
experimented with several different techniques for interpolating on this map to obtain 
reddening values for individual cluster stars, finally opting for a LOESS smoother
\citep{wc79,cd88}. 
That this provided very good results can readily be appreciated by comparing the 
differential reddening-corrected CMD (Fig.~\ref{fig:cmd}, right panel) with the 
original one. A zoomed-in plot around the HB region is shown in Figure~\ref{fig:hbcmd}. 

We also note the anomalous positions of stars NV225 (an RRc) and 
V23 (an RRab) in the CMD. Not only are these stars brighter and redder than other RR 
Lyrae stars in the cluster, but also~-- and importantly~-- they also present peculiarly 
large $A_B/A_V$ amplitude ratios. This strongly suggests that they are blended with 
redder companions.

\begin{figure}[!t]
\includegraphics[width=8.6cm]{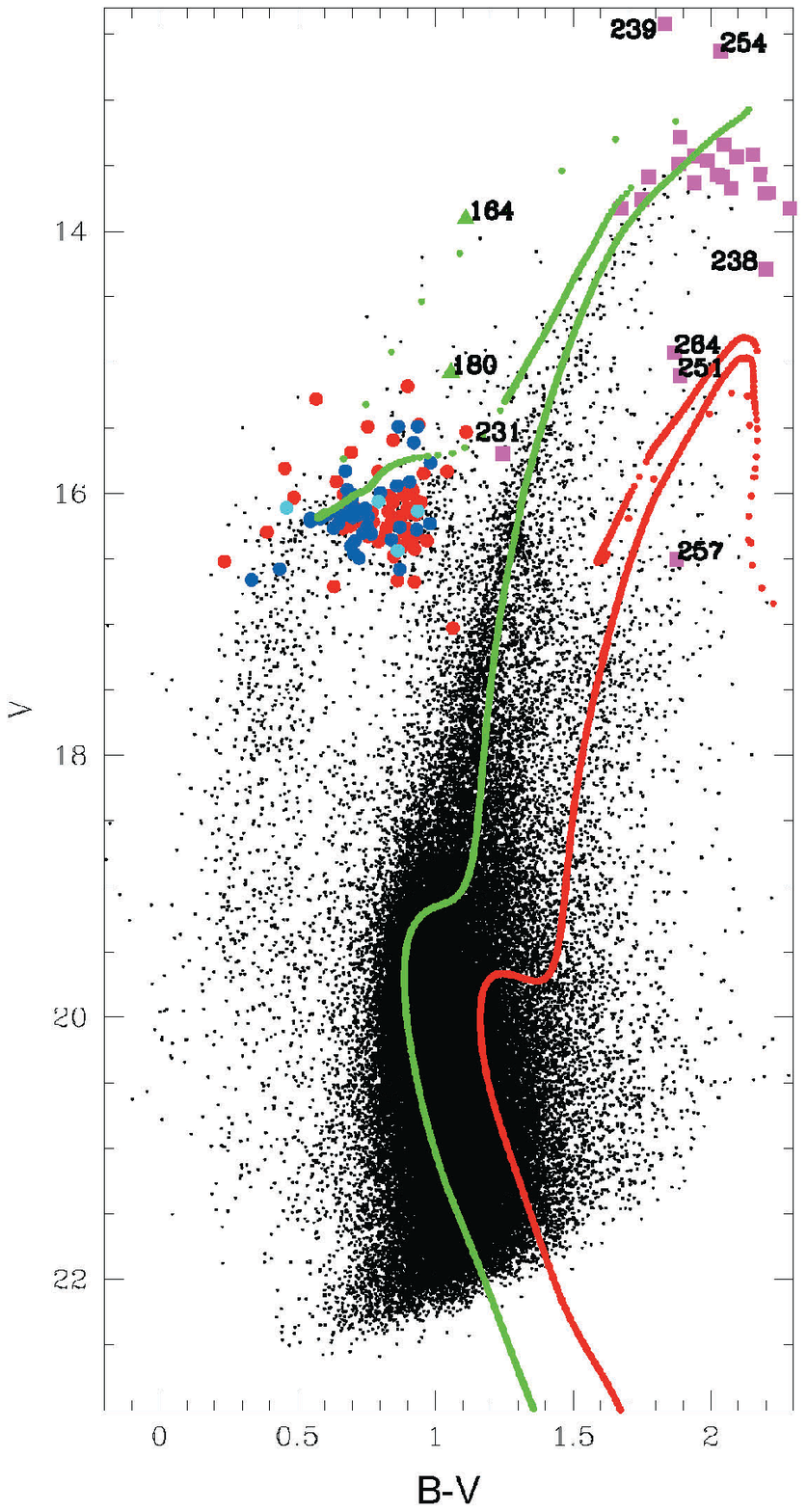}
\caption{As in Figure~\ref{fig:cmd}, but including the variable stars for which the average
 magnitudes and colors can only be roughly estimated, due to insufficient phase coverage. 
 Filled squares indicate candidate LPV stars. Isochrones for characteristic cluster 
 and bulge chemical compositions are shown as green and red lines, respectively.  
} 
\label{fig:cmdlpv}
\end{figure}

In order to verify whether those variable stars for which we were not able to obtain 
average magnitudes and colors over the full pulsation cycle belong to M62, we have included 
a third CMD in Figure~\ref{fig:cmdlpv}. In this case, the variable stars were simply identified 
in the photometry catalog and plotted in the CMD using mean magnitudes and colors computed as 
simple averages of the avaliable photometric data. While this necessarily leads to increased 
scatter in the derived CMD positions (as is particularly obvious around the RR Lyrae region 
of the CMD), it also allows 
us to investigate the likelihood that these stars may be cluster members. To further aid 
us in this direction, we overplot in Figure~\ref{fig:cmdlpv} two model isochrones from 
the \citet{apea06} set, computed for a chemical composition consistent with that of the 
cluster (in green, reddened and vertically shifted in order to match the HB of the cluster)
and for a chemical composition consistent with a bulge field at the position of the cluster
(in red, plotted using the same distance modulus as obtained for the cluster). 
From their CMD positions, it appears that most of the LPV stars discovered in 
this paper (i.e., 18 out of 25) 
are likely cluster members, with only a few LPV candidates likely belonging 
to the bulge. Note also that NV231, which we originally classified as an LPV candidate, 
may actually be more properly classified as a background type II Cepheid, judging from its
position in the CMD.

Finally, we note that all those RR Lyrae stars for which we derived metallicities using 
Fourier decomposition, and which are located inside the cluster's tidal radius, present chemical 
abundances that are compatible with M62 membership, except for the already cited case of 
V26~-- thus suggesting that most of the variable star candidates in the cluster outskirts 
are indeed cluster members.

%CMDcombine

\section{On the Oosterhoff Type of M62}
The Oosterhoff phenomenon is of great astrophysical importance, given the information 
that it carries on the early formation history of the Milky Way and its neighboring 
galaxies \citep[e.g.,][ and references therein]{ckea08,mc09,mimea09}, and (increasingly) 
in the Andromeda system \citep[e.g.,][ and references therein]{rcea08,gcea09,gfea09,asea09}. 
As recently summarized by \citet{mc09}, there is a general tendency for bona-fide 
Galactic globular clusters to present the so-called Oosterhoff dichotomy, i.e., a 
sharp division between Oosterhoff type I (OoI) systems, with 
$\langle P_{ab}\rangle \approx 0.55$~d, and Oosterhoff type II (OoII) systems, with 
$\langle P_{ab}\rangle \approx 0.65$~d, with 
exceedingly few Galactic globulars occupying the range between 
$0.58 \lesssim \langle P_{ab}({\rm d})\rangle \lesssim 0.62$. On the contrary, nearby 
extragalactic globular clusters and dwarf galaxies occupy {\em preferentially} the 
latter average period interval, thus clearly revealing a difference in (early) 
formation history between bona-fide Galactic and nearby extragalactic systems. 

As discussed by \citet{rcea05}, there is at present some debate as to whether the 
Oosterhoff type of a globular cluster is determined chiefly by the morphology of the 
HB \citep{cs99}, or whether metallicity plays an important role as well~-- as would 
be supported by theoretical calculations that indicate different evolutionary paths
for HB stars of different metallicities but similar zero-age HB (ZAHB) temperatures, 
and thus a different efficiency of production of stars evolved away from the ZAHB as 
a function of metallicity \citep[see \S5.7 in][ and references therein]{bpea02}. As 
noted by \citeauthor{rcea05}, M62 provides a near-ideal test of the relative 
importance of HB morphology and metallicity in defining the Oosterhoff type of 
a globular cluster, given that the cluster possesses a predominantly blue HB, as 
in the case of most OoII clusters, but is also a fairly metal-rich object, as in 
the case of most OoI clusters. 

Here we confirm the preliminary results by \citet{rcea05}, finding that the mean 
periods of the ab-type RR Lyrae in M62 support an OoI classification for the cluster, 
thus clearly showing that, at least in the case of M62, metallicity is the dominant
factor that defines the Oosterhoff type. 
Indeed, let us assume, as a first approximation, that all of our detected variables
are cluster members. In this case, and taking our homogeneous sample of 133
RRab's and 76 RRc's into account, we derive for the cluster average pulsation
periods of $\langle P_{ab}\rangle=0.547$~d and 
$\langle P_{c}\rangle=0.302$~d, thus confirming the preliminary values reported
by \citeauthor{rcea05}, which are quite typical for OoI systems. 
If the 5 RRab and 3 RRc with uncertain classification~-- namely, NV112, NV120, NV137, 
NV187, NV194 and NV134, NV149~-- are removed, we obtain $\langle P_{ab}\rangle=0.548$~d
and $\langle P_{c}\rangle=0.301$~d. As we have seen, the membership status for 
the RRab stars V26 and NV227 and the RRc variable NV224 is also questionable; if
we further remove these stars from the final tally, we obtain 
$\langle P_{ab}\rangle=0.550$~d (126 RRab stars) and 
$\langle P_{c}\rangle=0.302$~d (73 RRc stars). 

As discussed by Catelan et al. (2010, in preparation), 
$\langle P_{ab}\rangle$ and $P_{ab,min}$ are the two quantities that most strongly 
define the Oosterhoff type. For M62, the shortest-period RRab is NV188, and thus 
$P_{ab,min} = 0.436$~d~-- which again clearly indicates an OoI classification. 

\begin{figure}[!t]
\begin{center}
\includegraphics[width=8.7cm]{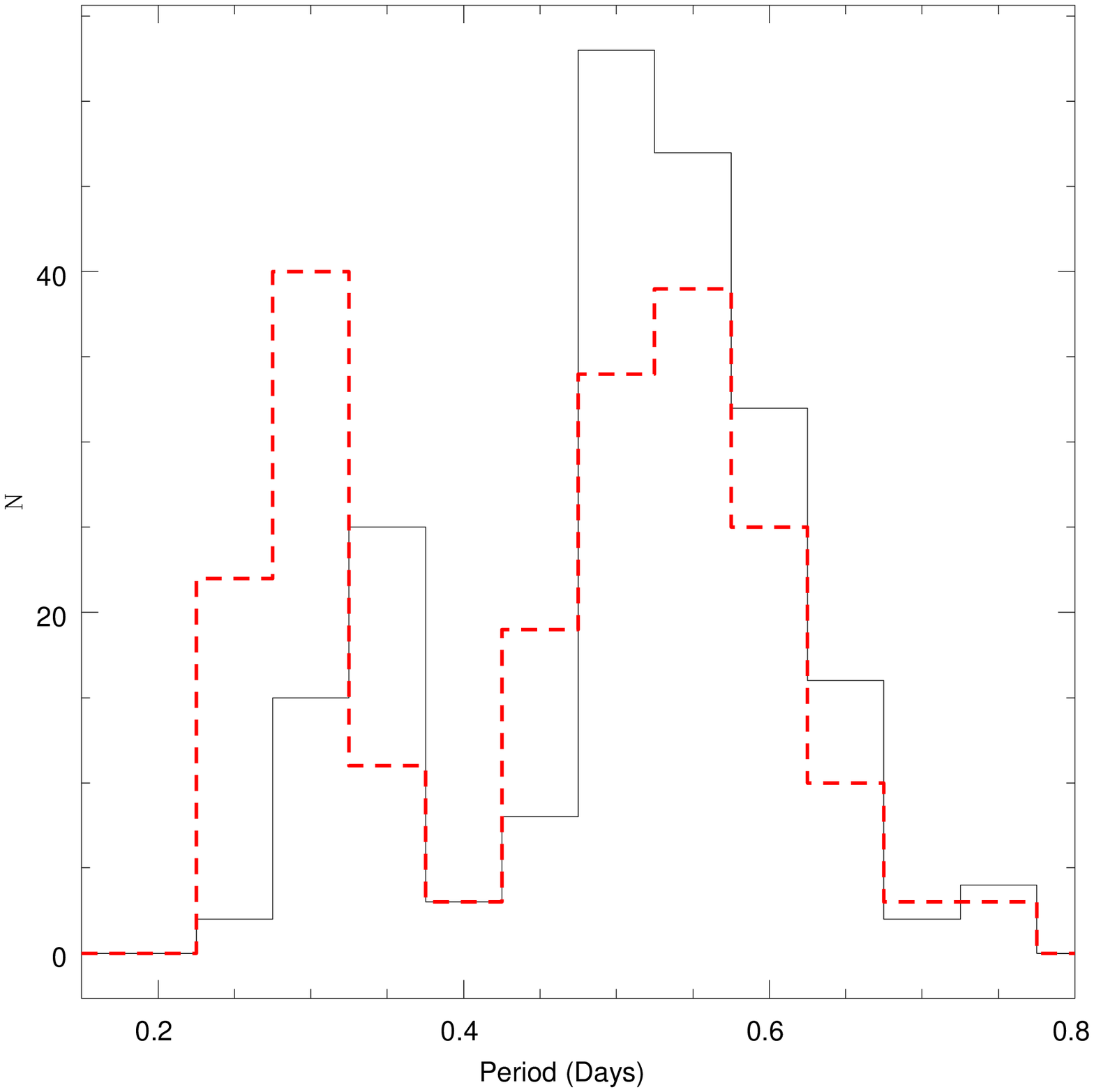}
\caption{Period histogram for the RR Lyrae stars in M62 (dashed line) and 
 for the RR Lyrae stars in the prototypical OoI globular cluster M3 (solid line).}
\label{fig:hist}
\end{center}
\end{figure}

As a 
matter of fact, as shown in Figure~\ref{fig:hist}, the detailed period distribution 
is quite similar for both the prototypical OoI globular cluster M3 and 
M62, with the main differences being a somewhat shorter mean period for the RRc 
stars in M62 and a slightly broader distribution of ab-type periods. 
The period-amplitude diagram may also provide further insight into these 
differences, in addition to useful information regarding the 
Oosterhoff classification of stellar systems \citep[e.g.,][ and references therein]{ccea05}. 
How does this diagram look in the case of M62, once those RRab stars identified as 
peculiar (i.e., with $D_{m} > 5.0$) have been removed?

\begin{figure}[!t]
\begin{center}
\includegraphics[width=8.86cm]{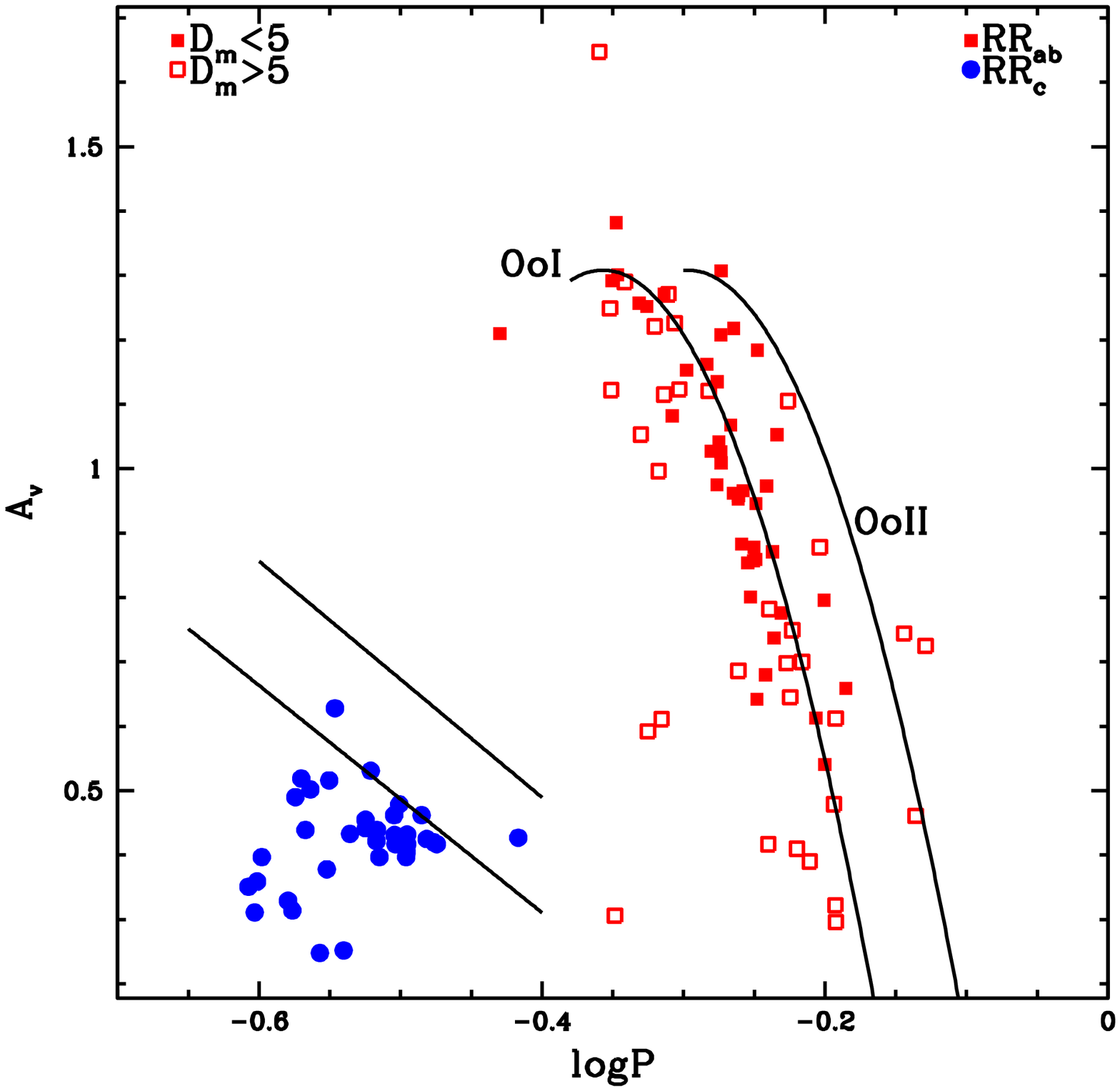}
\includegraphics[width=8.86cm]{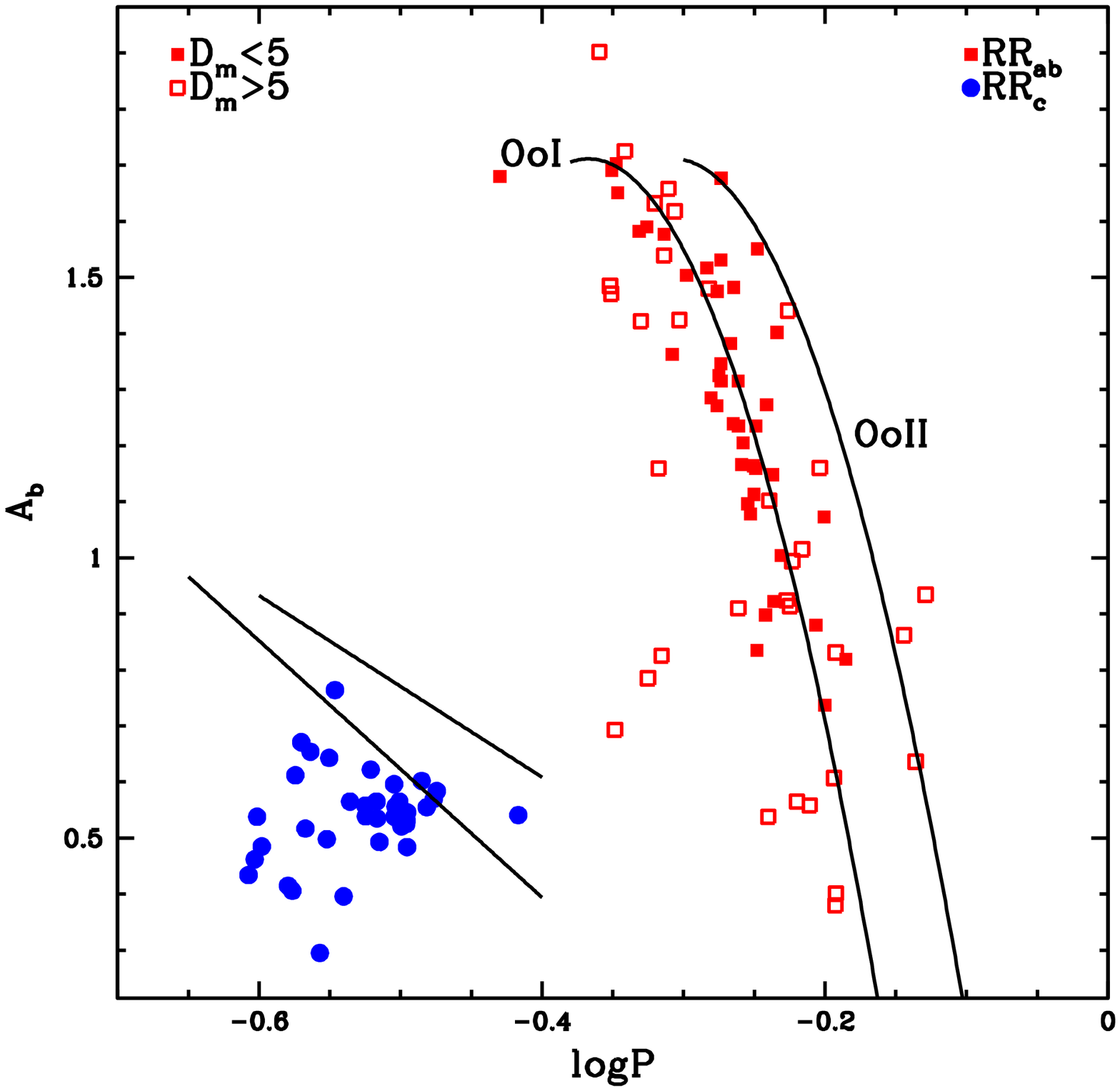}
\caption{Bailey (period-amplitude) diagram for the RR Lyrae stars in M62, 
 compared with reference lines for RR Lyrae stars in OoI and OoII globular clusters. 
 Top: $A_B - \log P$ diagram; bottom: $A_V - \log P$ plane.}
\label{fig:bailey}
\end{center}
\end{figure}

The answer is provided in Figure~\ref{fig:bailey}, where both the $A_V - \log P$ 
(upper panel) and $A_B - \log P$ (lower panel) planes are shown. In these figures, 
we also provide reference lines for OoI and OoII globular clusters, as derived by 
\citet{ccea05} and summarized in eqs.~(10)-(15) in \citet{mzea10}. Clearly, there 
is a tendency for most of the ab-type RR Lyrae to fall around the OoI line in this 
diagram, which again is fully consistent with an OoI classification for the cluster. 
The RRc's, on the other hand, appear to have shorter periods, at a given amplitude, 
than indicated by the reference OoI line, which in turn is based on the M3 RR Lyrae
\citep{ccea05}, which is consistent with the pattern observed in Figure~\ref{fig:hist}. 
A possible interpretation for these differences has been provided by \citet{cs99}, 
who pointed out that, in the period-amplitude diagram, well-behaved RRab stars 
of different metallicities seemed to follow a fairly universal mean locus, defined 
solely by their Oosterhoff types, whereas the RRc's, on the contrary, presented 
systematic deviations towards shorter periods (at a given amplitude) with 
increasing metallicity. Given that M62 is more metal-rich than M3, this provides  
a reasonable explanation for our results.  

There is, however, one aspect of the M62 variable star population that 
may not seem immediately compatible with an OoI classification, namely, the number 
fraction of c-type variables $f_c$. It has long been thought that the latter 
quantity is a strong discriminator of Oosterhoff type, with $f_c \simeq 0.17$ 
for the OoI systems, and $f_c \simeq 0.44$ for OoII systems 
\citep[see, e.g., Table~3.2 in][]{hs95}. In the case of M62, we find 
$f_c = 0.363$, which is intermediate between these two reference values, 
but closer to the one for OoII systems. However, as discussed more recently 
by Catelan et al. (2010, in preparation), $f_c$ is actually {\em not} a 
particularly reliable indicator of Oosterhoff type, with known OoI systems 
covering a wide range in $f_c$ values, from $f_c \approx 0$ up to 0.65 (with 
most of the objects falling in the range $0.2 \lesssim f_c \lesssim 0.4$), 
and likewise known OoII systems covering the range from $f_c \approx 0.1$ up 
to 0.6 (with most of the objects falling in the range $0.3 \lesssim f_c \lesssim 0.55$). 
We thus conclude that the $f_c$ value for M62 is not inconsistent with an OoI type
classification for the cluster; the fact that it is slightly larger than for most
OoI systems is likely due to the fact that M62 also has one of the bluest HB types 
among OoI globulars.

%Hist

%Bailey

As discussed by (e.g.) \citet{mcea03}, the average physical parameters of the c- and 
ab-type RR Lyrae, as derived on the basis of Fourier decomposition of their light 
curves, can also provide a useful consistency check of the derived Oosterhoff type. 
In Tables~\ref{tab:meanrc} and \ref{tab:meanrab} we accordingly compare some of 
the physical parameters that we derived on the basis of the Fourier decomposition 
method (\S\ref{sec:fourier}) with those similarly derived for other clusters in the 
literature, for the RRc and RRab stars, respectively. As can be seen from these 
tables, the Fourier-based physical parameters that we derived for M62 are again 
entirely consistent with an OoI classification for the cluster.

\begin{deluxetable*}{lccccccc}
\tablewidth{0pc} 
\tabletypesize{\scriptsize}
\tablecaption{Mean Parameters for RRc Stars in Globular Clusters (from Fourier Decomposition)}
\tablehead{\colhead{ID} & \colhead{Oo Type} & 
\colhead{${\rm [Fe/H]}_{\rm H03}$}\tablenotemark{a} & 
\colhead{No. of Stars} &
\colhead{$M/M_{\odot}$} & 
\colhead{$\log(L/L_{\odot})$} & \colhead{$T_{\rm eff}({\rm K})$} & \colhead{$y$} }
\startdata
NGC~6362\tablenotemark{b}          &  I   & $-0.95$ &  14  & 0.53  &  1.66  &  7429  &  0.29\\
NGC~6171 (M107)\tablenotemark{c}   &  I   & $-1.04$ &  6   & 0.53  &  1.65  &  7447  &  0.29\\
NGC~5904 (M5)\tablenotemark{d}     &  I   & $-1.27$ &  14  & 0.54  &  1.69  &  7353  &  0.28\\
NGC~6266 (M62)                     &  I   & $-1.29$ &  21  & 0.53  &  1.66  &  7413  &  0.29\\
NGC~6229\tablenotemark{e}          &  I   & $-1.43$ &  9   & 0.56  &  1.69  &  7332  &  0.28\\
NGC~6934\tablenotemark{f}          &  I   & $-1.54$ &  4   & 0.63  &  1.72  &  7290  &  0.27\\
NGC~5272 (M3)\tablenotemark{g}     &  I   & $-1.57$ &  5   & 0.59  &  1.71  &  7315  &  0.27\\
NGC~7089 (M2)\tablenotemark{h}     &  II  & $-1.62$ &  2   & 0.54  &  1.74  &  7215  &  0.27\\
NGC~5286\tablenotemark{i}          &  II  & $-1.67$ &  12  & 0.60  &  1.72  &  7276  &  0.27\\
NGC~6809 (M55)\tablenotemark{j}    &  II  & $-1.81$ &  5   & 0.53  &  1.75  &  7193  &  0.27\\
NGC~4147\tablenotemark{k}          &  I   & $-1.83$ &  9   & 0.55  &  1.69  &  7335  &  0.28\\
NGC~2298\tablenotemark{l}          &  II  & $-1.85$ &  2   & 0.59  &  1.75  &  7200  &  0.26\\
NGC~4590 (M68)\tablenotemark{m}    &  II  & $-2.06$ &  16  & 0.70  &  1.79  &  7145  &  0.25\\
NGC~7078 (M15)\tablenotemark{n}    &  II  & $-2.26$ &  8   & 0.76  &  1.81  &  7112  &  0.24\\
NGC~6341 (M92)\tablenotemark{o}    &  II  & $-2.28$ &  3   & 0.64  &  1.77  &  7186  &  0.26
\enddata
\tablenotetext{a}{From the \citet{wh96} catalog (Feb. 2003 issue).}
\tablenotetext{b}{From \citet{aoea01}.}
\tablenotetext{c}{From \citet{jkea00}.}
\tablenotetext{d}{From \citet{jkea00}.}
\tablenotetext{e}{From \citet{jbea01}.}
\tablenotetext{f}{From \citet{jkea01}.}
\tablenotetext{g}{From \citet{jkea98}.}
\tablenotetext{h}{From \citet{clea06}.}
\tablenotetext{i}{From \citet{mzea10}.}
\tablenotetext{j}{From \citet{aoea99}.}
\tablenotetext{k}{From \citet{aafea04}.}
\tablenotetext{l}{From \citet{ccea95}.}
\tablenotetext{m}{From \citet{cs97}.}
\tablenotetext{n}{From \citet{aafea06}.}
\tablenotetext{o}{From \citet{clea06}.}
\label{tab:meanrc}
\end{deluxetable*}							

\begin{deluxetable*}{lccccccc}
\tablewidth{0pc} 
\tabletypesize{\scriptsize}
\tablecaption{Mean Physical Parameters for RRab Stars in Globular Clusters 
(from Fourier Decomposition)\tablenotemark{a}}
\tablehead{\colhead{ID} & \colhead{Oo Type} & 
\colhead{${\rm [Fe/H]}_{\rm H03}$} & \colhead{No. of Stars} &
\colhead{${\rm [Fe/H]}_{\rm ZW84}$} & 
\colhead{${\rm [Fe/H]}_{\rm J95}$} & \colhead{$T_{\rm eff}^{\langle V-K \rangle}({\rm K})$} & 
\colhead{$M_{V}$} }
\startdata
NGC~6362          &  I   & $-0.95$ &  14  &   $-1.26$ &  $-0.93$ &   6555 &   0.86\\
NGC~6171 (M107)   &  I   & $-1.04$ &  3   &   $-1.25$ &  $-0.91$ &   6619 &   0.85\\
NGC~1851\tablenotemark{b}
                  &  I   & $-1.22$ &  7   &   $-1.43$ &  $-1.17$ &   6494 &   0.80\\
NGC~5904 (M5)     &  I   & $-1.27$ &  26  &   $-1.47$ &  $-1.23$ &   6465 &   0.81\\
NGC~6266 (M62)    &  I   & $-1.29$ &  39  &   $-1.31$ &  $-0.99$ &   6501 &   0.83\\
NGC~6229          &  I   & $-1.43$ &  9   &   $-1.60$ &  $-1.41$ &   6383 &   0.81\\
NGC~6934          &  I   & $-1.54$ &  24  &   $-1.53$ &  $-1.31$ &   6455 &   0.81\\
NGC~5272 (M3)     &  I   & $-1.57$ &  17  &   $-1.60$ &  $-1.42$ &   6438 &   0.78\\
NGC~7089 (M2)     &  II  & $-1.62$ &  9   &   $-1.64$ &  $-1.47$ &   6276 &   0.71\\
NGC~5286          &  II  & $-1.67$ &  12  &   $-1.68$ &  $-1.52$ &   6266 &   0.72\\
NGC~6809 (M55)    &  II  & $-1.81$ &  3   &   $-1.77$ &  $-1.65$ &   6333 &   0.67\\
NGC~4147          &  I   & $-1.83$ &  5   &   $-1.46$ &  $-1.22$ &   6633 &   0.80\\
NGC~7078 (M15)    &  II  & $-2.26$ &  11  &   $-1.92$ &  $-1.87$ &   6237 &   0.67\\
NGC~6341 (M92)    &  II  & $-2.28$ &  5   &   $-1.92$ &  $-1.87$ &   6160 &   0.67
\enddata
\tablenotetext{a}{References are the same as in Table~\ref{tab:meanrc}, except as noted.}
\tablenotetext{b}{From \citet{jkea00}.}
\label{tab:meanrab}
\end{deluxetable*}

\section{The $A$-Parameter and the He Abundance in M62}\label{sec:ahe}
Recently, several authors have suggested that He abundance enhancements may be 
quite commonplace among globular clusters \citep[e.g.,][]{dc08}. In this scenario, 
globular clusters with predominantly blue HB morphologies are suggested to be 
helium rich, and thus it is worthwhile to check the RR Lyrae stars in M62 in 
search for evidence of He enhancement that might help explain its blue HB. 

As is well known, the ``$A$-method'' of \citet{cc75} can provide strong constraints 
on the presence (or otherwise) of He enrichment among RR Lyrae stars. From the 
period-mean density relation of stellar pulsation theory \citep{vab71}, one finds

\begin{equation}
\log P = 11.497 + 0.84 \, A - 3.481 \, \log T_{\rm eff}, 
\end{equation}

\noindent where 

\begin{equation}
A \equiv \log(L/L_{\odot}) - 0.81 \, \log(M/M_{\odot}), 
\label{eq:apar}
\end{equation}

\noindent with the period in days and the temperature in K. Therefore, 

\begin{equation}
A = 13.353 - 1.19 \, \log P - 4.058 \, \log T_{\rm eff}.
\label{eq:apar2}
\end{equation}

\noindent Similarly, on the basis of the more recent models by \citet*{fcea98}, 
\citet{ccea05} obtain 

\begin{equation}
A = 13.687 - 1.19 \, \log P - 4.144 \, \log T_{\rm eff}.
\label{eq:apar3}
\end{equation}

Thus defined, the $A$-parameter can therefore 
be easily computed on the basis solely of period measurements 
and estimates of the stellar temperatures. As already mentioned, this parameter 
is strongly sensitive to the He abundance; in particular, according to the 
ZAHB models of \citet{sc98} for $Z = 0.002$, $A$ depends on $Y$ according to 
%$(\Delta A/\Delta Y)_{T_{\rm eff}, Z} \simeq 1.56$, in the range of $Y$ 
$(dA/dY)_{T_{\rm eff}, Z} \simeq 1.56$, in the range of $Y$ 
between 0.23 and 0.28. For comparison, the dependence on $Z$ at fixed $Y$ is 
much milder, the same models indicating, in the range between $Z = 0.0005$ 
%and $Z = 0.002$, a slope $(\Delta A/\Delta \log Z)_{T_{\rm eff}, Y} \simeq -0.05$.
and $Z = 0.002$, a slope $(dA/d\log Z)_{T_{\rm eff}, Y} \simeq -0.05$.

Here we provide a comparison with the globular cluster M3, which has been 
extensively studied previously, and which has a metallicity fairly similar 
to M62's. In particular, $A$-parameter values can be derived for the ab-type 
RR Lyrae stars in M3, based on the temperatures derived by \citet{jkea98} 
from their Fourier decomposition of the $V$-band light curves. 
Their procedure is essentially identical to the one adopted in our paper to 
derive the temperatures listed in the last column of Table~\ref{tab:pfrrab}, 
and thus $A$-parameter values for M62 RR Lyrae derived on the basis of these
temperatures can be directly compared with those for M3 RR Lyrae stars, 
based on the temperatures derived by Kaluzny and co-workers.  
As a result, we find for M3 a 
$\langle A \rangle = 1.803 \pm 0.023$, and for M62 a  
$\langle A \rangle = 1.806 \pm 0.024$, implying a difference of 
$\Delta A = 0.003 \pm 0.033$ between M62 and M3. (The standard deviation of 
the means are indicated.) If due to a difference in He abundance, these 
values suggest that M62 is more He-rich than M3, but by only about 0.002 in 
$Y$. Within the errors, this comparison suggests that the RR Lyrae stars in 
M3 and M62 have closely the same He abundance. 

\begin{figure}[!t]
\begin{center}
\includegraphics[width=8.75cm]{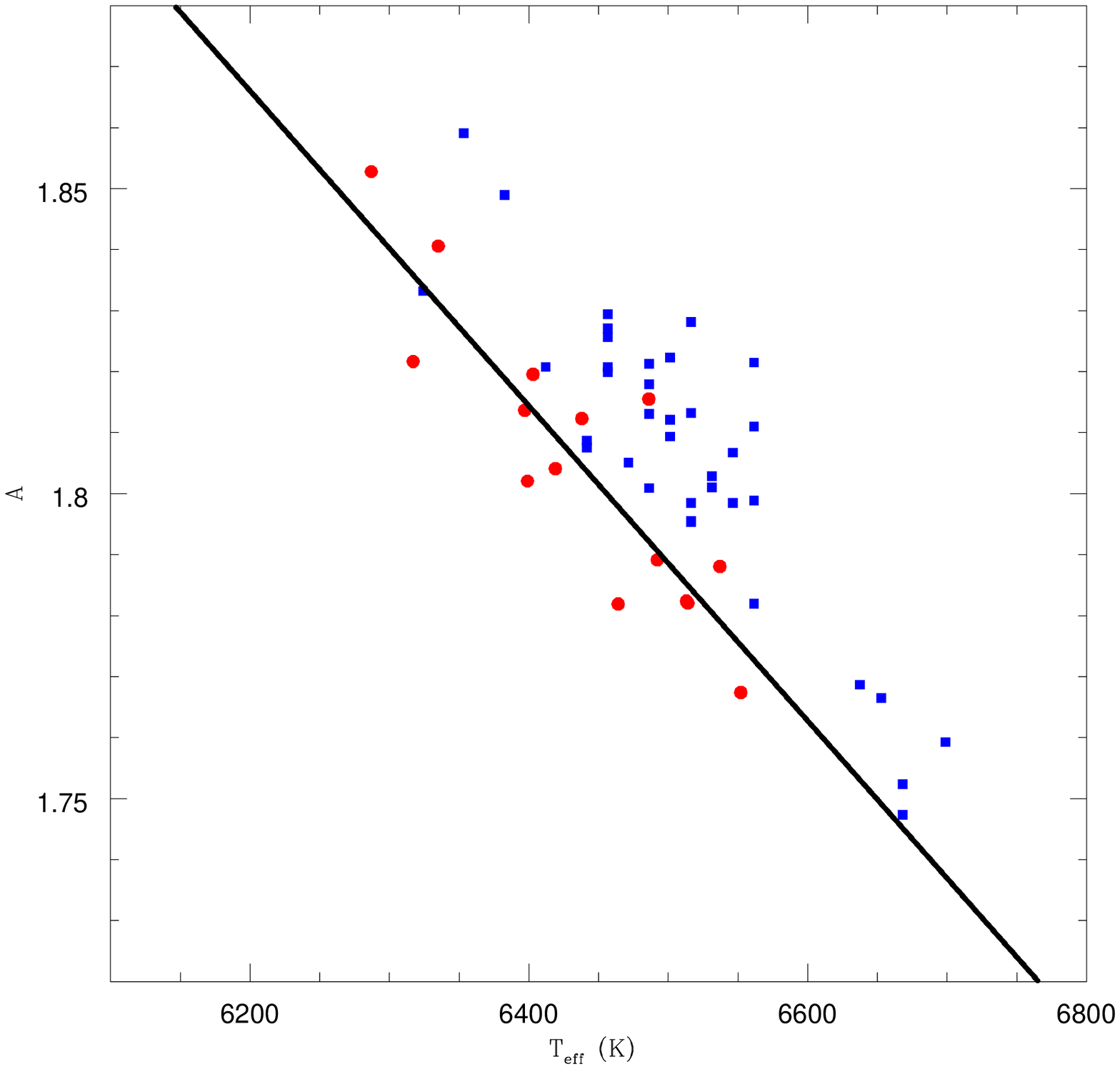}
\caption{In this figure, $A$-parameter values, as derived on the basis of 
equation~\ref{eq:apar} for ab-type RR Lyrae stars, are plotted as a function 
of temperature, as derived based on Fourier decomposition ($V-K$ colors).
Squares indicate M62 stars, whereas circles represent M3 stars. The straight
line is a least-squares fit to the M3 data. See text for more details.}
\label{fig:apar}
\end{center}
\end{figure}

On the other hand, the fact that the average $A$ values are closely the same
for both M3 and M62 does not necessarily imply that, at any given temperature, 
no offset between the two clusters in present. To check for the presence of 
such possible offsets, we compare in Figure~\ref{fig:apar} the derived 
distributions. Intriguingly, and in contrast with what was found in 
Figure~\ref{fig:bailey},\footnote{Recall
that the reference OoI line in Figure~\ref{fig:bailey}, from \citet{ccea05}, 
indicates the locus occupied by the presumably ``unevolved'' RR Lyrae stars 
in M3.} there does appear to be an offset between the 
two clusters, with the deviation of M62 datapoints from the M3 regression 
line in the $A-T_{\rm eff}$ plane (solid line in Fig.~\ref{fig:apar})
amounting to $\Delta A_{T_{\rm eff}} = 0.020 \pm 0.012$. If due to a 
difference in He abundance, this would imply that the M62 RR Lyrae stars 
are more He-rich than their M3 counterparts, by about $0.013 \pm 0.008$ in 
$Y$. This result is confirmed if, instead of the \citet{jkea98} Fourier 
parameters for M3 RR Lyrae stars we use those more recently derived by 
\citet{ccea05}. Further work on the temperatures of M62 RR Lyrae 
stars will be required before we are in a position to conclusively settle 
this issue. 

%A-parameter

\section{Summary}
In this paper we have provided a detailed account of the time-series observations 
that we have collected for the Galactic globular cluster M62, first reported on 
in \citet{rcea05}. Our results indicate that M62 is one of the most RR Lyrae-rich
(in the sense of total number of RR Lyrae stars present) 
globular clusters known in the galaxy, and it is actually not unlikely that future 
studies will reveal that it is {\em the} most RR Lyrae-rich globular cluster known. 
In like vein, M62 appears to be the globular cluster with the largest known number
of LPV stars in the Milky Way, thus making it a very attractive benchmark object
for future RR Lyrae and LPV studies alike. 

Discussing the distribution of variable stars in the cluster's CMD, we find that 
most of the detected variables are likely cluster members. The CMD of the cluster 
is, however, severely affected by differential reddening; we have accordingly taken
benefit of the large number of RR Lyrae variables that are present in the cluster to 
build a 2-D reddening map for the cluster, which allowed us to present a ``corrected'' 
CMD that is much less strongly affected by differential reddening. A full analysis 
of the cluster CMD will be presented in a forthcoming paper (Contreras et al. 2010, 
in preparation). 

From an analysis of the pulsation periods of the detected RR Lyrae stars, we provide
an updated metallicity (${\rm [Fe/H]_{ZW84}} = -1.31$, based on Fourier decomposition 
of the RR Lyrae light curves) and distance modulus [$(m-M)_V = 15.58$, based on the 
recent $M_V({\rm HB})-{\rm [Fe/H]}$ calibration by \citeauthor{cc08} \citeyear{cc08}]
estimates for the cluster. In addition, we discuss a variety 
of Oosterhoff indicators, including the mean periods, period distribution, and 
Bailey diagram, and conclude that the cluster is an OoI object, in spite of its
blue HB morphology but consistent with its moderately high metallicity. Therefore, 
metallicity does play an important role in defining Oosterhoff type, at least in 
the case of M62 \citep[see also][]{rcea05}. Finally, based on an application of the 
``$A$-method,'' we conclude that the M62 RR Lyrae stars likely have a similar He 
abundance as M3, although more work on the temperatures of the M62 RR Lyrae is 
needed before this result can be conclusively established.

\acknowledgments Support for M.C. is provided by MIDEPLAN's Programa
    Inicativa Cient\'{i}fica Milenio through grant P07-021-F, awarded to
	The Milky Way Millennium Nucleus;
    by Proyecto Basal PFB-06/2007; by FONDAP Centro de Astrof\'{i}sica 15010003; 
	and by Proyecto FONDECYT Regular \#1071002. 
	H.A.S. would like to acknowledge the National Science Foundation for support under 
	grants AST0607249 and AST0707756. J.B. acknowledges support from MIDEPLAN's grant 
	P07-021-F and Proyecto FONDECYT	Regular \#1080086. We thank an anonymous 
	referee for useful comments that helped improve the presentation of our results.

%\end{document}

\newpage
\appendix 

\section{Sample Light Curves}

Here we show a representative sample of light curves for the newly discovered variable stars in M62. 
The full set of derived light curves, including our light curves for the previously known variables in the
cluster, can be found in the electronic version of this paper.

\begin{figure}[ht]
  \begin{center}
\includegraphics[width=.325\textwidth]{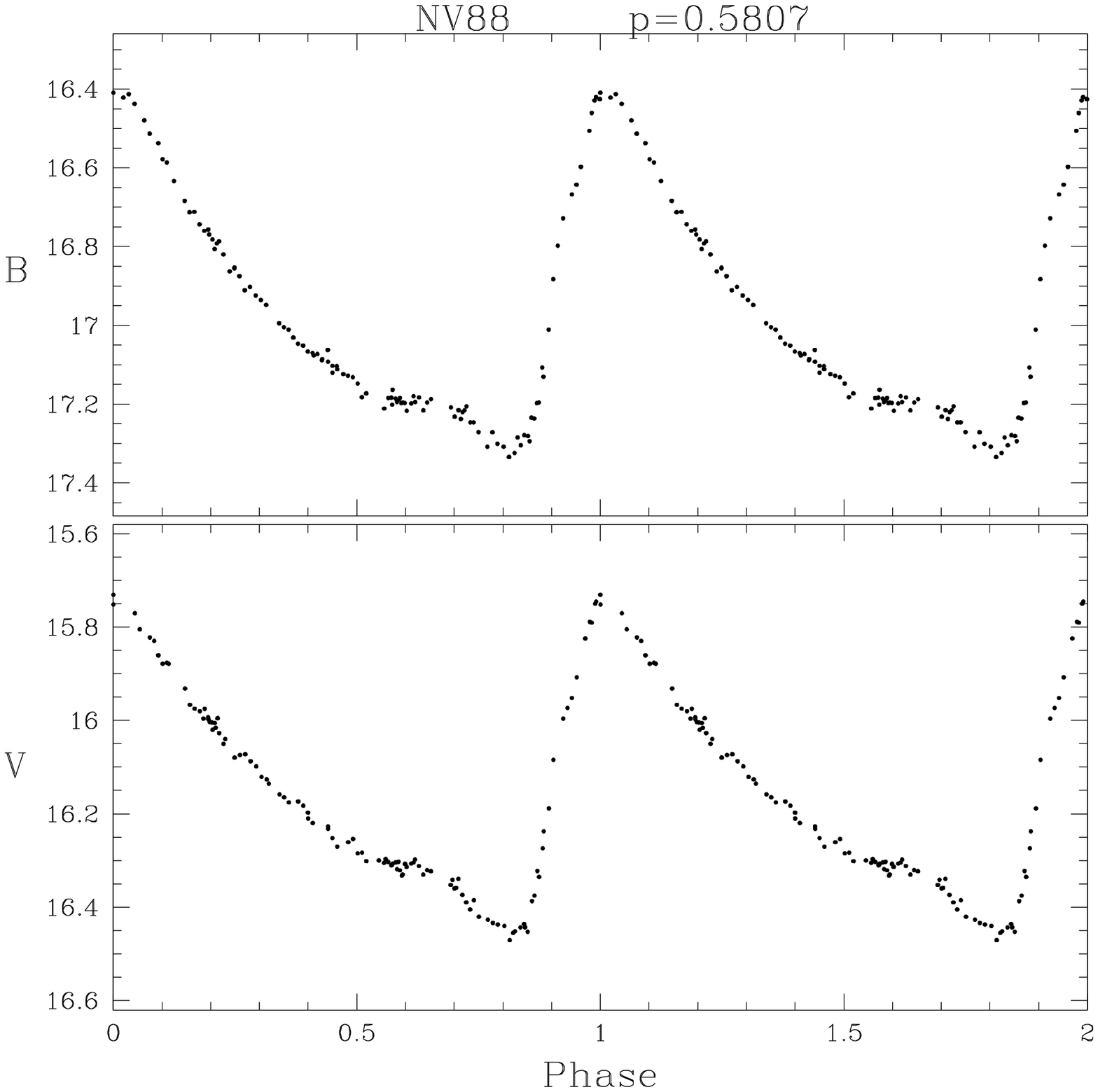}
\includegraphics[width=.325\textwidth]{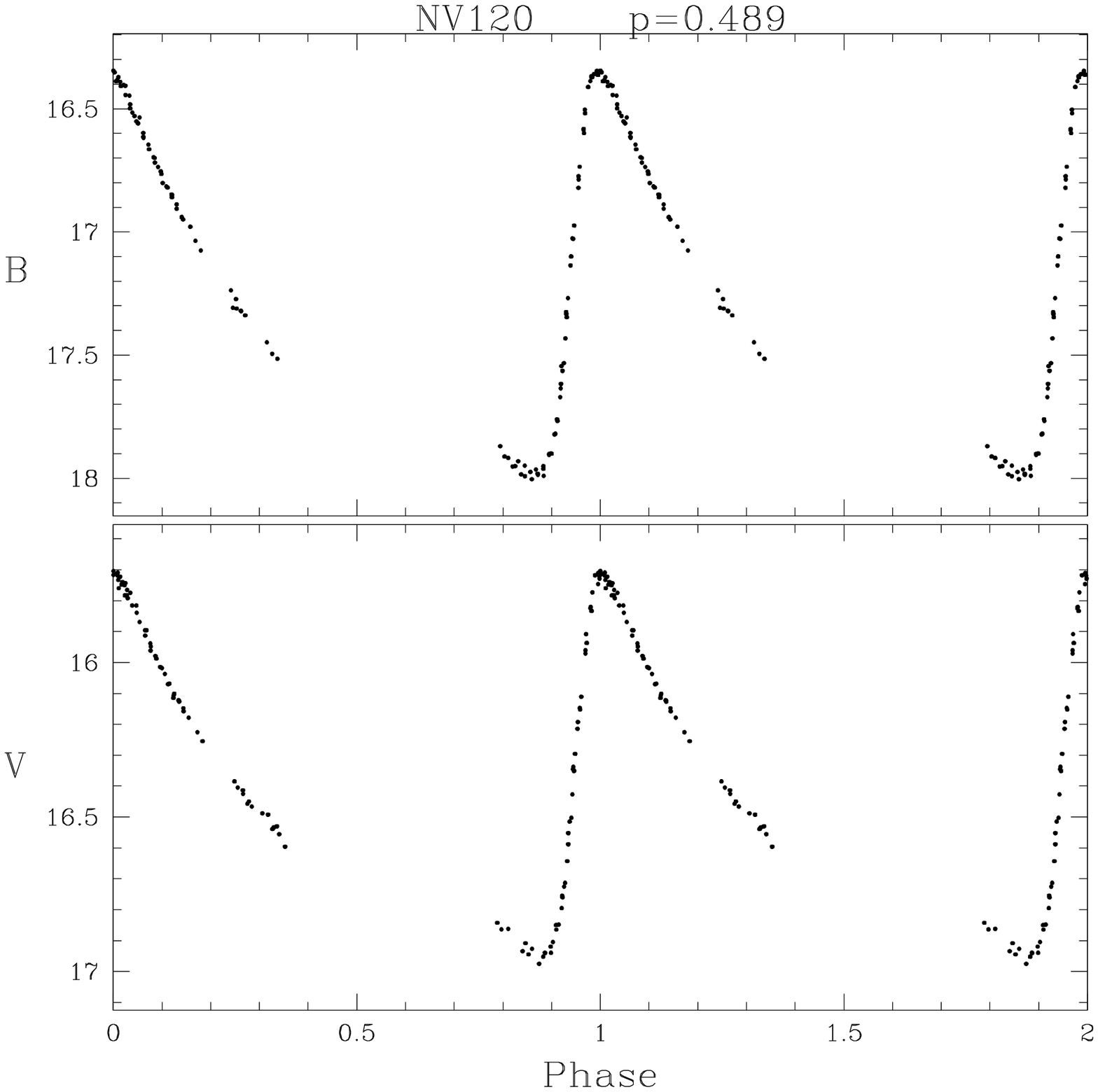}
\includegraphics[width=.325\textwidth]{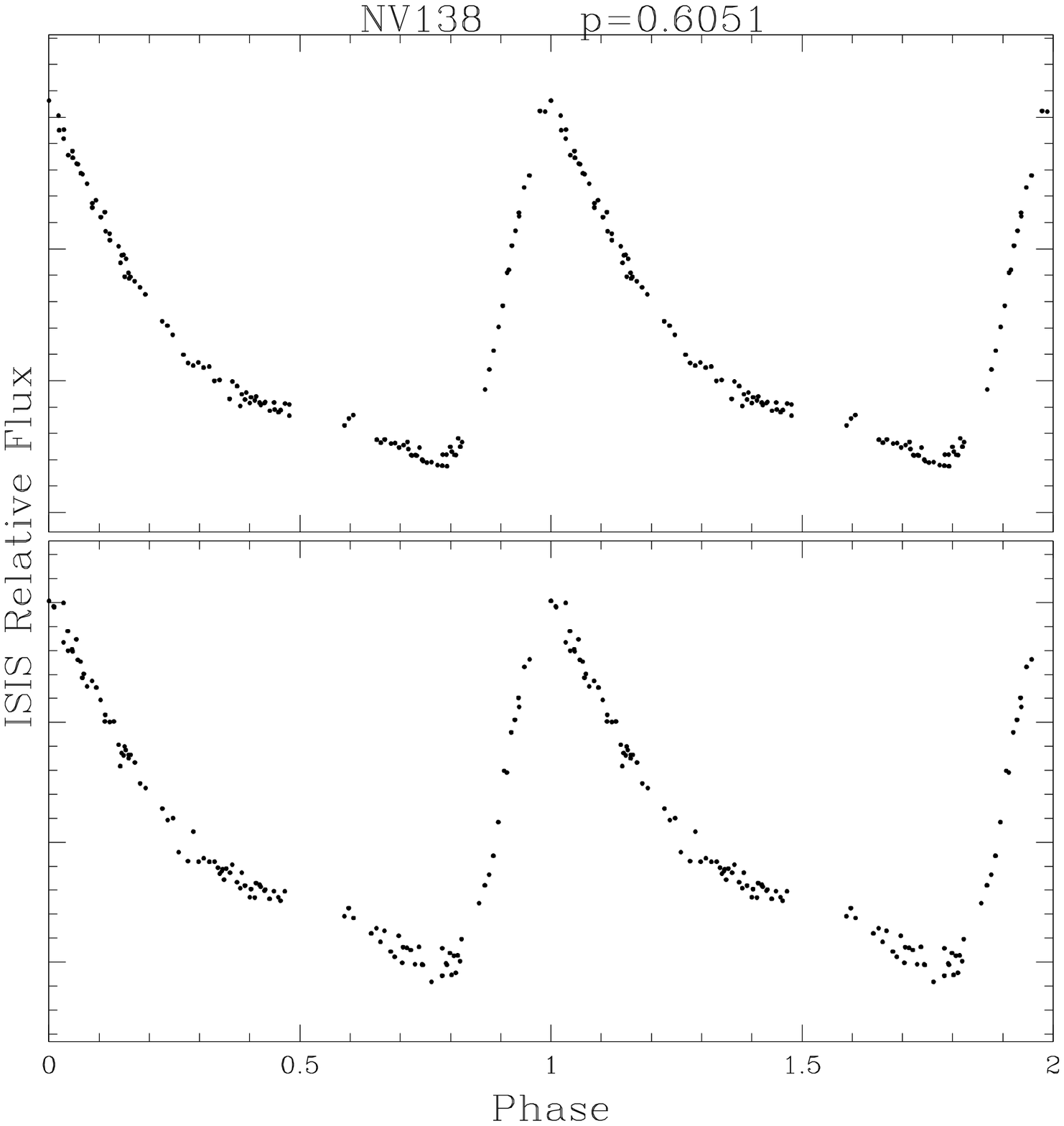}
\includegraphics[width=.325\textwidth]{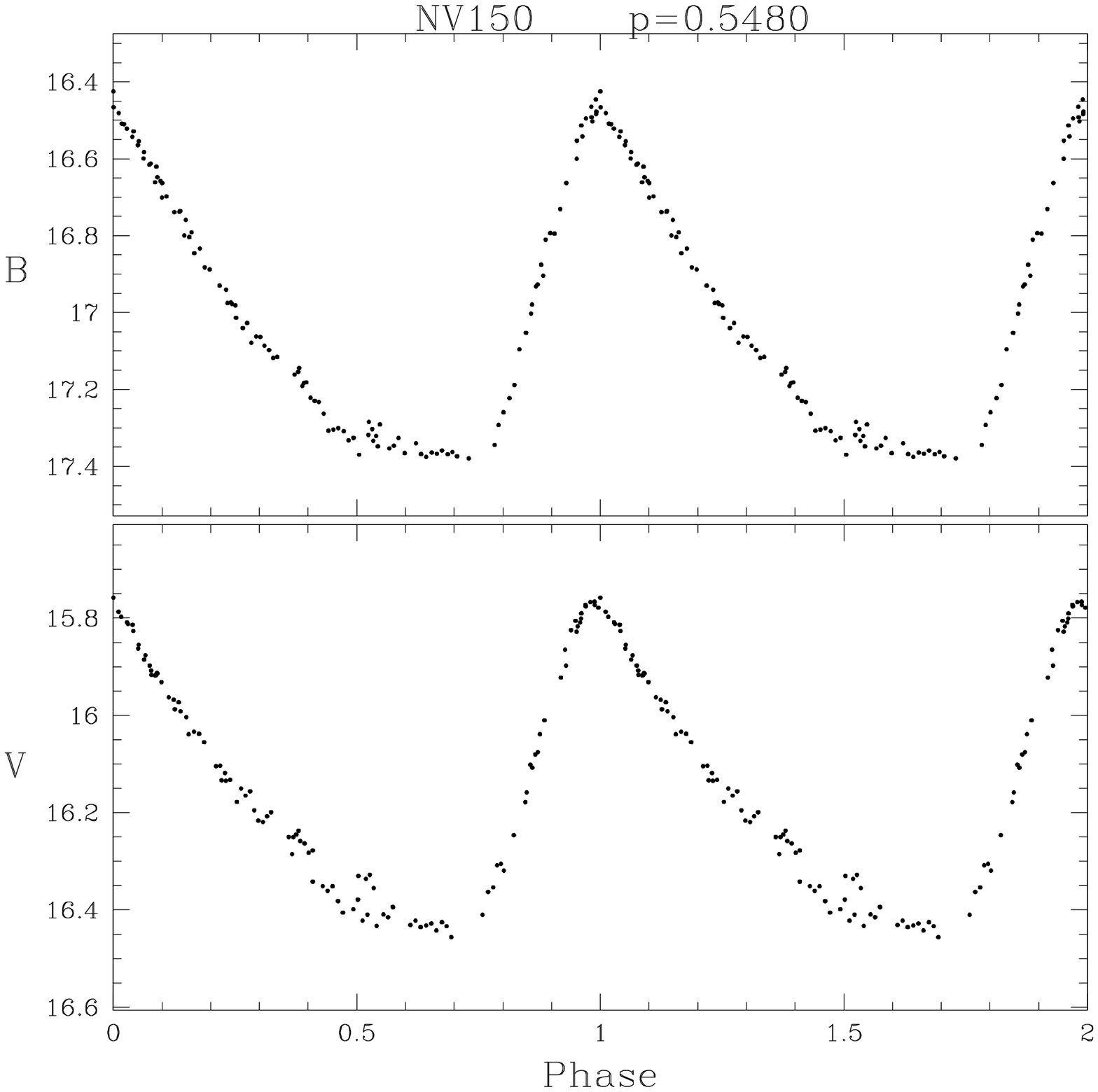}
\includegraphics[width=.325\textwidth]{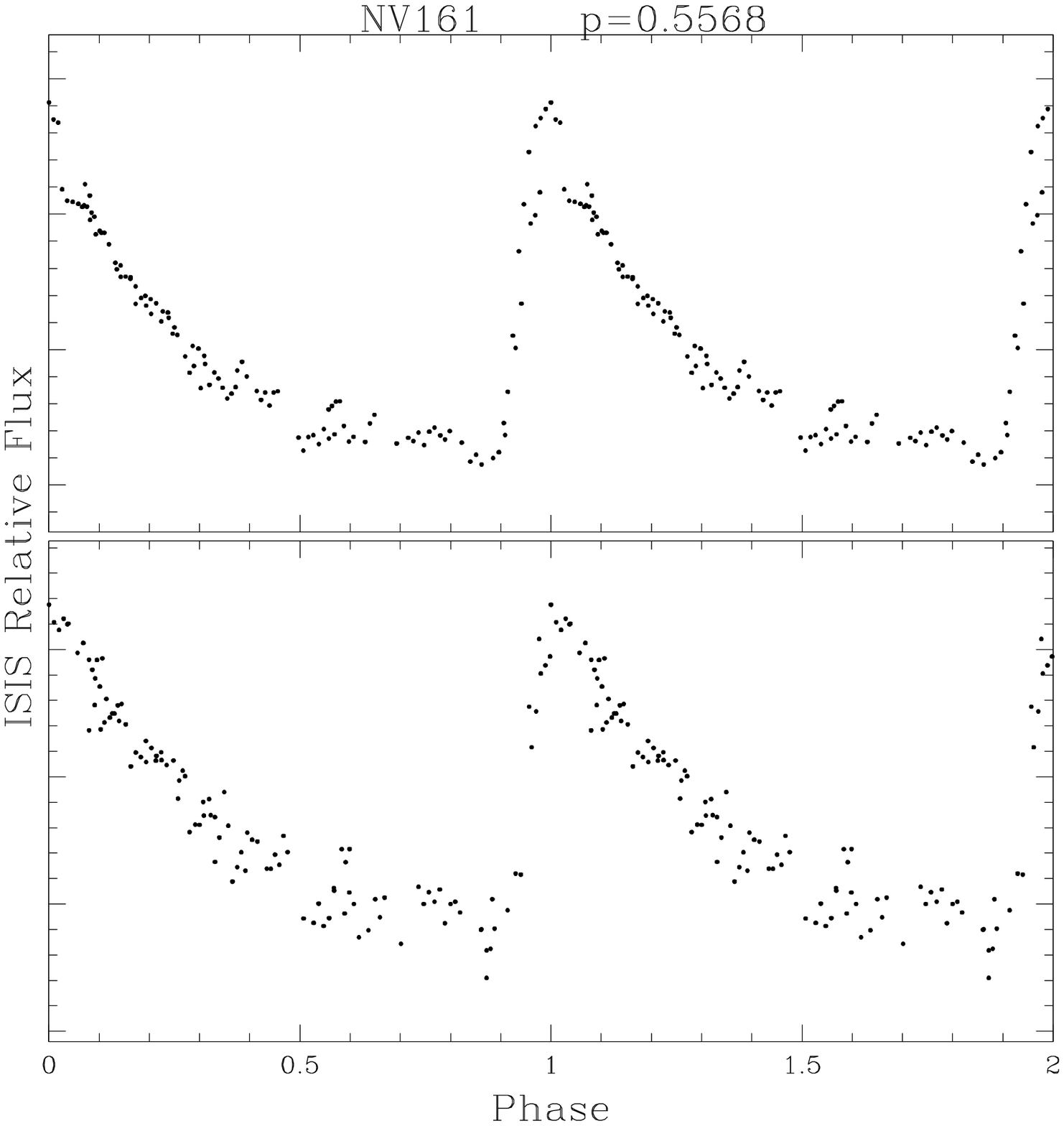}
\includegraphics[width=.325\textwidth]{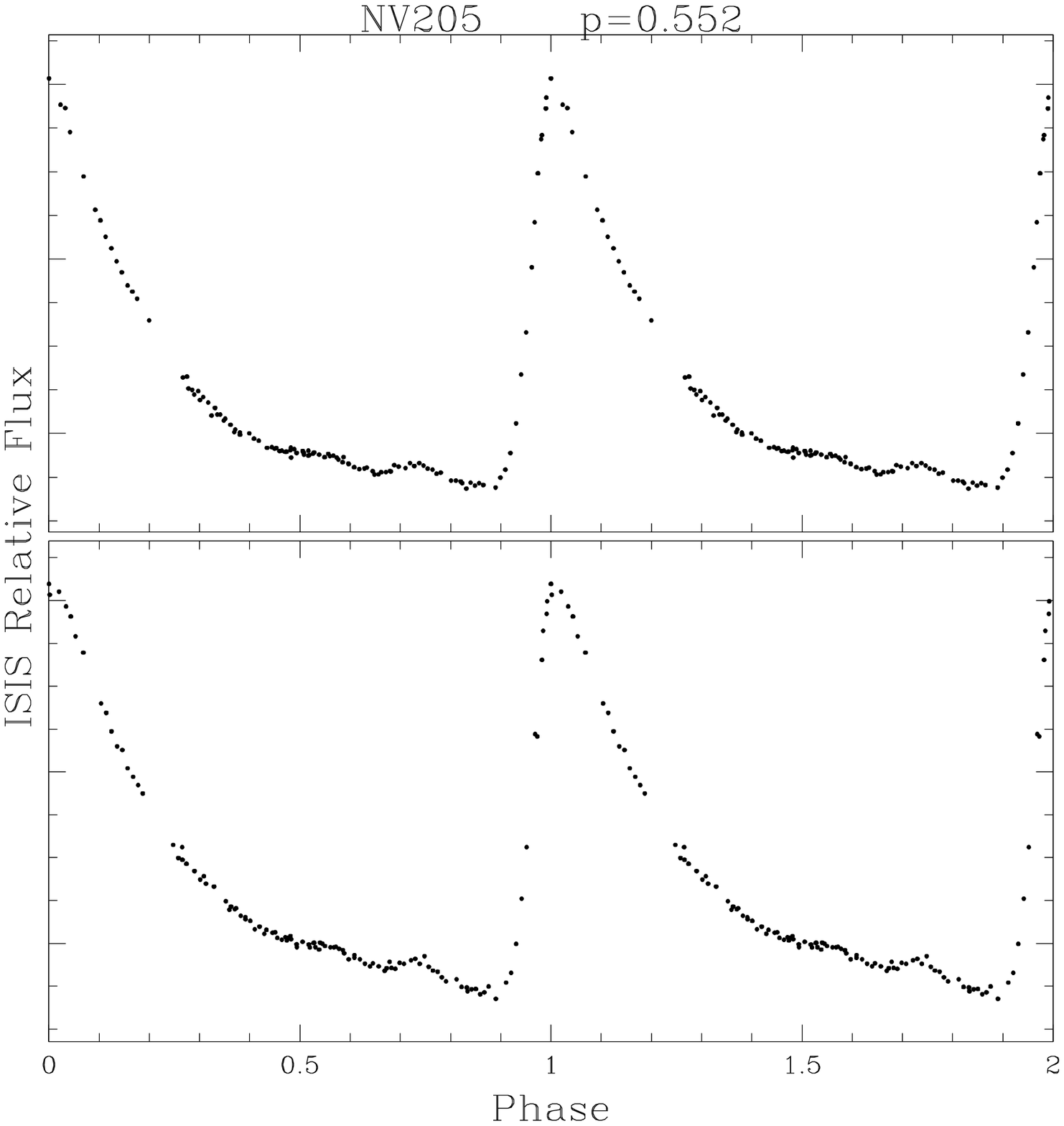}
\includegraphics[width=.325\textwidth]{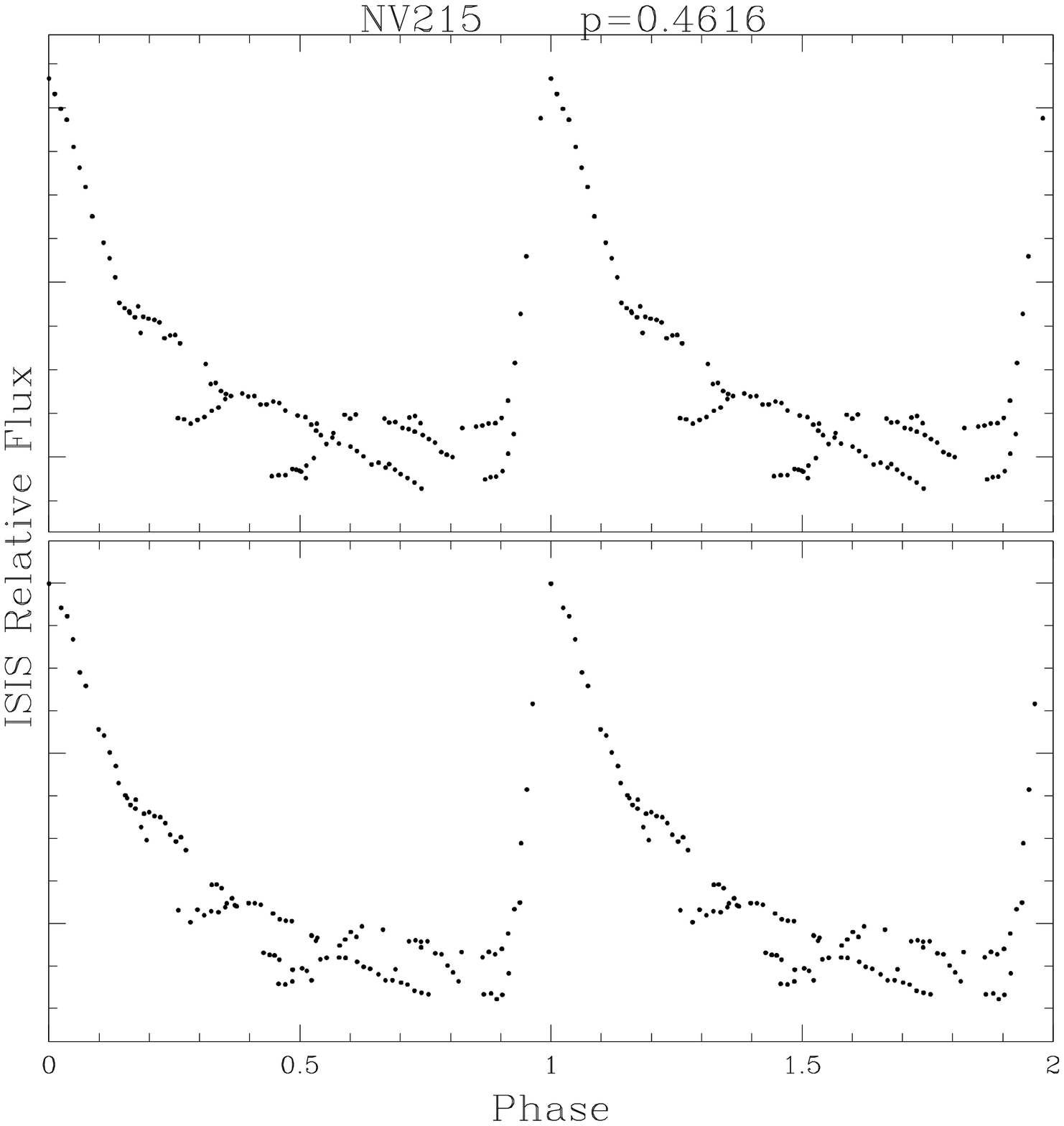}
\includegraphics[width=.325\textwidth]{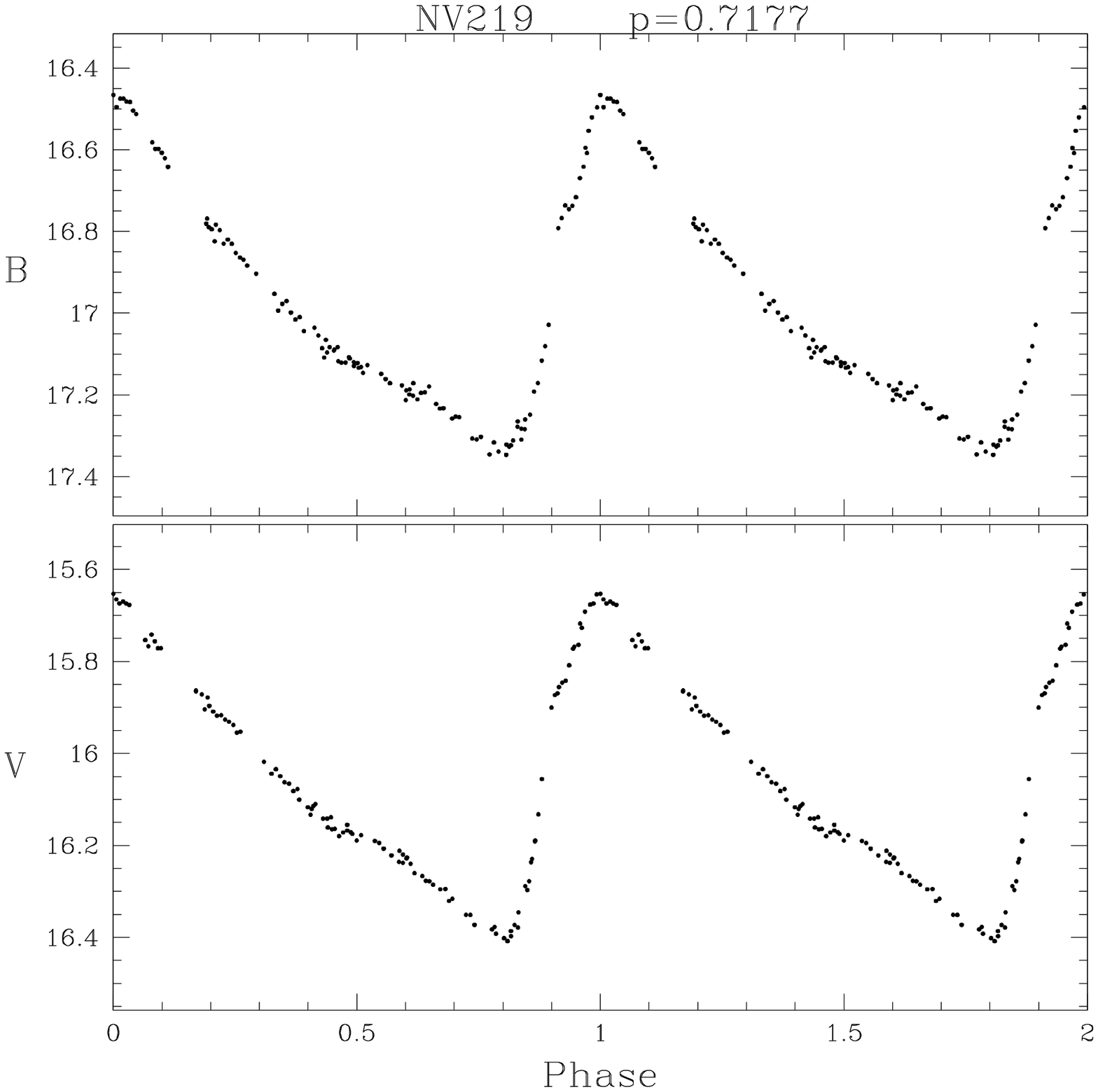}
\includegraphics[width=.325\textwidth]{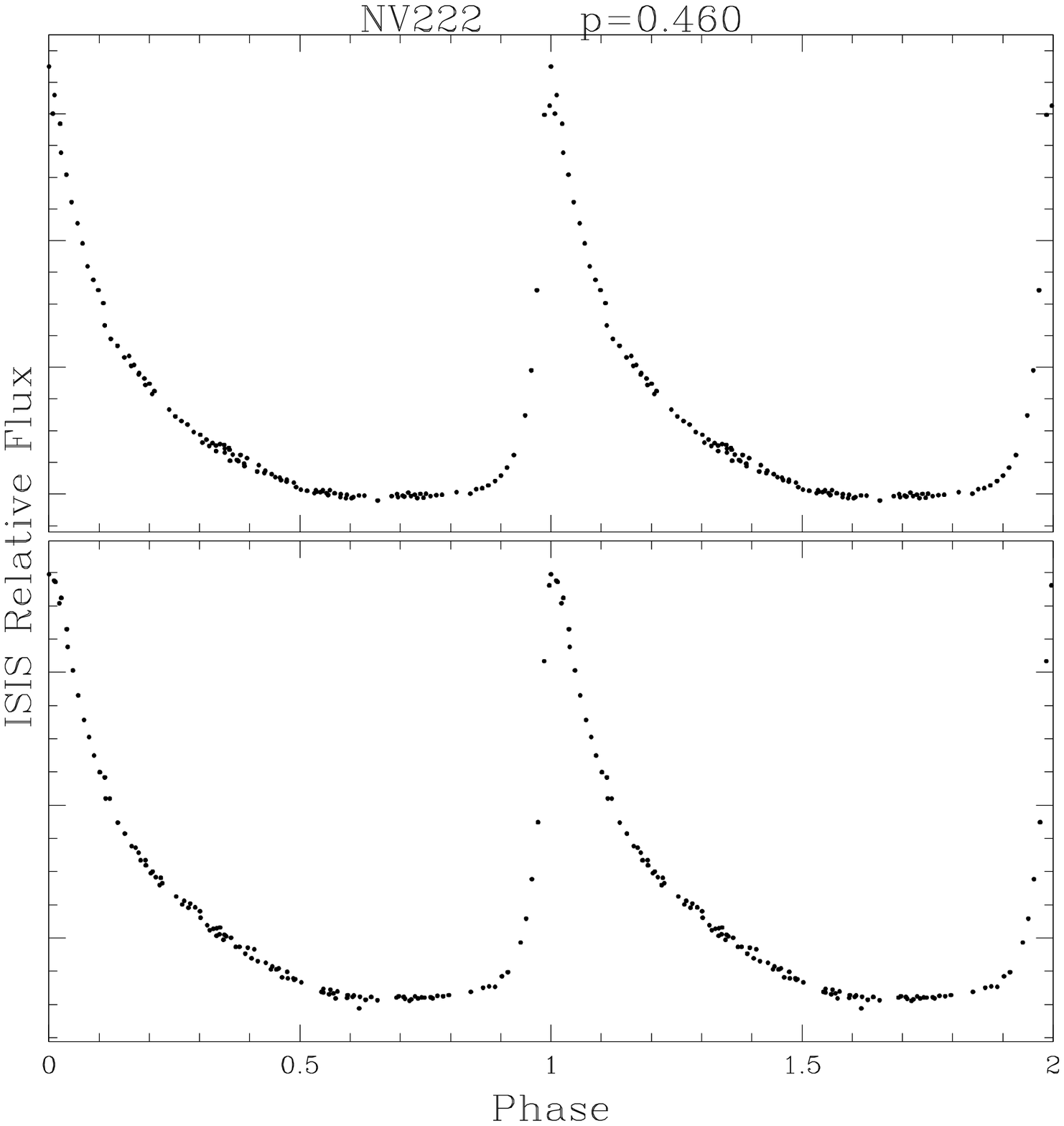}
\caption{Sample light curves for newly-discovered ab-type RR Lyrae stars. (The full set 
can be found in the electronic version of this paper.)}
\label{fig:rrab}
  \end{center}
\end{figure}

\begin{figure}[ht]
  \begin{center}
\includegraphics[width=.325\textwidth]{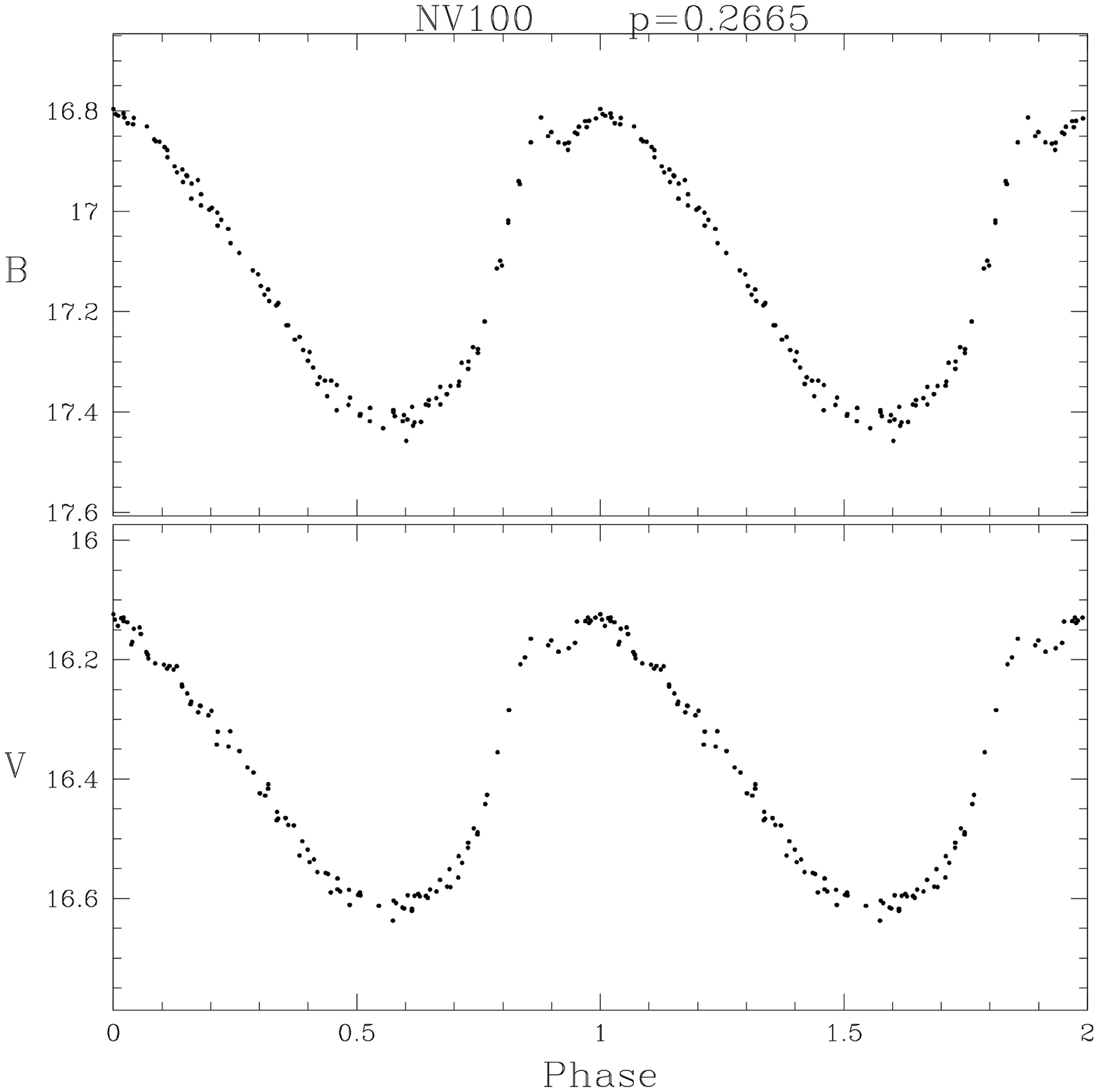}
\includegraphics[width=.325\textwidth]{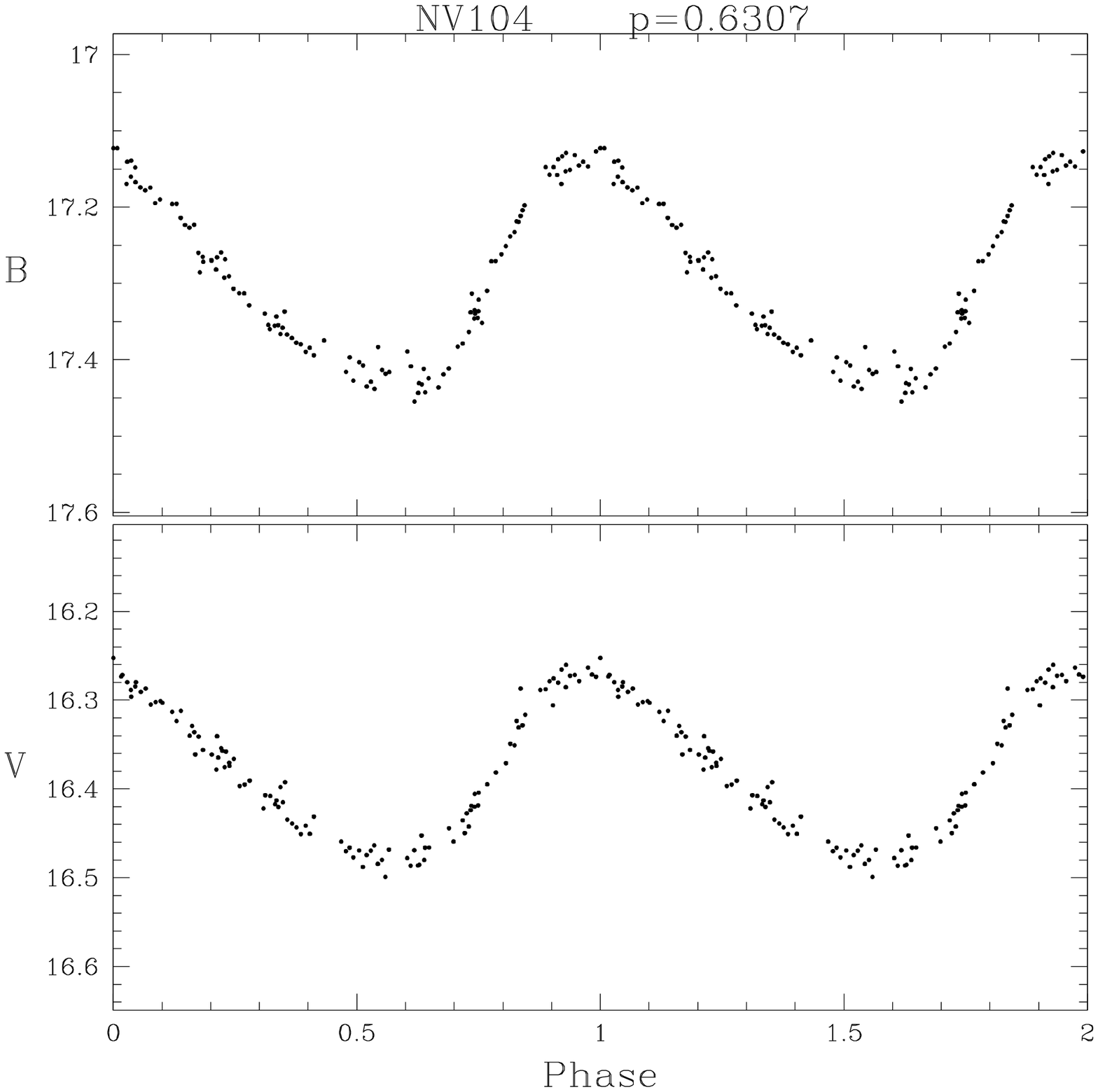}
\includegraphics[width=.325\textwidth]{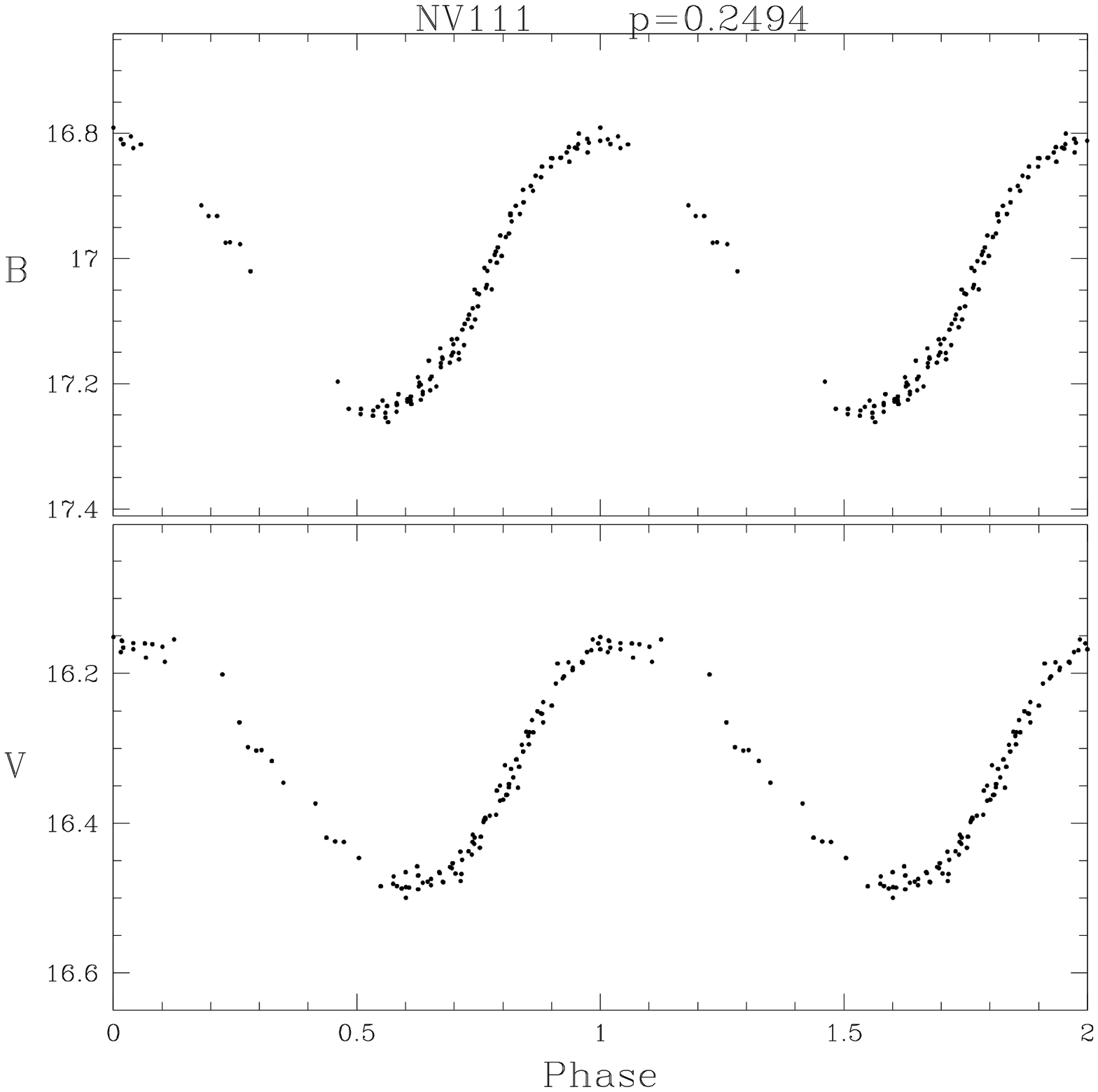}
\includegraphics[width=.325\textwidth]{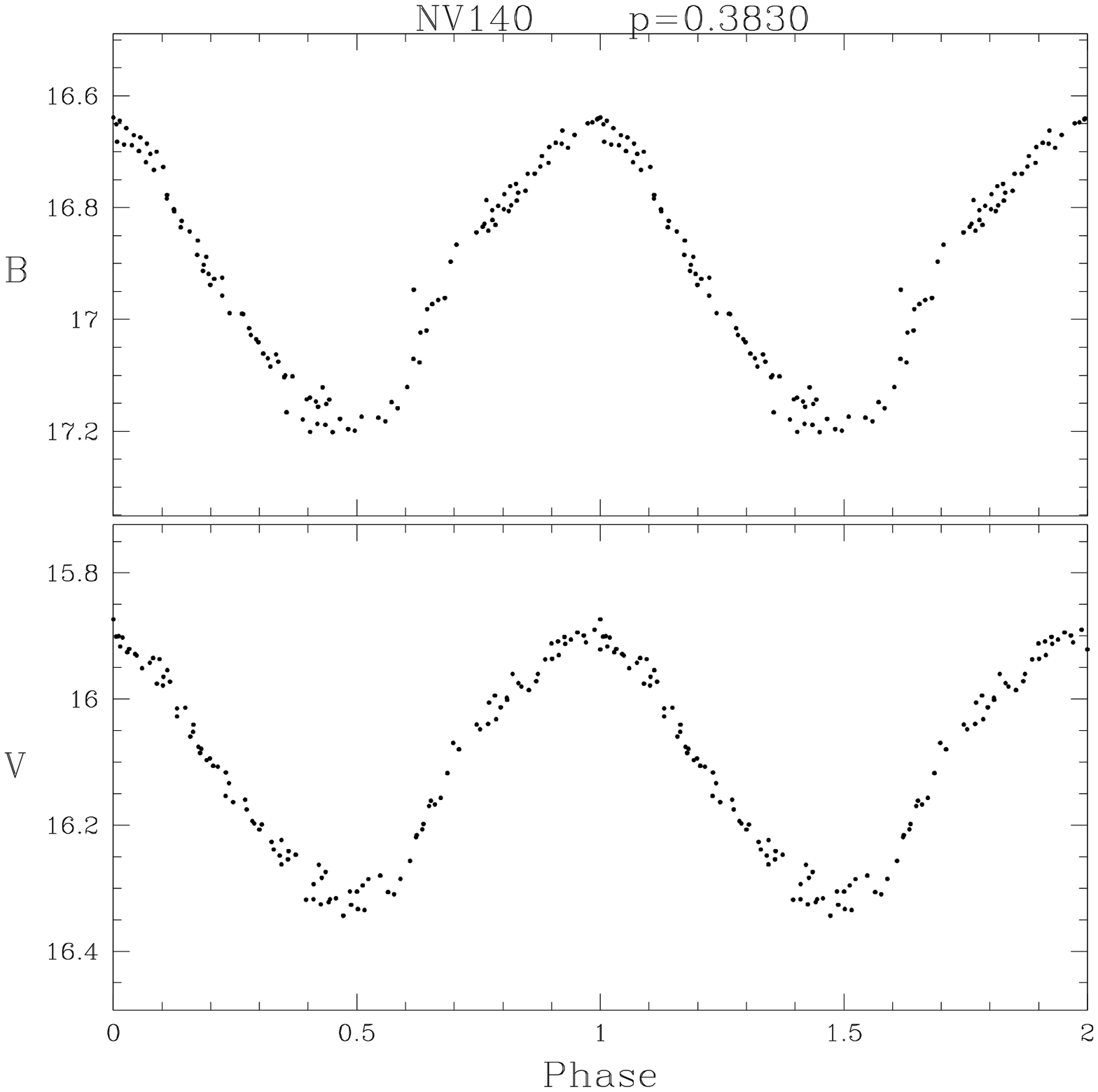}
\caption{Sample light curves for newly-discovered c-type RR Lyrae stars. (The full set 
can be found in the electronic version of this paper.)}
\label{fig:rrc}
  \end{center}
\end{figure}

\begin{figure}[ht]
  \begin{center}
\includegraphics[width=.325\textwidth]{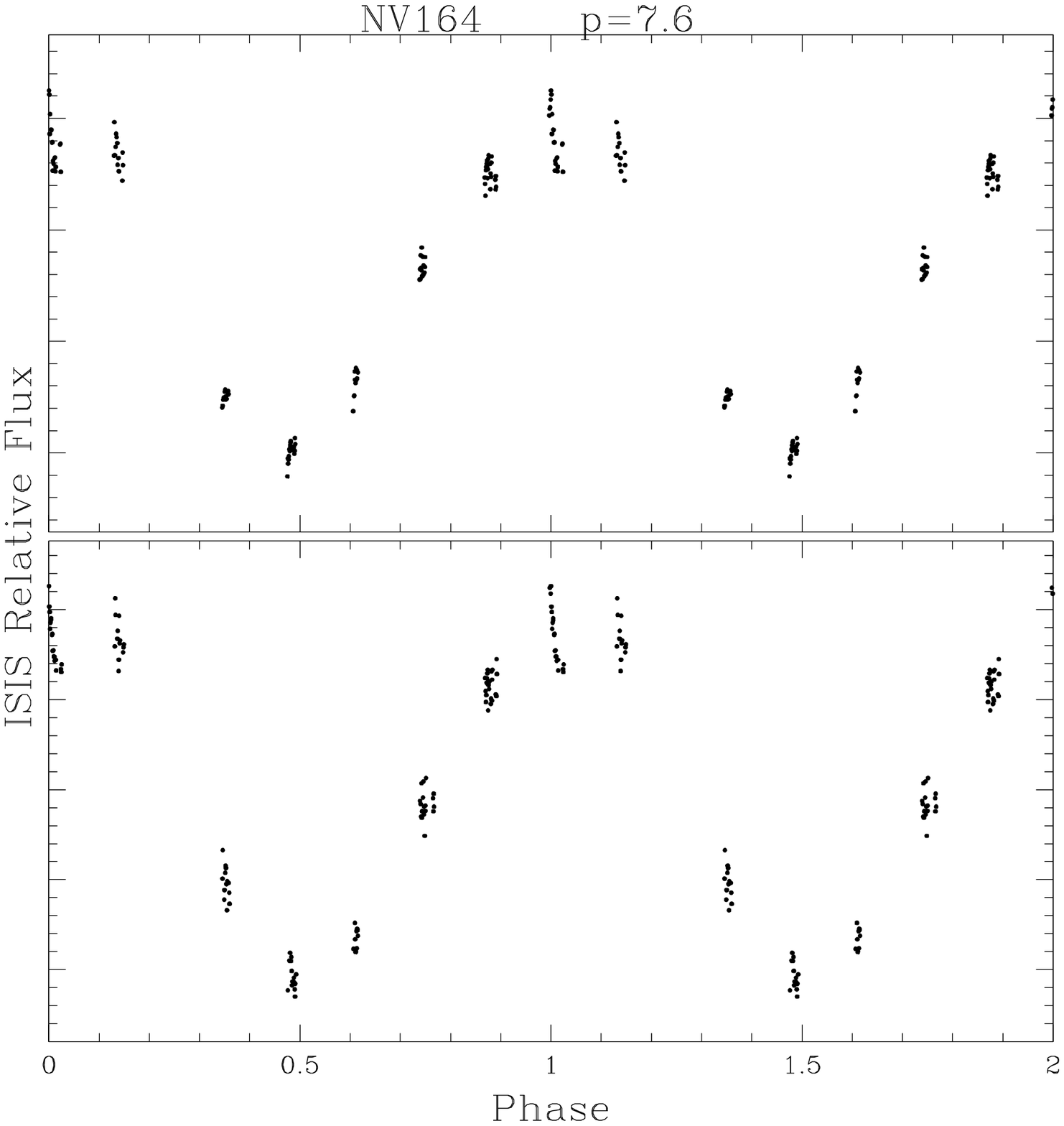}
\includegraphics[width=.325\textwidth]{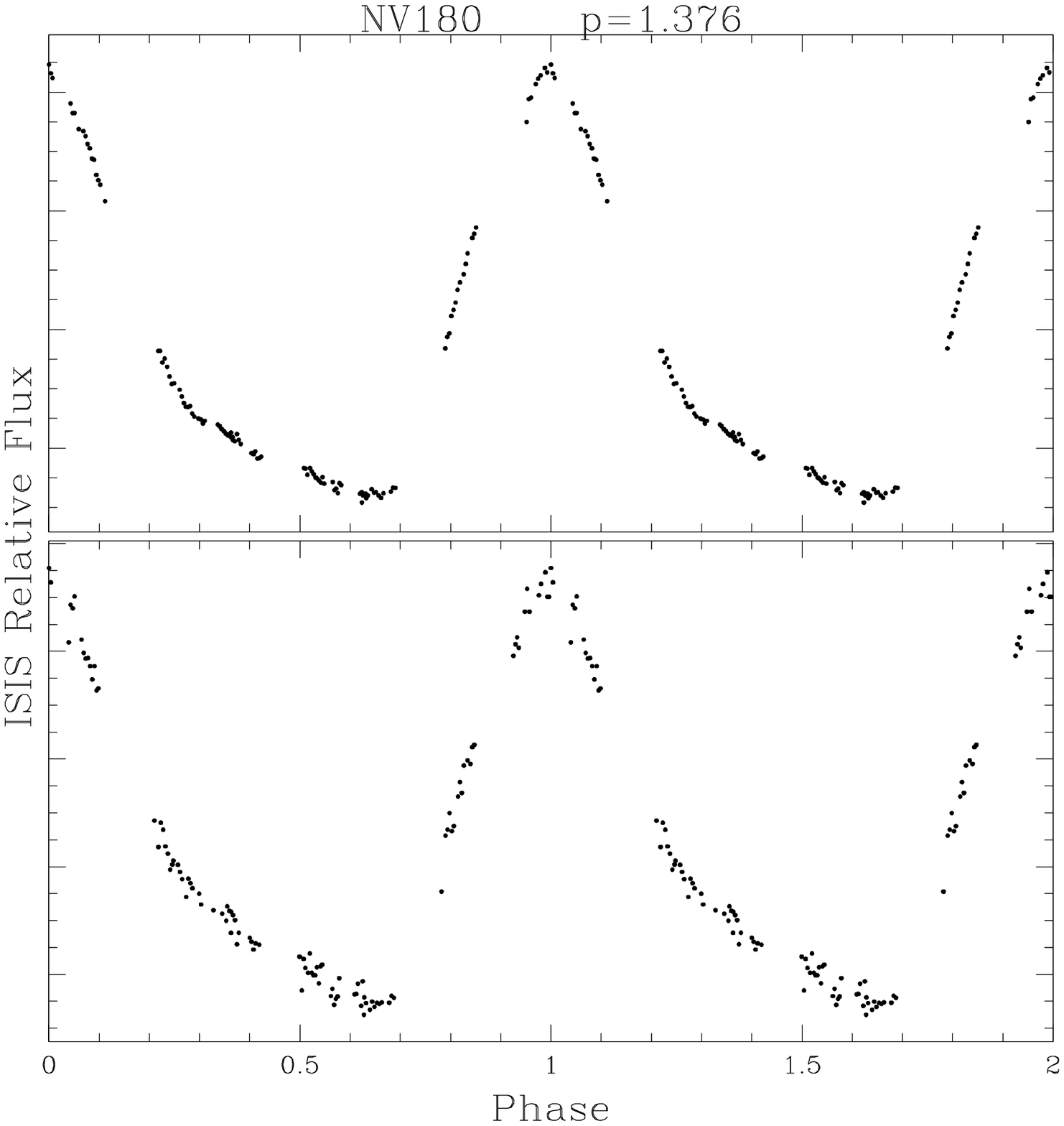}
\caption{Sample light curves for newly-discovered type II Cepheids. (The full set 
can be found in the electronic version of this paper.)}
\label{fig:cepII}
  \end{center}
\end{figure}

\begin{figure}[ht]
  \begin{center}
\includegraphics[width=.325\textwidth]{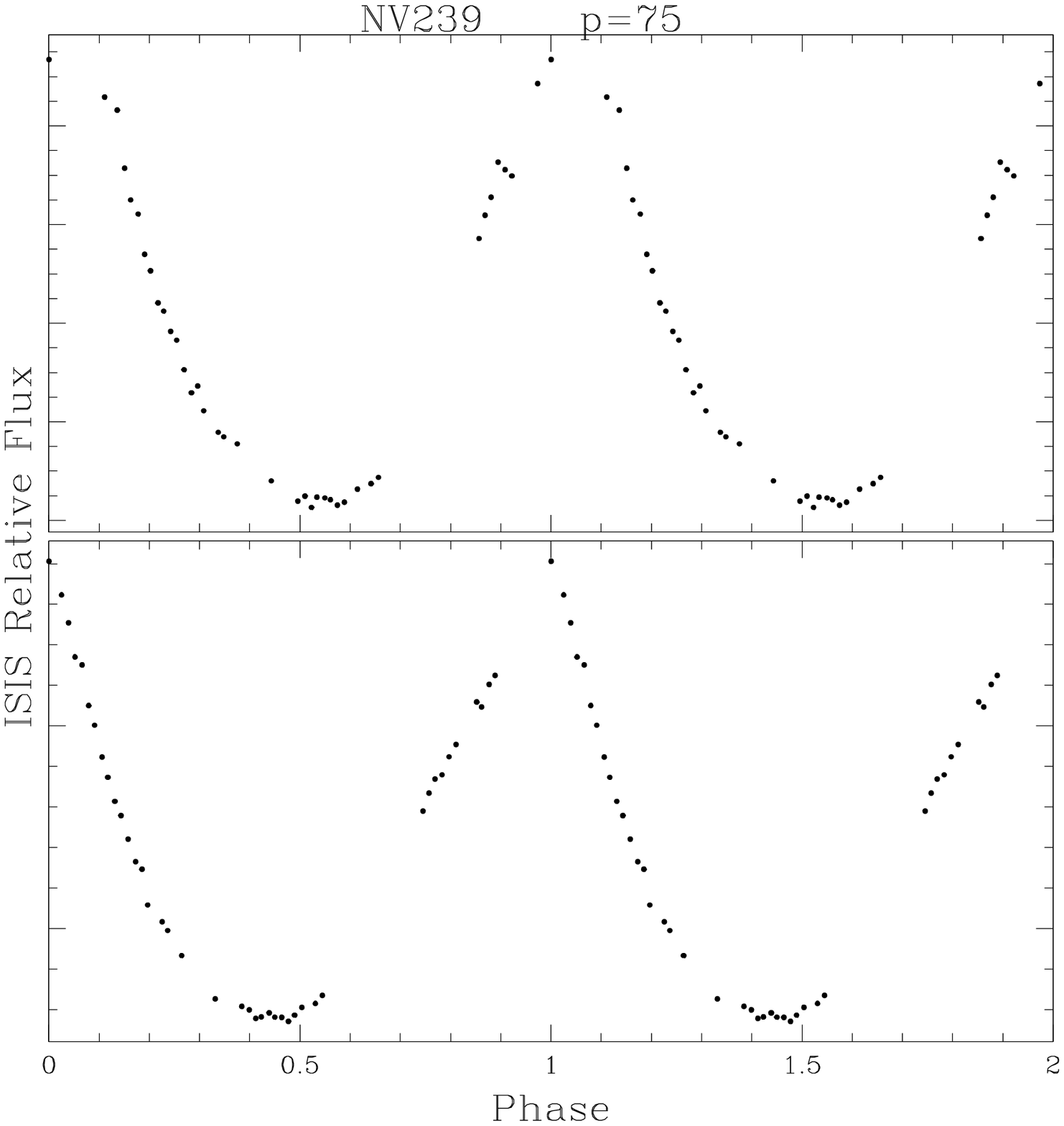}
\includegraphics[width=.325\textwidth]{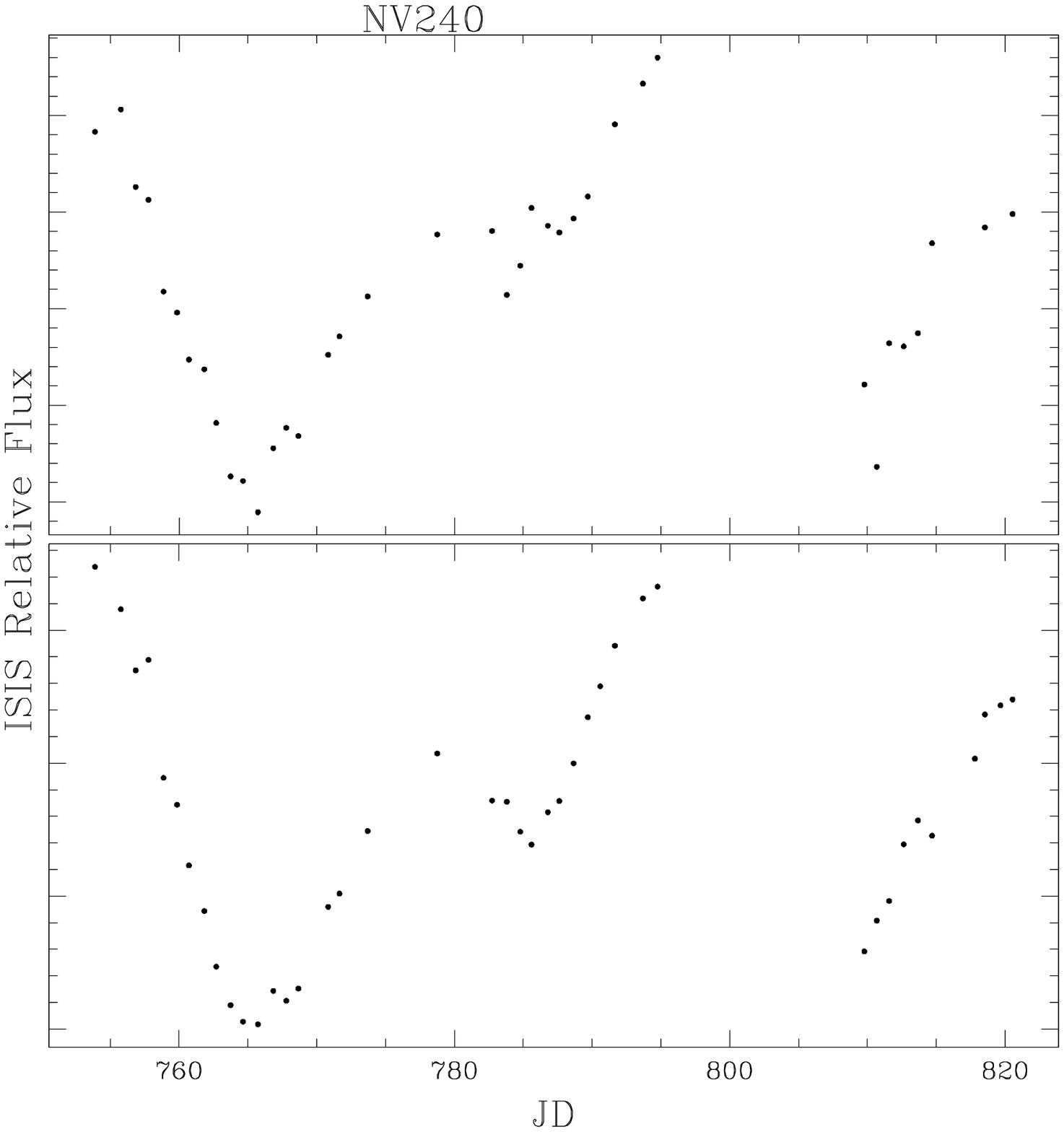}
\caption{Sample light curves for newly-discovered long-period variables. (The full set 
can be found in the electronic version of this paper.)}
\label{fig:lpv}
  \end{center}
\end{figure}

\begin{figure}[ht]
  \begin{center}
\includegraphics[width=.325\textwidth]{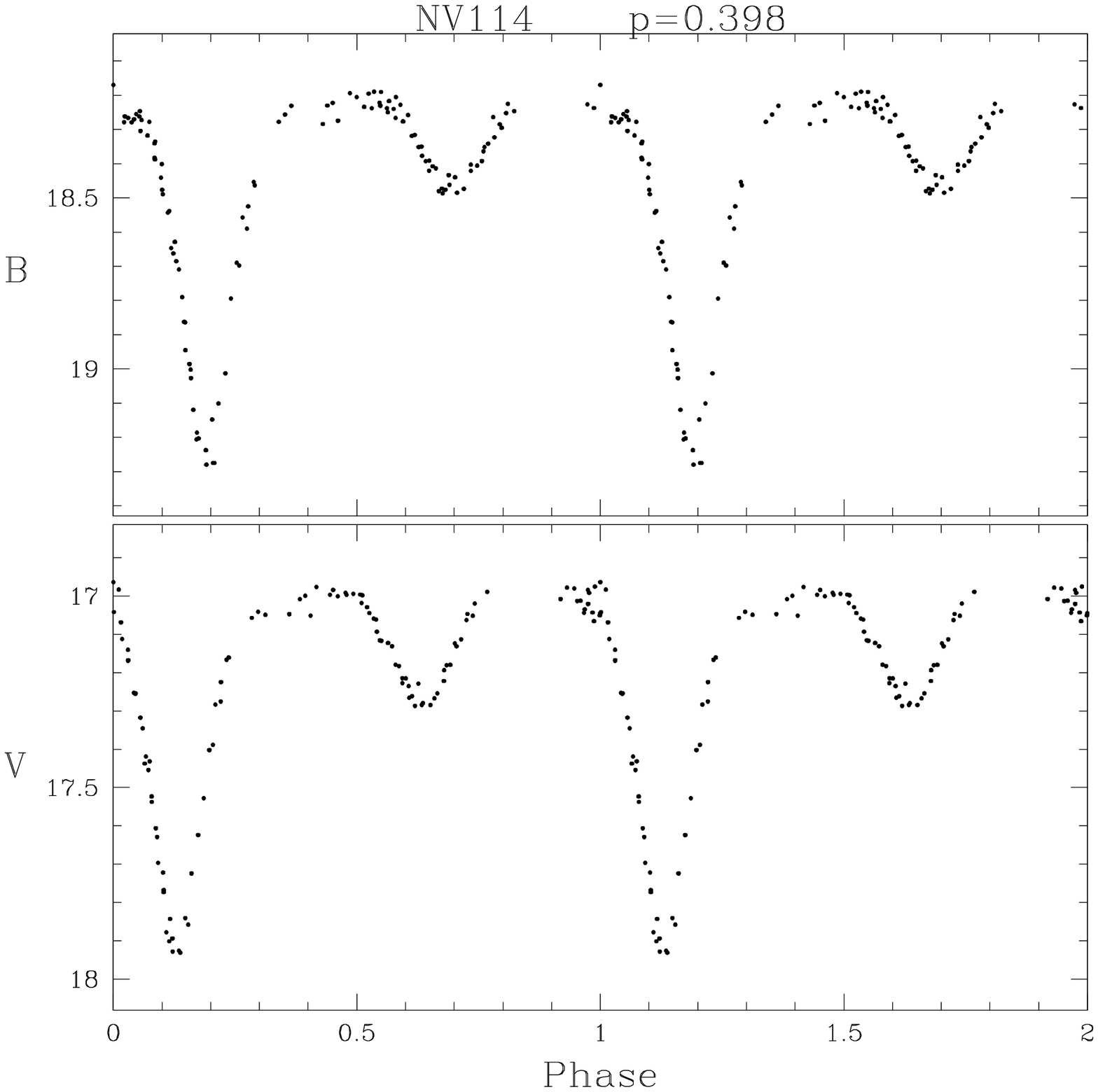}
\caption{Light curves for NV114, a newly discovered eclipsing binary.}
\label{fig:eb}
  \end{center}
  \end{figure}

\begin{figure}[ht]
  \begin{center}
\includegraphics[width=.325\textwidth]{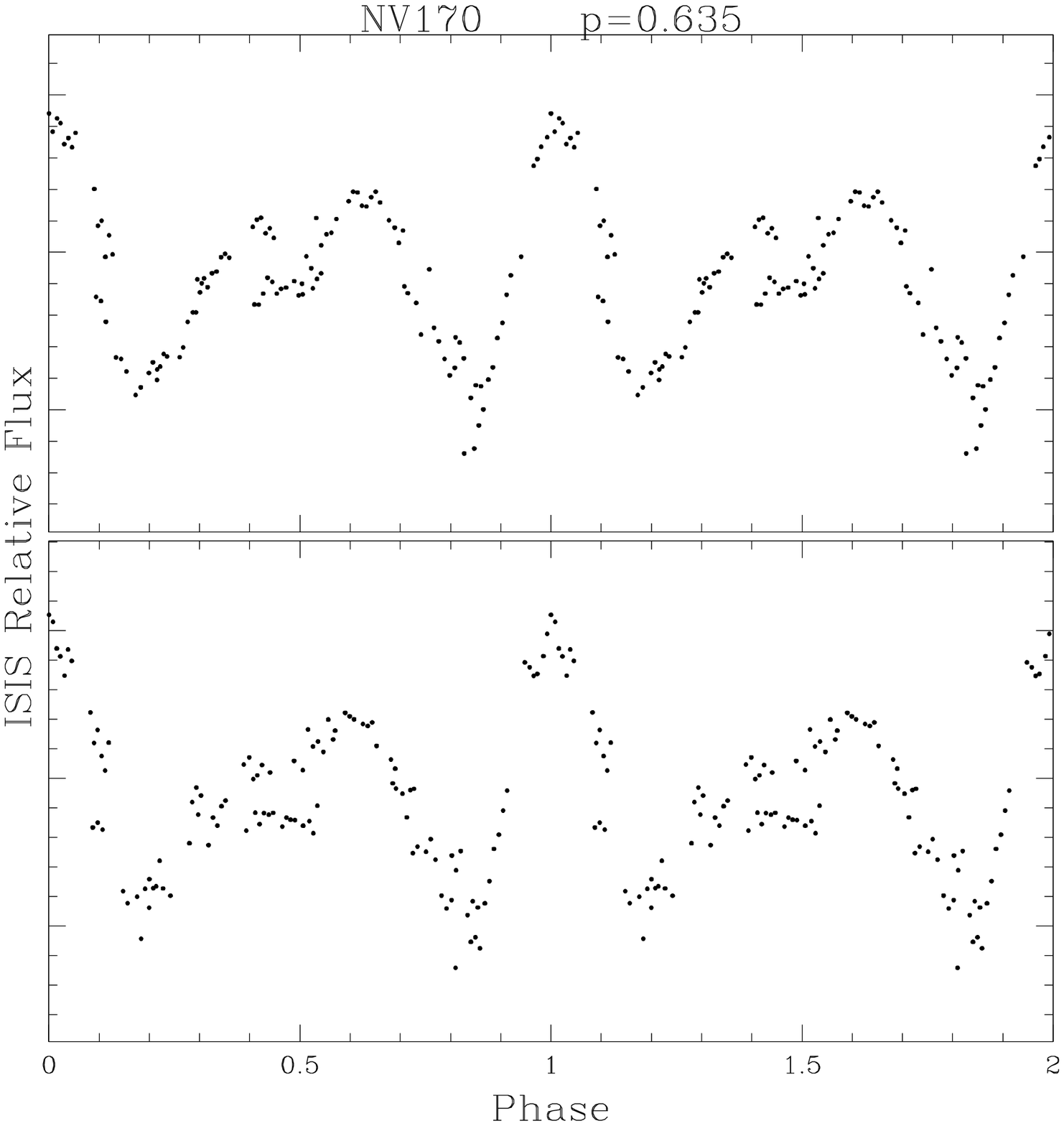}
\includegraphics[width=.325\textwidth]{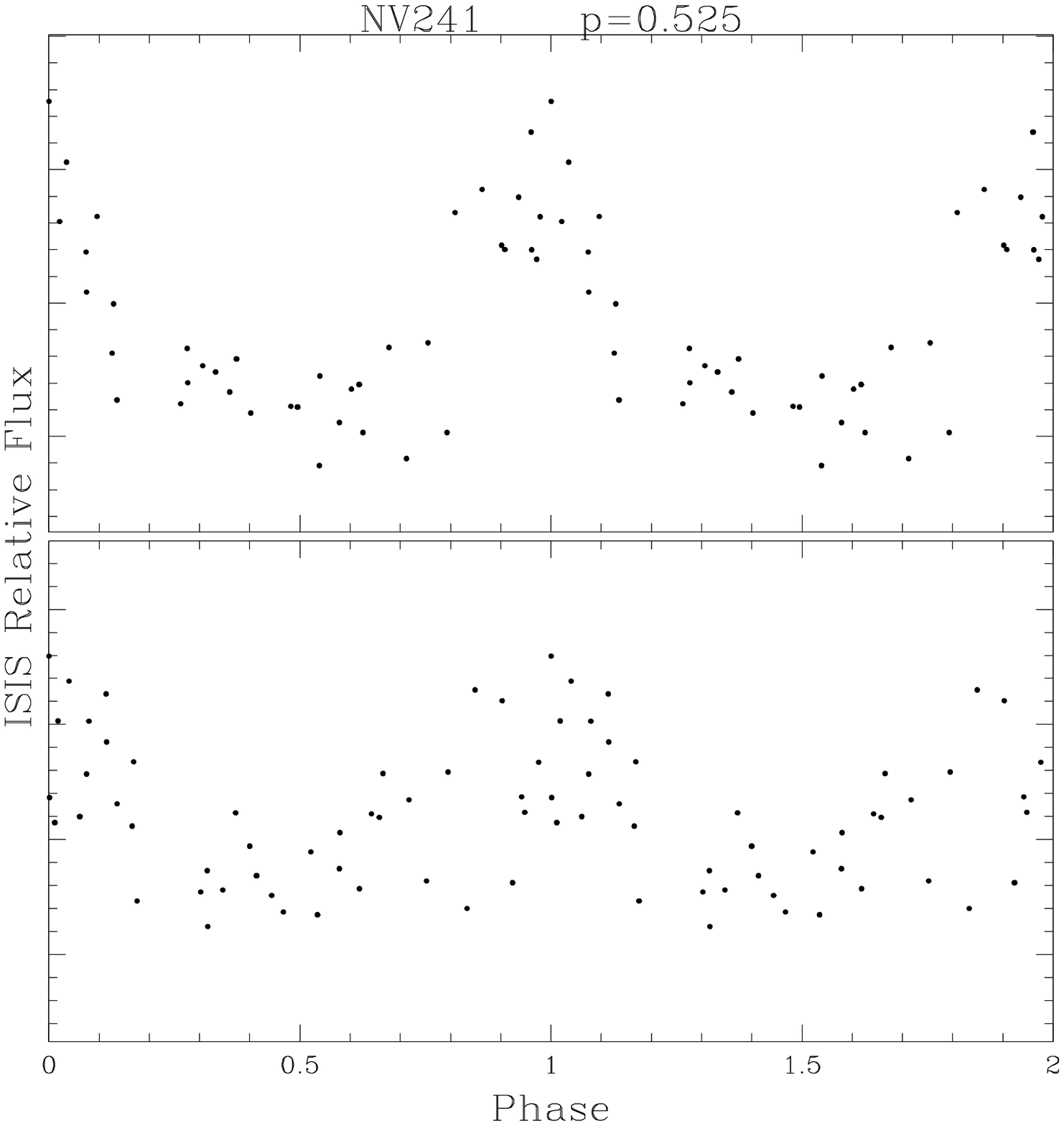}
\caption{Sample light curves for newly-discovered variable stars whose classification remains 
unclear. (The full set can be found in the electronic version of this paper.)}
\label{fig:nc}
  \end{center}
  \end{figure}

\end{document}